\documentclass[12pt,fleqn]{article}
\usepackage{mcite,amsfonts}
\usepackage{style}
\usepackage{mfpic}
\usepackage{diagrams}
\setdefaultlengths
\begin{document}
\begin{titlepage}
\pub{09}{96}
\title{Higher order QCD corrections to the transverse and longitudinal
  fragmentation functions in electron-positron annihilation}
    {P.J. Rijken and W.L. van Neerven}
    {September 1996}
\abstract{
We present the calculation of the order \alphastwo~corrections to the coefficient functions
contributing to the longitudinal ($F_L(x,Q^2)$) and transverse fragmentation functions
($F_T(x,Q^2)$) measured in electron-positron annihilation. The effect of these higher
order QCD corrections on the behaviour of the fragmentation functions and the
corresponding longitudinal ($d\sigma_L(x,Q^2)/dx$) and transverse cross sections
($d\sigma_T(x,Q^2)/dx$) are studied. In particular we investigate the dependence of the
above quantities on the mass factorization scale ($M$) and
the various parameterizations chosen for the parton fragmentation densities $D_p^H(x,M^2)$
($p=q,g$; $H=\pi^\pm, K^\pm, P, \bar{P}$). Our analysis reveals that the order \alphastwo~
contributions to $F_L(x,Q^2)$ are large whereas these contributions to $F_T(x,Q^2)$ are
small. From the above fragmentation functions one can also compute the integrated
cross sections $\sigma_L$ and $\sigma_T$ in an independent
way. The sum $\sigma_{\rm tot}=\sigma_L+\sigma_T$, corrected up to order \alphastwo, agrees
with the well known result in the literature providing us with an independent check
an our calculations.
}
\end{titlepage}
%

\section{Introduction}
Semi-leptonic processes represented by electron-positron annihilation into hadrons,
deep inelastic lepton-hadron scattering and the Drell-Yan process have provided us with
the most valuable testing grounds for perturbative quantum chromodynamics (QCD).
Perturbative calculations in next-to-leading order, and in some cases even to higher order,
give
a good explanation of numerous quantities measured in various experiments \cite{Zer92}. The
reason for these successes originates from the experimental as well as theoretical
characteristics of the above reactions. From the experimental viewpoint semi-leptonic
reactions
provide us with an overwhelming amount of data and in the case of electron-positron
annihilation into hadrons and deep inelastic lepton-hadron scattering the background
is fully under control. Therefore the systematical and statistical errors are very small.
From the theoretical viewpoint we want to mention the following features. First, the Born
approximation to semi-leptonic cross sections is of purely electroweak origin so that
it is independent of the strong coupling constant \alphas. Since the electroweak
standard model is tested up to about a few promille by the LEP1-experiments \cite{Mni96}
each deviation
from the Born approximation is due to the strong interactions. Second, if one limits oneself
to the computation of semi-inclusive or inclusive quantities, like structure functions
or total cross sections, the final hadronic state is completely integrated over and we do not
have to care about problems as jet definition or hadronization effects.
The third feature is that it is possible to extend the calculation of the QCD corrections
to the above integrated quantities beyond next-to-leading order. Examples are the
order \alphastwo~contributions to the coefficient functions corresponding to the
Drell-Yan cross section $d\sigma/dQ^2$ \cite{Mat90,-Ham91,*Nee92} and the deep inelastic
structure functions
$F_k(x,Q^2)$ \cite{Zij92,-Zij92b} where $Q^2$ denotes the virtuality of the
electroweak vector bosons
$\gamma,Z,W$. Order $\alpha_s^3$ corrections are even known for sum rules
$\int_0^1\,dx\,x^{n-1}F_k(x,Q^2)$ ($n\leq 10$) \cite{Lar91,*Lar91b,*Lar94b}
and the total cross section $\sigma_{\rm tot}(e^+e^-\rightarrow {\rm ``hadrons"}$)
\cite{Gor91,*Sur91,*Che94,*Lar94,-Lar95}. The reason that these
higher order corrections are much easier to compute than those encountered in e.g.
hadron-hadron collisions (except for the Drell-Yan process) can be attributed to the
simplicity of the phase space integrals and the virtual corrections appearing in semi-leptonic
processes. Moreover if one integrates in the latter processes over the total hadronic
state one can use alternative methods to evaluate the Feynman diagrams
(see e.g. \cite{Che81,*Kaz88}), which are not applicable to
hadron-hadron reactions or to more exclusive semi-leptonic processes.\\
In the past the order \alphastwo~contributions to the coefficient functions have been
calculated for the Drell-Yan cross section $d\sigma/dQ^2$ \cite{Mat90,-Ham91,*Nee92}
and the deep inelastic
structure functions $F_k(x,Q^2)$ \cite{Zij92,-Zij92b}. However the same corrections
were not computed
for the fragmentation functions showing up in the process $e^+e^-\rightarrow H+``X"$ where
$H$ is the detected hadron ($H=\pi^\pm, K^\pm, P, \bar{P}$) and $``X"$ stands for any
inclusive hadronic state. These corrections are needed
because of the large amount of data which have collected over the past ten years.
The above process has been studied over a wide range of energies of many different
$e^+e^-$-colliders. Data have been collected from DASP ($\sqrt{s}=5.2$ GeV) \cite{Bra79},
ARGUS ($\sqrt{s}=10$ GeV) \cite{Alb89}, TASSO ($\sqrt{s}=22, 35, 45$ GeV)
\cite{Bra89,-Bra90}, MARK II \cite{Pet88} and TPC/$2\gamma$ ($\sqrt{s}=29$ GeV)
\cite{Aih88}, CELLO ($\sqrt{s}=35$ GeV) \cite{Pod}, AMY ($\sqrt{s}=55$ GeV)
\cite{Kum90,*Li90} and the LEP experiments DELPHI \cite{Abr93}, ALEPH \cite{Bus95,-Bus95b},
OPAL \cite{Ake93,-Ake94,Ake95} ($\sqrt{s}=91.2$ GeV). In particular the last two
experiments
found a discrepancy between the measured longitudinal fragmentation function
$F_L(x,Q^2)$ and its theoretical prediction computed up to order \alphas.
We want to fill in this gap in our knowledge by presenting the order
\alphastwo~contributions to the longitudinal ($F_L(x,Q^2)$) and transverse fragmentation
($F_T(x,Q^2)$) functions and discuss their phenomenological implications. The coefficient
functions corrected up to order \alphastwo~are already computed for $F_L(x,Q^2)$ and
can be found in recent work \cite{Rij96}. Here we want to add the order \alphastwo~
contributions to the coefficient functions corresponding to $F_T(x,Q^2)$
which are much more complicate. The order
\alphastwo~corrections to the asymmetric fragmentation function will be postponed to a future
publication. Although a complete next-to-next-to-leading (NNLO) order analysis of the transverse
(and also asymmetric) fragmentation function is not possible, since we do not know the
three-loop order timelike DGLAP \cite{Gri72,*Alt77,*Dok77} splitting functions,
one can still study the effect
of the order \alphastwo~corrected coefficient functions. Furthermore one can obtain the
transverse cross section $\sigma_T$ for which analysis the DGLAP splitting functions
are not needed so that the former is factorization scheme independent.
The sum of the transverse ($\sigma_T$) and the longitudinal ($\sigma_L$)
cross sections yield $\sigma_{\rm tot}(e^+e^-\rightarrow {\rm ``hadrons"})$.
It turns out that the $\sigma_{\rm tot}$ presented in this paper is in agreement
with the order \alphastwo~corrected result quoted in the literature
\cite{Che79,*Din79,*Cel80} providing us with a very strong  check on our calculations.\\
This paper will be organized as follows. In section 2 we introduce our notations of the
fragmentation functions and the corresponding cross sections. In section 3 we give an outline
of the calculations of the parton subprocesses contributing to the process
$e^+e^-\rightarrow H+``X"$ up to order \alphastwo. In section 4 we perform the
renormalization and mass
factorization of the partonic quantities providing us with the longitudinal and
transverse coefficient functions. The discussion of our results will be presented
in section 5 and a comparison with data coming from recent and past experiments on
electron-positron annihilation will be made. The long expressions obtained for the order
\alphastwo~corrected coefficient functions are presented in the \MS-scheme and the
A- (annihilation) scheme in appendix A and appendix B respectively.
%

\newpage
\section{Single particle inclusive cross sections}
In this paper we want to study the QCD corrections to the single particle inclusive process
\begin{equation}
  \label{eq:2_1}
  e^+ + e^- \rightarrow \gamma,Z \rightarrow H + ``X",
\end{equation}
where $``X"$ denotes any inclusive final hadronic state and $H$ represents either a
specific charged outgoing hadron or a sum over all charged hadron species. The unpolarized
differential cross section of the above process is given by \cite{Nas94,Web94}
\begin{equation}
  \label{eq:2_2}
  \frac{d^2\sigma^H}{dx\,d\cos\theta} = \frac{3}{8}(1+\cos^2\theta)\frac{d\sigma_T^H}{dx}
  + \frac{3}{4}\sin^2\theta\frac{d\sigma_L^H}{dx} + \frac{3}{4}\cos\theta
  \frac{d\sigma_A^H}{dx}.
\end{equation}
The Bj{\o}rken scaling variable $x$ is defined by
\begin{equation}
  \label{eq:2_3}
  x = \frac{2pq}{Q^2},\hspace{8mm} q^2 = Q^2 > 0,\hspace{8mm} 0 < x \leq 1,
\end{equation}
where $p$ and $q$ are the four-momenta of the produced particle $H$ and the virtual vector
boson ($\gamma$, $Z$) respectively. In the centre of mass (CM) frame of the electron-positron
pair the variable $x$ can be interpreted as a fraction of the total CM energy carried away
by the hadron $H$. The variable $\theta$ denotes the angle of emission of particle $H$
with respect to the electron beam direction in the CM frame. The transverse, longitudinal
and asymmetric cross sections in \eref{eq:2_2} are defined by $\sigma_T^H$, $\sigma_L^H$, and
$\sigma_A^H$ respectively. The latter only shows up if the intermediate vector boson is given
by the $Z$-boson and is absent in purely electromagnetic annihilation.\\
In the QCD improved parton model which describes the production of the parton denoted by $p$
and its subsequent fragmentation into hadron $H$, the cross sections $\sigma_k^H$ ($k=T,L,A$)
can be expressed as follows
\begin{eqnarray}
  \label{eq:2_4}
  \lefteqn{\frac{d\sigma_k^H}{dx} = \int_x^1\,\frac{dz}{z}\BigLeftHook
    \sigma_{\rm tot}^{(0)}(Q^2)\,\BigLeftBrace
    D_\Singlet^H\left(\frac{x}{z},M^2\right)
    \mathbb{C}_{k,q}^\Singlet(z,Q^2/M^2)
    + D_g^H\left(\frac{x}{z},M^2\right)\cdot}\nonumber\\[2ex]
  && {}\cdot\mathbb{C}_{k,g}^\Singlet(z,Q^2/M^2)\BigRightBrace
  +\sum_{p=1}^{n_f}\,\sigma_p^{(0)}(Q^2)\,
  D_{\NonSinglet,p}^H\left(\frac{x}{z},M^2\right)
  \mathbb{C}_{k,q}^\NonSinglet(z,Q^2/M^2)\BigRightHook,
\end{eqnarray}
for $k=T,L$. In the case of the asymmetric cross section we have
\begin{equation}
  \label{eq:2_5}
  \frac{d\sigma_A^H}{dx} = \int_x^1\,\frac{dz}{z}\BigLeftHook\sum_{p=1}^{n_f}\,
  A_p^{(0)}(Q^2)
  D_{A,p}^H\left(\frac{x}{z},M^2\right)
  \mathbb{C}_{A,q}^{\NonSinglet}(z,Q^2/M^2)\BigRightHook.
\end{equation}
In the formulae \eref{eq:2_4} and \eref{eq:2_5} we have introduced the following notations.
The function $D_g^H(z,M^2)$ denotes the gluon fragmentation density corresponding to the
hadron of species $H$. The same notation holds for the quark and anti-quark fragmentation
densities which are given by $D_p^H(z,M^2)$ and $D_{\bar{p}}^H(z,M^2)$ respectively.
Further we have defined the singlet (S) and non-singlet (NS, A) combinations of quark
fragmentation densities. They are given by
\begin{eqnarray}
  \label{eq:2_5_1}
  \lefteqn{
    D_\Singlet^H(z,M^2) = \frac{1}{n_f}\,\sum_{p=1}^{n_f}\left(D_p^H(z,M^2) +
    D_{\bar{p}}^H(z,M^2)\right),}\\[2ex]
  \label{eq:2_5_2}
  \lefteqn{
    D_{\NonSinglet,p}^H(z,M^2) = D_p^H(z,M^2) + D_{\bar{p}}^H(z,M^2) - D_\Singlet^H(z,M^2),}\\[2ex]
  \label{eq:2_5_3}
  \lefteqn{
    D_{A,p}^H(z,M^2) = D_p^H(z,M^2) - D_{\bar{p}}^H(z,M^2).}
\end{eqnarray}
The index $p$ stands for the quark species and $n_f$ denotes the number of light flavours.
Assuming that the charm and the bottom quark can be treated as massless we can put
$n_f=5$ and the indices $p=1,2,3,4,5$ stand for $p=u,d,s,c,b$. Further the variable $M$
appearing in $D_p^H(z,M^2)$ stands for the mass factorization scale which for
convenience has been put equal to the renormalization scale. The pointlike cross
section of the process
\begin{equation}
  \label{eq:2_6}
  e^+ + e^- \rightarrow p + \bar{p},
\end{equation}
which shows up in \eref{eq:2_4} is equal to
\begin{eqnarray}
  \label{eq:2_7}
  \lefteqn{\sigma_p^{(0)}(Q^2) = \frac{4\pi\alpha^2}{3Q^2}N\,\BigLeftHook e_\ell^2 e_p^2 +
  \frac{2Q^2(Q^2-M_Z^2)}{\left|Z(Q^2)\right|^2}\,e_\ell e_p C_{V,\ell} C_{V,p} +
  \frac{(Q^2)^2}{\left|Z(Q^2)\right|^2}\cdot}\nonumber\\[2ex]
  &&\cdot(C_{V,\ell}^2 + C_{A,\ell}^2)(C_{V,p}^2 + C_{A,p}^2)\BigRightHook,\\[2ex]
  \label{eq:2_7_1}
  \lefteqn{
    \sigma_{\rm tot}^{(0)}(Q^2) = \sum_{p=1}^{n_f}\,\sigma_p^{(0)}(Q^2),}
\end{eqnarray}
with
\begin{equation}
  \label{eq:2_8}
  Z(Q^2) = Q^2 - M_Z^2 + iM_Z\Gamma_Z.
\end{equation}
Here $N$ stands for the number of colours ($N=3$) and $M_Z$, $\Gamma_Z$ denote the mass
and width of the $Z$-boson respectively. For the latter we have used the narrow width
approximation. Furthermore we have neglected all quark masses in \eref{eq:2_7}.
The charges of the lepton and the up and down quarks are given by
\begin{equation}
  \label{eq:2_9}
  e_\ell = -1,\hspace{8mm} e_u = \frac{2}{3},\hspace{8mm} e_d = -\frac{1}{3}.
\end{equation}
The vector- and axial-vector coupling constants of the $Z$-boson to the lepton and quarks
are equal to
\begin{equation}
  \label{eq:2_10}
  \begin{array}{ll}
    C_{A,\ell} = \displaystyle\frac{1}{2\sin2\theta_W}, &
    C_{V,\ell} = -C_{A,\ell}\,(1-4\sin^2\theta_W),\\[2ex]
    C_{A,u} = -C_{A,d} = -C_{A,\ell}, & \\[2ex]
    C_{V,u} = C_{A,\ell}\,\displaystyle(1-\frac{8}{3}\sin^2\theta_W), &
    C_{V,d} = -C_{A,\ell}\,\displaystyle(1-\frac{4}{3}\sin^2\theta_W),
  \end{array}
\end{equation}
where $\theta_W$ denotes the Weinberg angle.\\
The electroweak coupling constants also appear in the asymmetry factor
$A_p^{(0)}$ \eref{eq:2_5} which is given by
\begin{eqnarray}
  \label{eq:2_11}
  \lefteqn{A_p^{(0)} = \frac{4\pi\alpha^2}{3Q^2}N\,\BigLeftHook \frac{2Q^2(Q^2-M_Z^2)}
  {\left|Z(Q^2)\right|^2}\,e_\ell e_p C_{A,\ell} C_{A,p}}\nonumber\\[2ex]
  && {}+
  4\frac{(Q^2)^2}{\left|Z(Q^2)\right|^2}\,C_{A,\ell} C_{A,p} C_{V,\ell} C_{V,p}\BigRightHook.
\end{eqnarray}
The QCD corrections in \eref{eq:2_4}, \eref{eq:2_5} are described by the coefficient
functions $\mathbb{C}_{k,\ell}^r$
($k=T,L,A$; $\ell=q,g$) which can be distinguished with respect to the flavour group
$SU(n_f)$ in a singlet ($r=\Singlet$) and a non-singlet part ($r=\NonSinglet$). They depend on the
factorization scale $M$ and in order \alphastwo~on the number of flavours $n_f$.
As will be shown later on the
gluonic coefficient function only receives contributions from flavour singlet channel partonic
subprocesses so that we can drop the superscript $S$ on $\mathbb{C}_g$. However the quark
coefficient functions can be of flavour singlet as well as flavour non-singlet origin.
Up to first order in the strong coupling constant \alphas~it turns out that
$\mathbb{C}_{k,q}^\NonSinglet=\mathbb{C}_{k,q}^\Singlet$. However in higher order both quantities
start to deviate from each other. Hence we define the purely singlet coefficient function
$\mathbb{C}_{k,q}^\PureSinglet$ via
\begin{equation}
  \label{eq:2_12}
  \mathbb{C}_{k,q}^\Singlet = \mathbb{C}_{k,q}^\NonSinglet + \mathbb{C}_{k,q}^\PureSinglet.
\end{equation}
Like $\mathbb{C}_{k,g}$ the purely singlet coefficient function only receives contributions
from the flavour singlet channel partonic subprocesses which for the first time show up
in order $\alpha_s^2$.\\
Using charge conjugation invariance of the strong interactions
one can show that $\mathbb{C}_{A,q}^\NonSinglet = -\mathbb{C}_{A,\bar{q}}^\NonSinglet$
and $\mathbb{C}_{A,q}^\PureSinglet = \mathbb{C}_{A,g} = 0$. This implies
that to $\sigma_A^H$ \eref{eq:2_5} only non-singlet channel partonic subprocesses can
contribute. Another important property of the coefficient function is that they do not
depend on the probe $\gamma$ or $Z$ or on the electroweak couplings given in \eref{eq:2_9},
\eref{eq:2_10} so that one can extract the overall pointlike cross section $\sigma_p^{(0)}$
\eref{eq:2_7} or the asymmetry factor $A_p^{(0)}$ \eref{eq:2_11}. However this is only
true if all quark masses are equal to zero and if one sums over all quark members in one
family provided the latter appear in the inclusive state of the partonic subprocess (see
section 4).\\
From \eref{eq:2_2} we can derive the total hadronic cross section
\begin{equation}
  \label{eq:2_13}
  \sigma_{\rm tot}(Q^2) = \frac{1}{2}\sum_H\,\int_0^1dx\,\int_{-1}^1 d\cos\theta\,
  \left(x\frac{d^2\sigma^H}{dx\,d\cos\theta}\right) = \sigma_T(Q^2) + \sigma_L(Q^2),
\end{equation}
with
\begin{equation}
  \label{eq:2_14}
  \sigma_k(Q^2) = \frac{1}{2}\sum_H\,\int_0^1dx\,x\frac{d\sigma_k^H}{dx},
  \hspace{4mm} (k=T,L,A),
\end{equation}
where one has summed over all types of outgoing hadrons $H$. Hence we obtain the result
\begin{equation}
  \label{eq:2_15}
  \sigma_{\rm tot}(Q^2) = R_{ee}\,\sigma_{\rm tot}^{(0)}(Q^2),
\end{equation}
where $R_{ee}$ represents the QCD corrections to the pointlike total cross section
$\sigma_{\rm tot}^{(0)}(Q^2)$. At this moment the perturbation series of $R_{ee}$
is already known up to order $\alpha_s^3$ \cite{Gor91,*Sur91,*Che94,*Lar94,-Lar95}.
Up to order $\alpha_s^2$ it reads \cite{Che79,*Cel80}
\begin{eqnarray}
  \label{eq:2_16}
  \lefteqn{R_{ee} = 1 + \frac{\alpha_s}{4\pi}\,C_F\,[3] + \left(\frac{\alpha_s}{4\pi}\right)^2
  \BigLeftHook C_F^2\left\{-\frac{3}{2}\right\} + C_A C_F \BigLeftBrace-11\ln\frac{Q^2}{M^2}
  - 44 \zeta(3)}\nonumber\\[2ex]
  && {}+ \frac{123}{2}\BigRightBrace + n_f C_F T_f\left\{4\ln\frac{Q^2}{M^2} + 16 \zeta(3) - 22\right\}
  \BigRightHook.
\end{eqnarray}
In section 4 we also want to present the coefficient functions $\mathbb{C}_{k,\ell}$
up to order $\alpha_s^2$ and show that they lead to the same $R_{ee}$ as calculated
in the literature (see section 5).\\
Finally we also define the transverse, longitudinal and asymmetric fragmentation functions
$F_k^H(x,Q^2)$\footnote{Notice that we make a distinction in nomenclature between
the fragmentation densities $D_p^H$ and the fragmentation functions $F_k^H$.}
\begin{equation}
  \label{eq:2_17}
  F_k^H(x,Q^2) = \frac{1}{\sigma_{\rm tot}^{(0)}(Q^2)}\,\frac{d\sigma_k^H}{dx},
  \hspace{4mm} k=(T,L,A).
\end{equation}
Further the total fragmentation function is given by
\begin{equation}
  \label{eq:2_17_1}
  F^H(x,Q^2) = F_L^H(x,Q^2) + F_T^H(x,Q^2).
\end{equation}
In the case the virtual photon dominates the annihilation process \eref{eq:2_1} one
observes that, apart from the charge squared $e_p^2$ in \eref{eq:2_7} ($p=u,d$),
the above structure
functions are just the timelike photon analogues of the ones measured in deep inelastic
electron-proton scattering.
When the $Z$-boson contributes we will define in section 4 for each combination of the
electroweak coupling constants
in \eref{eq:2_7} a separate structure function. However for the discussion
of our results in section 5
this distinction will not be needed.
%

\newpage
\section{Fragmentation coefficient functions in $e^+\,e^-$ annihilation up to
  order \alphastwo}
In this section we will give an outline of the calculation of the order
\alphastwo~corrections to the fragmentation coefficient functions. The procedure
is analogous to
the one presented for the calculation of the Drell-Yan process in
\cite{Mat90,*Ham91,*Nee92} and the
deep inelastic lepton-hadron reaction in \cite{Zij92}. The coefficient functions originate
from the following reaction
\begin{equation}
  \label{eq:3_1}
  V(q) \rightarrow ``p(k_0)" + p_1(k_1) + p_2(k_2) + \cdots + p_\ell(k_\ell),
\end{equation}
where $V=\gamma,Z$ and $``p"$ denotes the detected parton which fragments into
the hadron $H$. The process \eref{eq:3_1} is inclusive with respect to the partons
$p_i$ ($i=1,2,\ldots,\ell$) so that one has to integrate over all momenta indicated
by $k_i$. Notice that the first part of reaction \eref{eq:2_1} i.e.
$e^+\,e^-\rightarrow V$
is not relevant for the determination of the coefficient function.\\
Up to order \alphastwo~all parton subprocesses represented by \eref{eq:3_1} are
listed in table~\ref{tab:t1}. From the amplitude $M_\mu(\ell)$ describing process
\eref{eq:3_1} one obtains the parton structure tensor (indicated by a hat)
\begin{equation}
  \label{eq:3_2}
  \hat{W}_{\mu\nu}^{(V,V')}(p,q) = \frac{z^{n-3}}{4\pi}\,\sum_{\ell=1}^\infty\,
  \int{\rm dPS}^{(\ell)}\,M_\mu^V(\ell)\,M_\nu^{V'}(\ell)^\ast.
\end{equation}
Here $\int\,{\rm dPS}^{(k)}$ denotes the $k$-body phase space integral defined by
\begin{eqnarray}
  \label{eq:3_3}
  \lefteqn{
    \int\,{\rm dPS}^{(\ell)} = \left\{\prod_{j=1}^\ell\,\int\,\frac{d^nk_j}
    {(2\pi)^{n-1}}\,\delta^+(k_j^2)\right\}\,(2\pi)^n\,\delta^{(n)}
    (q-k_0-\prod_{i=1}^\ell\,k_i),}\\[2ex]
  \label{eq:3_4}
  \lefteqn{
    \delta^+(k_j^2) = \theta(k_j^0)\,\delta(k_j^2),}
\end{eqnarray}
and $\mu$ and $\nu$ stand for the Lorentz indices of the vector bosons $V$ and $V'$
respectively with $V=\gamma,Z$ and $V'=\gamma,Z$. Further we have defined the partonic
scaling variable
\begin{equation}
  \label{eq:3_4_1}
  z = \frac{2k_oq}{Q^2},
\end{equation}
and the factor $z^{n-3}$ in \eref{eq:3_2} originates from the $n$-dimensional phase
space of the detected parton $p$ \eref{eq:3_1}. It appears in the definition
of the cross sections $d\hat{\sigma}_{k,p}/dz$ which are the partonic analogues
of the hadronic cross sections in \eref{eq:2_2}. The former are proportional
to the functions $\hat{\cal F}_{k,p}$ defined below.\\
To regularize the ultraviolet (U), infrared(IR) and collinear (C) divergences showing
up in expression \eref{eq:3_2} we have chosen the method of $n$-dimensional
regularization. Therefore the phase space integral in \eref{eq:3_3} is generalized
to $n$ dimensions so that the above divergences show up as pole terms of the type
$(1/\varepsilon)^m$ with $\varepsilon=n-4$. The calculation of the matrix elements
$M_\mu(k)\,M_\nu(k)^\ast$ was performed in $n$ dimensions using the algebraic
manipulation program FORM \cite{Ver}. After having computed the traces we have to integrate
the matrix elements over all internal loop and final state momenta where the momentum
$k_0$ of the detected parton is kept fixed. In this paper we take all partons to be
massless.
The case of massive quarks is discussed in \cite{Nas94} where their contributions are
presented up to order \alphas.\\
The parton structure tensor in \eref{eq:3_2} can be also written as
\begin{equation}
  \label{eq:3_5}
  \hat{W}_{\mu\nu}^{(V,V')}(k_0,q) = \frac{1}{4\pi}\,\sum_{\ell=1}^\infty\,\int\,{\rm dPS}^{(\ell)}
  \,\langle0|\,\hat{J}_\mu^{(V)}(0)\,|p,\{p_\ell\}\rangle\,\langle p,\{p_\ell\}|\,
  \hat{J}_\nu^{(V')}(0)\,|0\rangle,
\end{equation}
where $\hat{J}_\mu^{(V)}$ is the electroweak partonic current corresponding to the vector
boson $V$. Using Lorentz covariance and CP invariance \eref{eq:3_5} can be
written as follows
\begin{eqnarray}
  \label{eq:3_6}
  \lefteqn{
    \hat{W}_{\mu\nu}^{(V,V')}(k_0,q) = (v_{q_1}^{(V)}v_{q_2}^{(V')} + a_{q_1}^{(V)}
    a_{q_2}^{(V')})\,\BigLeftHook ({k_0}_\mu - \frac{k_0q}{q^2}q_\mu)({k_0}_\nu
    - \frac{k_0q}{q^2}q_\nu)
    \,\frac{q^4}{(k_0q)^3}\cdot}\nonumber\\[2ex]
  && \cdot\hat{\cal F}_{L,p}(z,Q^2) - \BigLeftParen g_{\mu\nu} - \frac{1}{k_0q}
  (k_0^\mu q^\nu+q^\mu k_0^\nu) + \frac{q^2}{(k_0q)^2}{k_0}_\mu {k_0}_\nu\BigRightParen\,
  \frac{q^2}{2k_0q}\cdot\nonumber\\[2ex]
  && \cdot\hat{\cal F}_{T,p}(z,Q^2)\BigRightHook
  - (v_{q_1}^{(V)}a_{q_2}^{(V')} +
  a_{q_1}^{(V)}v_{q_2}^{(V')})\,i\epsilon_{\mu\nu\alpha\beta}k_0^\alpha q^\beta
  \frac{q^2}{2(k_0q)^2}\,\hat{\cal F}_{A,p}(z,Q^2).
\end{eqnarray}
We will call $\hat{\cal F}_{k,p}(z,Q^2)$ ($p=q,g$) the parton fragmentation functions
which describe the Born reaction plus the higher order QCD corrections represented
by the parton subprocesses in table \ref{tab:t1}. The vector and axial-vector couplings
of the quark $q$
interacting with the vector boson $V$ are given by $v_q^{(V)}$ and $a_q^{(V)}$
respectively. In the standard model they read
\begin{equation}
  \label{eq:3_7}
  \begin{array}{ll}
    v_u^{(\gamma)} = \frac{2}{3}, & a_u^{(\gamma)} = 0,\\[2ex]
    v_d^{(\gamma)} = -\frac{1}{3}, & a_d^{(\gamma)} = 0,\\[2ex]
    v_u^{(Z)} = \frac{1}{2} - \frac{4}{3}\sin^2\theta_W, & a_u^{(Z)} = \frac{1}{2},\\[2ex]
    v_d^{(Z)} = -\frac{1}{2} + \frac{2}{3}\sin^2\theta_W, & a_u^{(Z)} = -\frac{1}{2}.
  \end{array}
\end{equation}
As we have already mentioned in section 2 below \eref{eq:2_12}
in the case of massless quarks the electroweak
factors can be completely factorized out of the radiative corrections according to
\eref{eq:3_6} so that $\hat{\cal F}_{k,p}$ ($k=T, L, A$) do not depend on them.
Therefore we can also put them all equal to $1/\sqrt{2}$ without affecting the parton
fragmentation functions. Hence the latter are obtained via the following
projections
\begin{eqnarray}
  \label{eq:3_8}
  \lefteqn{
    \hat{\cal F}_{T,p}(z,Q^2) = \frac{1}{n-2}\,\left(-\frac{2k_0q}{q^2}\hat{W}^\mu_{\cov\mu}
    - \frac{2}{k_0q}k_0^\mu k_0^\nu\hat{W}_{\mu\nu}\right),}\\[2ex]
  \label{eq:3_9}
  \lefteqn{
    \hat{\cal F}_{L,p}(z,Q^2) = \frac{1}{k_0q}k_0^\mu k_0^\nu\hat{W}_{\mu\nu},}\\[2ex]
  \label{eq:3_10}
  \lefteqn{
    \hat{\cal F}_{A,p}(z,Q^2) = -\frac{2}{q^2}\frac{1}{(n-2)(n-3)}i\epsilon
    ^{\mu\nu\alpha\beta} {k_0}_\alpha q_\beta\hat{W}_{\mu\nu},}
\end{eqnarray}
where according to the prescription in \cite{Lar93} we have contracted the Levi-Civita
tensors $\epsilon^{\mu\nu\alpha\beta}$ in $n$-dimensions. In this paper we will
compute the transverse and longitudinal fragmentation functions only and leave the
calculation of the asymmetric fragmentation to a future publication. The
computation of the latter involves the prescription of the $\gamma_5$-matrix and the
Levi-Civita tensor in $n$-dimensions which is quite intricate.\\
We will now discuss the QCD corrections order by order in perturbation theory. In
zeroth order in \alphas~(see fig.~\ref{fig:f2}) we obtain the simple parton model results
\begin{equation}
  \label{eq:3_11}
  \hat{\cal F}_{T,q}^{(0)}=\delta(1-z),\hspace*{4mm}
  \hat{\cal F}_{T,g}^{(0)}=0;\hspace*{8mm}
  \hat{\cal F}_{L,q}^{(0)} = \hat{\cal F}_{L,g}^{(0)} = 0.
\end{equation}
The first order corrections denoted by $\hat{\cal F}_{k,i}^{(1)}$ ($k=T, L; i=q, g$) have been
calculated in the literature \cite{Bai79,Alt79b}. In the case of $n$-dimensional regularization
they are computed up to finite terms in the limit
$\varepsilon\rightarrow 0$ and can be found in \cite{Nas94,Alt79b}. Since the mass factorization has to
be carried out up to order \alphastwo~one also needs those terms in
$\hat{\cal F}_{k,i}^{(1)}(z,Q^2,\varepsilon)$ which are proportional to $\varepsilon$.
Therefore we have repeated the calculation of the graphs in fig.~\ref{fig:f3} and
~\ref{fig:f4} and the results can be presented in the following form
\begin{eqnarray}
  \label{eq:3_12}
  \lefteqn{
    \hat{\cal F}_{L,q}^{(1)} = \left(\frac{\hat{\alpha}_s}{4\pi}\right)\,S_\varepsilon\,
    \left(\frac{Q^2}{\mu^2}\right)^{\varepsilon/2}\,\BigLeftHook
      \bar{c}_{L,q}^{(1)} + \varepsilon\,a_{L,q}^{(1)}\BigRightHook,}\\[2ex]
  \label{eq:3_13}
  \lefteqn{
    \hat{\cal F}_{T,q}^{(1)} = \left(\frac{\hat{\alpha}_s}{4\pi}\right)\,S_\varepsilon\,
    \left(\frac{Q^2}{\mu^2}\right)^{\varepsilon/2}\,\BigLeftHook
      P_{qq}^{(0)}\frac{1}{\varepsilon} +
      \bar{c}_{T,q}^{(1)} + \varepsilon\,a_{T,q}^{(1)}\BigRightHook,}\\[2ex]
  \label{eq:3_14}
  \lefteqn{
    \hat{\cal F}_{L,g}^{(1)} = \left(\frac{\hat{\alpha}_s}{4\pi}\right)\,S_\varepsilon\,
    \left(\frac{Q^2}{\mu^2}\right)^{\varepsilon/2}\,\BigLeftHook
      \bar{c}_{L,g}^{(1)} + \varepsilon\,a_{L,g}^{(1)}\BigRightHook,}\\[2ex]
  \label{eq:3_15}
  \lefteqn{
    \hat{\cal F}_{T,g}^{(1)} = \left(\frac{\hat{\alpha}_s}{4\pi}\right)\,S_\varepsilon\,
    \left(\frac{Q^2}{\mu^2}\right)^{\varepsilon/2}\,\BigLeftHook
      2P_{gq}^{(0)}\frac{1}{\varepsilon} +
      \bar{c}_{T,g}^{(1)} + \varepsilon\,a_{T,g}^{(1)}\BigRightHook.}
\end{eqnarray}
The pole terms $1/\varepsilon$ stand for the collinear divergence in the final state
and $\mu^2$ and $S_\varepsilon$ are artefacts of $n$-dimensional regularization. The mass
parameter $\mu$ originates from the dimensionality of the gauge coupling constant in
$n$ dimensions and should not be confused with the renormalization scale $R$ and the
mass factorization scale $M$. The spherical factor $S_\varepsilon$ is defined by
\begin{equation}
  \label{eq:3_16}
  S_\varepsilon = \exp\left[ \frac{1}{2}\varepsilon(\gamma_E - \ln 4\pi)\right].
\end{equation}
Further $\hat{\alpha}_s$ denotes the bare coupling constant and $P_{ij}^{(0)}$ (
$i,j=q, \bar{q}, g$) stand for the lowest order contribution to the
DGLAP splitting functions \cite{Gri72,*Alt77,*Dok77}. Using our convention they are presented in
eqs. (2.13)-(2.16) of \cite{Mat90,*Ham91,*Nee92}. Notice that in lowest order there is no difference
in the expressions for $P_{ij}^{(0)}$ found for the deep inelastic structure
functions (spacelike process) and those appearing in the fragmentation functions
(timelike process). In next-to leading order the DGLAP splitting functions are different for
spacelike and timelike processes as will be shown later on.\\
The coefficients $\bar{c}_{k,i}^{(1)}$, presented in the \MS-scheme,
are already calculated in the literature \cite{Bai79,Alt79b} (see also appendix A).
Furthermore we also have to compute the coefficients $a_{k,i}^{(1)}$
(proportional to $\varepsilon$), since they are needed for the mass factorization
which has to be carried out up to order \alphastwo. The results are
\begin{eqnarray}
  \label{eq:3_17}
  \lefteqn{
    a_{L,q}^{(1)} = C_F\,\left\{-1 + \ln(1-z) + 2\ln z\right\},}\\[2ex]
  \label{eq:3_18}
  \lefteqn{
    a_{T,q}^{(1)} = C_F\,\BigLeftBrace D_2(z) - \frac{3}{2}D_1(z) + \BigLeftParen
    \frac{7}{2}-3\zeta(2)\BigRightParen D_0(z)
    - \frac{1}{2}(1+z)\ln^2(1-z)}\nonumber\\[2ex]
  && {}+ 2\frac{1+z^2}{1-z}\,\ln z\ln(1-z) + 2\frac{1+z^2}{1-z}\,
  \ln^2z - 3\frac{1}{1-z}\,\ln z + \frac{3}{2}(1-z)\ln(1-z)\nonumber\\[2ex]
  &&{}+ 3(1-z)\ln z
  - \frac{3}{2} + \frac{5}{2}z + \frac{3}{2}(1+z)\zeta(2) + \delta(1-z)(9-\frac{33}{4}\zeta(2))
  \BigRightBrace,\\[2ex]
  \label{eq:3_19}
  \lefteqn{
    a_{L,g}^{(1)} = C_F\,\left\{ 4\frac{1-z}{z}(\ln(1-z) + 2\ln z - 2)\right\},}\\[2ex]
  \label{eq:3_20}
  \lefteqn{
    a_{T,g}^{(1)} = C_F\,\BigLeftBrace(\frac{2}{z}-2+z)(\ln^2(1-z) + 4\ln z\ln(1-z) + 4\ln^2 z
    - 3\zeta(2))}\nonumber\\[2ex]
  && {}- 4\frac{1-z}{z}(\ln(1-z) + 2\ln z - 3) + 4z\BigRightBrace.
\end{eqnarray}
The calculation of the order \alphastwo~corrections proceeds in the following way. First we
have the two-loop
corrections to the quark-vector boson vertex represented by the graphs in fig.~\ref{fig:f5} which
only contribute to $\hat{\cal F}_{T,q}^{(2)}$. The two-loop vertex correction  can be found
in eq. (2.49) of \cite{Mat88a} (see also appendix A of \cite{Mat89}). The result agrees with the
one
quoted in \cite{Kra87}. Notice that the first graph in fig.~\ref{fig:f5} does not contribute
for $V=\gamma$ because of Furry's theorem. It only plays a role in the case $V=Z$ provided
one sums over all flavours in a quark family in order to cancel the anomaly which originates
from the triangle fermion sub-loop. Since all quarks are massless the final result for this
graph is zero too even in the case of $V=Z$.\\
Next we have to compute the one-loop virtual corrections to the radiative process in
fig.~\ref{fig:f4} which contribute to $\hat{\cal F}_{k,q}^{(2)}$ as well as
$\hat{\cal F}_{k,g}^{(2)}$ ($k=T, L$). The corresponding graphs are shown in fig.~\ref{fig:f6}.
Notice that we have omitted the diagrams with the self energy insertions on the external
quark and gluon legs. Their contributions vanish because of the method of $n$-dimensional
regularization and the on-mass shell conditions $k_0^2 = k_\ell^2 = 0$. Another vanishing
contribution happens for the last graph in fig.~\ref{fig:f6} when $V=\gamma$ because of
Furry's theorem. In the case of $V=Z$ it only contributes when the quarks are massive. However
here one has to sum over all members of a quark family in order to cancel the anomaly
originating from the triangle fermion loop.\\
The amplitude of the parton subprocesses in fig.~\ref{fig:f6} will be denoted by $M(2)$
(see \eref{eq:3_1} where $\ell=2$). The momenta of the incoming vector boson $V$ and
the outgoing partons are parameterized like
\begin{eqnarray}
  \lefteqn{q = \sqrt{s}\,(1,{\bf 0}_{n-1}),}\nonumber\\[2ex]
  \lefteqn{k_0 = \frac{s-s_{12}}{2\sqrt{s}}\,(1,1,{\bf 0}_{n-2}),}\nonumber\\[2ex]
  \label{eq:3_21}
  \lefteqn{k_1 = \frac{s-s_2}{2\sqrt{s}}\,(1,\cos\theta_1,\sin\theta_1,{\bf 0}_{n-3}),
    \hspace*{8mm}k_2=q-k_0-k_1,}
\end{eqnarray}
where ${\bf 0}_n$ stands for the $n$-dimensional null vector.
The phase space integral in \eref{eq:3_2}, \eref{eq:3_3} becomes
\begin{eqnarray}
  \lefteqn{
    \int\,{\rm dPS}^{(2)}\,\left|M(2)\right|^2 = \frac{1}{8\pi}\,\frac{1}
    {\Gamma(1+\frac{1}{2}\varepsilon)}\,\frac{1}{(4\pi)^{\varepsilon/2}}\,
    s^{\varepsilon/2}(1-z)^{\varepsilon/2}\cdot}\nonumber\\[2ex]
  \label{eq:3_22}
  && \cdot\int_0^1dy\,y^{\varepsilon/2}(1-y)^{\varepsilon/2}\,\left|M(2)\right|^2. \\[2ex]
\end{eqnarray}
Here we have defined the Lorentz invariants
\begin{equation}
  \label{eq:3_23}
  s = Q^2,\hspace*{4mm}s_1=(k_0+k_1)^2,\hspace*{4mm}s_2=(k_0+k_2)^2,\hspace*{4mm}
  s_{12}=(k_1+k_2)^2,
\end{equation}
with $s=s_1+s_2+s_{12}$. The parameterization of \eref{eq:3_22} follows from momentum
conservation and the on-shell condition $k_0^2=k_\ell^2=0$. Hence we get
\begin{eqnarray}
  \label{eq:3_24}
  \lefteqn{
    \cos\theta_1=\frac{s_2s_{12}-s_1s}{(s-s_{12})(s-s_2)},\hspace*{4mm}
    s_1=z(1-y)s,\hspace*{4mm}s_{12}=(1-z)s,\hspace*{4mm}s_2=zys.}\nonumber\\[2ex]
  &&
\end{eqnarray}
The Feynman integrals corresponding to the one-loop graphs which contribute to $M(2)$ in
\eref{eq:3_22} contain loop momenta in the numerator. They can be reduced to scalar one-loop
integrals using an $n$-dimensional extension of the reduction program in \cite{Pas79,*Bee89b}.
The expressions
for these scalar integrals which are valid for all $n$ can be found in appendix D of \cite{Mat89}.
The phase space integrals which emerge from the computation of
$\left|M(2)\right|^2$ are very numerous so that we cannot present them in this paper. They are
calculated algebraically using the program FORM \cite{Ver}.\\
The most difficult and laborious part of the calculation can be attributed to the parton
subprocesses \eref{eq:3_1} where one has to integrate over three partons in the final state
(see also table~\ref{tab:t1}). These parton subprocesses are depicted in figs.~\ref{fig:f7},
\ref{fig:f8} providing us with the amplitude $M(3)$ (see \eref{eq:3_1} where $\ell=3$). The
graphs in fig.~\ref{fig:f7} determine $\hat{\cal F}_{k,q}^{(2)}$ as well as
$\hat{\cal F}_{k,g}^{(2)}$ ($k=T, L$) whereas the graphs in fig.~\ref{fig:f8}, which only
contain quarks and anti-quarks in the final state, contribute to $\hat{\cal F}_{k,q}^{(2)}$
only. For the computation of the three body phase space integrals we choose the following
parameterization for the momenta of the virtual vector boson $V$ and the outgoing partons
(see \cite{Kra87})
\begin{eqnarray}
  \label{eq:3_25}
  \lefteqn{
    q = \sqrt{s}\,(1,{\bf 0}_{n-1}),}\nonumber\\[2ex]
  \lefteqn{
    k_0 = \left(\frac{s_{023}-s_{23}}{2\sqrt{s_{23}}}\right)(1,1,{\bf 0}_{n-2}),
    }\nonumber\\[2ex]
  \lefteqn{
    k_1 = \left(\frac{s_{123}-s_{23}}{2\sqrt{s_{23}}}\right)
    (1,\cos\chi,\sin\chi,{\bf 0}_{n-3}),
    }\nonumber\\[2ex]
  \lefteqn{
    k_2 = \frac{1}{2}\sqrt{s_{23}}(1,\cos\theta_1,\sin\theta_1\cos\theta_2,
    \sin\theta_1\sin\theta_2,{\bf 0}_{n-4}),}\nonumber\\[2ex]
  \lefteqn{
    k_3 = \frac{1}{2}\sqrt{s_{23}}(1,-\cos\theta_1,-\sin\theta_1\cos\theta_2,
    -\sin\theta_1\sin\theta_2,{\bf 0}_{n-4}),}
\end{eqnarray}
where we have defined the invariants
\begin{equation}
  \label{eq:3_26}
  s_{ij}=(k_i+k_j)^2,\hspace*{4mm}s_{ijm}=(k_i+k_j+k_m)^2,\hspace*{4mm}s=q^2.
\end{equation}
From momentum conservation and the on-mass shell conditions one can derive
\begin{equation}
  \label{eq:3_27}
  1-\cos\chi = \frac{2s_{23}(s+s_{23}-s_{023}-s_{123})}{(s_{023}-s_{23})(s_{123}-s_{23})}.
\end{equation}
The three-body phase space integral in \eref{eq:3_2}, \eref{eq:3_3} can be expressed as
\begin{eqnarray}
  \label{eq:3_28}
  \lefteqn{
    \int\,{\rm dPS}^{(3)}\,\left|M(3)\right|^2 = \frac{1}{2^8\pi^4}\,\frac{1}
    {\Gamma(1+\varepsilon)}\,\frac{1}{(4\pi)^\varepsilon}
    \,s^{1+\varepsilon}\,z^{1+\varepsilon/2}(1-z)^{1+\varepsilon}\cdot}\nonumber\\[2ex]
  && \cdot\int_0^1dy_1\,\int_0^1dy_2\,y_1^{1+\varepsilon}(1-y_1)^{\varepsilon/2}\,
  y_2^{\varepsilon/2}(1-y_2)^{\varepsilon/2}\,(1-y_2(1-z))^{-\varepsilon-2}\cdot\nonumber\\[2ex]
  && \cdot\int_0^\pi d\theta_1\,(\sin\theta_1)^{1+\varepsilon}
  \int_0^\pi d\theta_2\,(\sin\theta_2)^\varepsilon\,\left|M(3)\right|^2,
\end{eqnarray}
where the invariants in \eref{eq:3_26} depend on $z$, $y_1$, and $y_2$ in the following way
\begin{eqnarray}
  \label{eq:3_29}
  \lefteqn{
    s_{023} = \frac{y_1zs}{1-y_2(1-z)},\hspace*{4mm}
    s_{123} = (1-z)s,\hspace*{4mm}
    s_{23} = \frac{z(1-z)y_1y_2s}{1-y_2(1-z)},}\nonumber\\[2ex]
  && 2k_0q = zs,\hspace*{4mm} s_{01} = z(1-y_1)s.
\end{eqnarray}
Before we can perform the angular integrations the matrix element $\left|M(3)\right|^2$
has to be decomposed via partial fractioning in terms which have the general form
\begin{eqnarray}
  \label{eq:3_30}
  \lefteqn{
    T^{n_1n_2n_3n_4} = (s_{i_1j_1})^{n_1}\,(s_{i_2j_2})^{n_2}\,(s_{i_3j_3})^{n_3}\,
    (s_{i_4j_4})^{n_4},}\nonumber\\[2ex]
  \lefteqn{
    n_i = \cdots,2,1,0,-1,-2,\cdots.}
\end{eqnarray}
The decomposition can be done in such a way that one invariant e.g. $s_{i_1j_1}$ in the
product \eref{eq:3_30} depends on the polar angle $\theta_1$ whereas an other invariant e.g.
$s_{i_2j_2}$ contains the polar angle $\theta_1$ as well as the azimuthal angle $\theta_2$.
The remaining invariants i.e. $s_{i_3j_3}$ and $s_{i_4j_4}$ do not depend on the angles.\\
Sometimes it happens that the azimuthal angle $\theta_2$ also appears in $s_{i_1j_1}$. In this
case one has to rotate the frame in \eref{eq:3_25} so that $s_{i_1j_1}$ becomes independent
of the azimuthal angle. This is always possible because the phase space integral \eref{eq:3_4}
is Lorentz invariant. The angular integrals take the form
\begin{equation}
  \label{eq:3_31}
  I_\varepsilon^{(i,j)} = \int_0^\pi d\theta_1\,\int_0^\pi d\theta_2\,
  \frac{(\sin\theta_1)^{1+\varepsilon}\,(\sin\theta_2)^\varepsilon}
  {(a+b\cos\theta_1)^i\,(A+B\cos\theta_1+C\sin\theta_1\cos\theta_2)^j},
\end{equation}
where $a$, $b$, $A$, $B$, and $C$ are functions of the kinematical invariants $s$,
$s_{123}$, $s_{023}$, $s_{23}$ \eref{eq:3_29}. These integrals can be found in appendix C
of \cite {Bee89}. However they have to be extended by including terms which are
proportional to $\varepsilon^k$ where the degree $k$ has to be larger than the one
appearing in the integrals of \cite{Bee89}. This is necessary because these terms
contribute due to the appearance of high power singularities $(1/\varepsilon)^k$
in the phase space integral \eref{eq:3_28}. The $n$-dimensional expression for
\eref{eq:3_31} becomes very cumbersome if $a^2\neq b^2$ and $A^2\neq B^2+C^2$.
Fortunately this situation can be avoided when one chooses the frame presented
in \eref{eq:3_25}. In this frame the worst case is given by $a^2\neq b^2$,
$A^2=B^2+C^2$ or $a^2=b^2$, $A^2\neq B^2+C^2$. These type of integrals have to be
partially done by hand before one can algebraically evaluate expression \eref{eq:3_28}
using the program FORM \cite{Ver}. The angular integrals are easy to perform
when $a^2=b^2$ and $A^2=B^2+C^2$ because they can be expressed into a hypergeometric
function ${}_2F_1(\alpha,\beta;\gamma;x)$ \cite{Erd53}. Inserting the latter in
\eref{eq:3_28} the remaining integrations are then again performed using the
algebraic manipulation program FORM.\\
Finally we would like to comment on a special type of term appearing in the transverse
parton fragmentation function $\hat{\cal F}_{T,q}(z,Q^2,\varepsilon)$. They only show up
in the non-singlet part and the order $\alpha_s^m$ contribution takes the form
\begin{equation}
  \label{eq:3_32}
  \hat{\cal F}_{T,q}^{(m)}(z,Q^2,\varepsilon) = \sum_{\ell=-1}^{2m-1}\,
  (1-z)^{\frac{m}{2}\varepsilon-1}\,\frac{f_\ell(z)}{\varepsilon^\ell},
\end{equation}
where $f_\ell(1)$ is finite. These type of terms originate from gluon bremsstrahlung
(figs.~\ref{fig:f4},\ref{fig:f6},\ref{fig:f7}) and gluon  splitting into a quark--anti-quark
pair (fig.~\ref{fig:f8}). In the limit $z\rightarrow 1$ all gluons become soft and the
angle between the quark and anti-quark pair goes to zero (collinear emission). In the next
section expression \eref{eq:3_32} has to be convoluted with the so-called bare fragmentation
densities $\hat{D}_q^H(z)$ (for the definition see section 4) which yields the integral
\begin{equation}
  \label{eq:3_33}
  \sum_{\ell=-1}^{2m-1}\,\int_x^1dz\,\hat{D}_q^H\left(\frac{x}{z}\right)\,(1-z)^{\frac{m}{2}
    \varepsilon
    -1}\frac{f_\ell(z)}{\varepsilon^\ell}.
\end{equation}
Inspection of the above integral reveals that at $z=1$ one gets an additional pole term which
means that we also have to compute $f_{-1}(1)$. Therefore for $z=1$ the phase space
integrals \eref{eq:3_28} have to be computed even up to one order higher in powers of
$\varepsilon$ than is needed for those which are integrable at $z=1$. Since
$f_\ell(z)-f_\ell(1)$ is integrable at $z=1$ we can replace in \eref{eq:3_32}
$f_\ell(z)$ by $f(1)$ and one only has to consider the integral
\begin{equation}
  \label{eq:3_34}
  \hat{\cal F}_{T,q}^{(m)} = \sum_{\ell=-1}^{2m-1}\,\int_x^1dz\,\hat{D}_q^H\left(\frac{x}{z}
  \right)\,
  (1-z)^{\frac{m}{2}\varepsilon-1}\,\frac{f_\ell(1)}{\varepsilon^\ell},
\end{equation}
which can be written as
\begin{eqnarray}
  \label{eq:3_35}
  \lefteqn{
    \hat{\cal F}_{T,q}^{(m)} = \sum_{\ell=-1}^{2m-1}\BigLeftHook
    \int_x^1dz\,(1-z)^{\frac{m}{2}\varepsilon-1}\,\frac{f_\ell(1)}{\varepsilon^\ell}
    \left\{\hat{D}_q^H\left(\frac{x}{z}\right)-\hat{D}_q^H(x)\right\}}\nonumber\\[2ex]
  && {}+\frac{2}{m}\varepsilon^{-\ell-1}\hat{D}_q^H(x)f_\ell(1)(1-x)^{\frac{m}{2}\varepsilon-1}
  \BigRightHook.
\end{eqnarray}
If we define the distribution (see \cite{Alt79})
\begin{equation}
  \label{eq:3_35_1}
    {\cal D}_i(z) = \left(\frac{\ln^i(1-z)}{1-z}\right)_+,
\end{equation}
by
\begin{equation}
  \label{eq:3_36}
    \int_0^1dz\,{\cal D}_i(z)\,g(z) = \int_0^1dz\,\frac{\ln^i(1-z)}{1-z}\,(g(z)-g(1)),
\end{equation}
one can rewrite \eref{eq:3_35} in the following way
\begin{eqnarray}
  \label{eq:3_37}
  \lefteqn{
    \hat{\cal F}_{T,q}^{(m)} = \int_x^1dz\BigLeftHook\BigLeftBrace\sum_{\ell=0}^{2m-1}\,
    \frac{f_\ell(1)}{\varepsilon^\ell}\,\hat{D}_q^H\left(\frac{x}{z}\right)\,\sum_{i=0}^\ell\,
    \frac{1}{i!}\,\left(\frac{1}{2}m\varepsilon\right)^i\,{\cal D}_i(z)\BigRightBrace}
  \nonumber\\[2ex]
  && {}+ \hat{D}_q^H\left(\frac{x}{z}\right)\,\hat{\cal F}_{T,q}^{(m),{\rm soft}}(z)\BigRightHook,
\end{eqnarray}
where $\hat{\cal F}_{T,q}^{\rm soft}$ stands for the soft gluon bremsstrahlung contribution
which is given by (see the definition in \cite{Hum81})
\begin{equation}
  \label{eq:3_38}
  \hat{\cal F}_{T,q}^{(m),{\rm soft}} = \delta(1-z)\,\sum_{\ell=-1}^{2m-1}\,
  \frac{2}{m}\varepsilon^{-\ell-1}\,f_\ell(1).
\end{equation}
In order \alphastwo~($m=2$) the highest order pole term which can occur in \eref{eq:3_38}
is represented by $1/\varepsilon^4$. The latter is cancelled by similar terms originating
from the virtual gluon contributions given by the two-loop vertex corrections in
fig.~\ref{fig:f5}.
Finally we want to emphasize that the type of singular terms in \eref{eq:3_32}
only occur in $\hat{\cal F}_{T,q}^\NonSinglet$ and $\hat{\cal F}_{A,q}^\NonSinglet$
and are absent in
$\hat{\cal F}_{L,p}^\PureSinglet$ ($p=q,g$) or $\hat{\cal F}_{k,g}$ \footnote{Notice that
$\hat{\cal F}_{A,q}^\PureSinglet = \hat{\cal F}_{A,g}=0$ because of charge conjugation
invariance of the strong interactions.} ($k=T,L$).\\
Adding all virtual-, soft- and hard-gluon contributions, the IR divergences
cancel while computing the parton structure functions
$\hat{\cal F}_{k,p}(z,Q^2,\varepsilon)$ which is in agreement with the
Bloch-Nordsieck theorem. The left-over divergences are removed by coupling constant
renormalization and the C-divergences are factorized out of
$\hat{\cal F}_{k,p}(z,Q^2,\varepsilon)$ leaving us with the coefficient functions
which are finite in the limit $\varepsilon\rightarrow 0$. These two procedures
will be carried out in the next section.
%

\newpage
\section{Determination of the coefficient functions in the \MS- and
  the annihilation scheme (A-scheme) \label{sec:s4}}
In this section we determine the coefficient functions of the process \eref{eq:2_1}
by applying coupling constant renormalization and mass factorization to the parton
fragmentation functions
$\hat{\cal F}_{k,p}$ ($p=q,g$) which are computed up to order \alphastwo~in the
last section. These coefficient functions have to satisfy renormalization group equations.
One can formally solve these equations order by order in \alphas~by writing the
renormalization group functions like the beta-function $\beta(\alpha_s)$ and the anomalous
dimension $\gamma_{ij}(\alpha_s)$ ($i,j=q,g$) as a power series in \alphas. In this way
one can algebraically express the coefficient functions into the coefficients of the power
series. Using the mass factorization theorem which holds in every renormalizable
field theory for all leading twist two contributions, one can also express the parton
fragmentation functions $\hat{\cal F}_{k,p}$ into the same coefficients. Our calculations
described in the last section have to satisfy the algebraic expressions of
$\hat{\cal F}_{k,p}$ at least up to pole terms $(1/\varepsilon)^m$ which is a minimal
requirement for the correctness of our results.\\
Before presenting the algebraic expressions for $\hat{\cal F}_{k,p}$  we have to
decompose them according to the flavour symmetry group. Convoluting the parton structure
tensor $\hat{W}_{\mu\nu}^{(V,V')}$ \eref{eq:3_6} with the bare parton fragmentation
densities $\hat{D}_p^H(z)$ we obtain the following functions
\begin{eqnarray}
  \label{eq:4_1}
  \lefteqn{
    F_k^{H,(V,V')}(x,Q^2) = \sum_{p=q,\bar{q},g}\,\sum_{q_1,q_2=1}^{n_f}\,
    (v_{q_1}^{(V)}v_{q_2}^{(V')} + a_{q_1}^{(V)}a_{q_2}^{(V')})\cdot}\nonumber\\[2ex]
  && \cdot\int_0^1\frac{dz}{z} \hat{D}_p^H\left(\frac{x}{z}\right)\hat{\cal F}_{k,p}(z,Q^2,
  \varepsilon),\hspace*{8mm}(k=T,L),\\[2ex]
  \label{eq:4_2}
  \lefteqn{
    F_A^{H,(V,V')}(x,Q^2) = \sum_{p=q,\bar{q}}\,\sum_{q_1,q_2=1}^{n_f}\,
    (v_{q_1}^{(V)}a_{q_2}^{(V')} + a_{q_1}^{(V)}v_{q_2}^{(V')})\cdot}\nonumber\\[2ex]
  && \int_0^1\frac{dz}{z} \hat{D}_p^H\left(\frac{x}{z}\right)\hat{\cal F}_{A,p}(z,Q^2,\varepsilon).
\end{eqnarray}
The reason that we call $\hat{D}_p^H$ `bare', originates from the fact that the
C-divergence which are removed from $\hat{\cal F}_{k,p}$ via mass factorization
will be absorbed by $\hat{D}_p^H$ so that the latter are dressed up to the phenomenological
fragmentation densities defined in \eref{eq:2_4}, \eref{eq:2_5}. The hadronic
fragmentation functions defined in \eref{eq:2_17} are obtained by contracting the
parton structure tensor $\hat{W}_{\mu\nu}^{(V,V')}$ \eref{eq:3_6}, after convolution
by $\hat{D}_p^H$, with the leptonic tensor due to the subprocess $e^+ + e^- \rightarrow
V(V')$ where one also has to include the vector boson propagators given by
$Z(Q^2)^{-1}$ in \eref{eq:2_8}.\\
The contributions to $\hat{\cal F}_{k,p}$ ($p=q,g$) can be distinguished in a flavour
singlet (S) and a flavour non-singlet (NS) part. Equations \eref{eq:4_1}, \eref{eq:4_2}
can then be written as
\begin{eqnarray}
  \label{eq:4_3}
  \lefteqn{
    F_k^{H,(V,V')}(x,Q^2) = \int_x^1\frac{dz}{z}\BigLeftHook\sum_{p=1}^{n_f}\,
    (v_p^{(V)}v_p^{(V')} + a_p^{(V)}a_p^{(V')}) (\hat{D}_p^H\left(\frac{x}{z}\right) +
    \hat{D}_{\bar{p}}^H\left(\frac{x}{z}\right))\cdot}\nonumber\\[2ex]
  && \cdot\hat{\cal F}_{k,q}^\NonSinglet(z,Q^2,\varepsilon) +
  \left(\sum_{q_1=1}^{n_f}\,a_{q_1}^{(V')}\,\sum_{p=1}^{n_f}\,a_p^{(V)} +
    \sum_{q_1=1}^{n_f}\,a_{q_1}^{(V)}\sum_{p=1}^{n_f}\,a_p^{(V')}\right)
  (\hat{D}_p^H\left(\frac{x}{z}\right) \nonumber\\[2ex]
  && {}+\hat{D}_{\bar{p}}^H\left(\frac{x}{z}\right))\,\hat{\cal F}_{k,q}^{',\NonSinglet}
  (z,Q^2,\varepsilon)
  +\sum_{q_1=1}^{n_f}\,(v_{q_1}^{(V)}v_{q_1}^{(V')}+a_{q_1}^{(V)}a_{q_1}^{(V')})\BigLeftBrace
  \frac{1}{n_f}\sum_{p=1}^{n_f}\,(\hat{D}_p^H\left(\frac{x}{z}\right) \nonumber\\[2ex]
  && {}+\hat{D}_{\bar{p}}^H\left(\frac{x}{z}\right))\,
  \hat{\cal F}_{k,q}^\PureSinglet(z,Q^2,\varepsilon) + \hat{D}_g^H\left(\frac{x}{z}\right)\,
  \hat{\cal F}_{k,g}(z,Q^2,\varepsilon)\BigRightBrace\BigRightHook,\hspace*{3mm}(k=T,~L),\\[2ex]
  \label{eq:4_4}
  \lefteqn{
    F_A^{H,(V,V')}(x,Q^2) = \int_x^1\frac{dz}{z}\BigLeftHook\sum_{p=1}^{n_f}\,
    (v_p^{(V)}a_p^{(V')} + a_p^{(V)}v_p^{(V')}) (\hat{D}_p^H\left(\frac{x}{z}\right) -
    \hat{D}_{\bar{p}}^H\left(\frac{x}{z}\right))\cdot}\nonumber\\[2ex]
  && \cdot\hat{\cal F}_{A,q}^\NonSinglet(z,Q^2,\varepsilon) +
  (\sum_{q_1=1}^{n_f}\,a_{q_1}^{(V')}\sum_{p=1}^{n_f}\,v_p^{(V)} +
  \sum_{q_1=1}^{n_f}\,a_{q_1}^{(V)}\sum_{p=1}^{n_f}\,v_p^{(V')})(\hat{D}_p^H\left(
  \frac{x}{z}\right)\nonumber\\[2ex]
  && {}- \hat{D}_{\bar{p}}^H\left(\frac{x}{z}\right))\,
  \hat{\cal F}_{A,q}^{',\NonSinglet}(z,Q^2,\varepsilon)\BigRightHook.
\end{eqnarray}
Here we use the same notation as introduced above \eref{eq:2_6} where $p=1,2,\ldots,n_f$
stands for $p=u,d,\ldots$. Further we have the relations
\begin{eqnarray}
  \label{eq:4_5}
  \lefteqn{
    \hat{\cal F}_{k,q}^{(r)} = \hat{\cal F}_{k,\bar{q}}^{(r)},\hspace*{4mm}
    (k=T,L; r=\NonSinglet,\PureSinglet),}\\[2ex]
  \label{eq:4_6}
  \lefteqn{
    \hat{\cal F}_{A,q}^\NonSinglet = -\hat{\cal F}_{A,\bar{q}}^\NonSinglet,\hspace*{4mm}
    \hat{\cal F}_{A,q}^\PureSinglet = \hat{\cal F}_{A,\bar{q}}^\PureSinglet=0,\hspace*{4mm}
    \hat{\cal F}_{A,g} = 0.}
\end{eqnarray}
Relations \eref{eq:4_5}, \eref{eq:4_6} follow from charge conjugation invariance of the
strong interactions. The parton fragmentation function $\hat{\cal F}_{k,q}^\PureSinglet$ is called
the purely singlet part for reasons we will explain below.\\
The function $\hat{\cal F}_{k,g}$ ($k=T,L$), describing process \eref{eq:3_1} where the
gluon is detected ($p=g$), receives contributions from the graphs in figs.~\ref{fig:f4},
\ref{fig:f6}, \ref{fig:f7}. Since the gluon is a flavour singlet $\hat{\cal F}_{k,g}$
belongs to the same representation. The quarks $q_1$ and $q_2$ in \eref{eq:4_1}, which
are directly coupled to the vector bosons $V$ and $V'$ respectively, automatically belong
to the inclusive state when $p=g$ in reaction \eref{eq:3_1} so that the sums over $q_1$, $q_2$,
and $p$ in \eref{eq:4_1} have to be separately performed.\\
The non-singlet part $\hat{\cal F}_{k,q}^\NonSinglet$ ($k=T,L,A$) describing process
\eref{eq:3_1}
where the quark or anti-quark is detected ($p=q$ or $p=\bar{q}$), is determined by the
graphs in figs.~\ref{fig:f2}-\ref{fig:f8} except for the combinations $C^2$, $D^2$ and
$AD$, $BC$ (see below). Notice that groups B and D only contribute when the anti-quarks
$q_1$ and $q_2$ are identical. In the case of the
non-singlet contribution the quarks $q_1$ and $q_2$ can be identified with $p$ (i.e.
$p=q_1=q_2$) so that the sums over $q_1$, $q_2$, and $p$ in \eref{eq:4_1} are now connected.
The above diagrams also contribute to $\hat{\cal F}_{k,q}^\Singlet$ in the case of $k=T,L$
when they
are projected on the singlet channel and the result is the same as the one obtained for
$\hat{\cal F}_{k,q}^\NonSinglet$ so that we can set
$\hat{\cal F}_{k,q}^\Singlet = \hat{\cal F}_{k,q}^\NonSinglet$.
The groups $C^2$ and $D^2$ in fig.~\ref{fig:f8} only survive if they are projected
on the singlet channel. This is because the detected quark $p$ is only connected with the
vector bosons $V$ and $V'$ via the exchange of a gluon which is a flavour singlet. To show
this more explicitly we have drawn the cut graphs contributing to the parton structure tensor
$\hat{W}_{\mu\nu}^{(V,V')}$ which originate from groups $C$ and $D$ in fig.~\ref{fig:f9}.
Because of the purely singlet nature the groups $C$ and $D$ only contribute to
$\hat{\cal F}_{k,q}^\Singlet$ and their contribution will be called
$\hat{\cal F}_{k,q}^\PureSinglet$.
Like
in the case of $\hat{\cal F}_{k,g}$ the quarks $q_1$, $q_2$ belong to the inclusive state
since $p\neq q_1$, $p\neq q_2$. Therefore one can separately sum over $p$ and $q_1$, $q_2$ which
determines the factor of $\hat{\cal F}_{k,q}^\PureSinglet$ in \eref{eq:4_3}. Finally we have
a special non-singlet contribution which we will call $\hat{\cal F}_{k,q}^{'\NonSinglet}$
\eref{eq:4_3} ($k=T,L$) and $\hat{\cal F}_{A,q}^{''\NonSinglet}$ \eref{eq:4_4}. The latter
originates from the combinations $AD$ and $BC$ in fig.~\ref{fig:f8} which only appear
in the case when the anti-quarks $p_1$ and $p_2$ are identical. The corresponding
cut graphs are drawn in fig.~\ref{fig:f10}. If one removes the dashed line, which indicates the
integration over the momenta cut by that line, one obtains a closed fermion loop.
This fermion loop, to which are attached two gluons and one vector boson $V$ ($V'$), has
the same properties as the triangular fermion loops
inserted in the virtual diagrams of \frefs{fig:f5},~\ref{fig:f6}. In fig.~\ref{fig:f10}
we have taken the example that
the vector boson $V'$ couples to the cut fermion loop via the quark $q_1$ whereas
$V$ couples to the detected quark $p$ (see also \eref{eq:4_3},\eref{eq:4_4}). Like
in the case of the triangle fermion loops in figs.~\ref{fig:f5},\ref{fig:f6} only the
axial vector current can couple to the cut fermion-loop which rules out $V'=\gamma$ so
that only $V'=Z$ remains. Since $a_p^{(\gamma)}=a_{q_1}^{(\gamma)}=0$ we have in
\eref{eq:4_3} $V=Z$
whereas in \eref{eq:4_4} we either can get $V=\gamma$ or $V=Z$. Only when the above condition
is satisfied the parton fragmentation functions $\hat{\cal F}_{k,q}^{'\NonSinglet}$ ($k=T,L$) and
$\hat{\cal F}_{A,q}^{''\NonSinglet}$ can contribute to $F_k^{(Z,Z)}$ \eref{eq:4_3} and
$F_A^{(V,Z)}$ ($V=\gamma,Z$) respectively. If we now in addition sum in fig.~\ref{fig:f10}
over all quark flavours $q_1$ belonging to one family one gets
$\sum_{q_1=u,d}\,a_{q_1}^{(Z)}=0$ (see \eref{eq:3_7}) so that in this case the above contributions
due to $\hat{\cal F}_{k,q}^{'\NonSinglet}$, $\hat{\cal F}_{A,q}^{''\NonSinglet}$ will vanish.
Since one has to sum over all members of one family anyhow in order to cancel
the anomaly appearing in the triangle fermion-loops in figs.~\ref{fig:f5},\ref{fig:f6}
we will do the same for the graphs in fig.~\ref{fig:f10}. Therefore we do not have to
calculate $\hat{\cal F}_{k,q}^{'\NonSinglet}$ and $\hat{\cal F}_{k,q}^{''\NonSinglet}$
and they will
not be included in our phenomenological analysis in this paper.\\
Summarizing the above the singlet fragmentation function $\hat{\cal F}_{k,q}^\Singlet$ ($k=T,L$)
receives two kinds of contributions and it can be written as
\begin{equation}
  \label{eq:4_7}
  \hat{\cal F}_{k,q}^\Singlet = \hat{\cal F}_{k,q}^\NonSinglet + \hat{\cal F}_{k,q}^\PureSinglet,
  \hspace*{4mm}(k=T,L).
\end{equation}
After having specified
the various parts to the parton fragmentation functions we will now list them below.
Starting with the non-singlet part the parton fragmentation function expanded
in the bare coupling constant $\hat{\alpha}_s$ read as follows
\begin{eqnarray}
  \label{eq:4_8}
  \lefteqn{
    \hat{\cal F}_{L,q}^{\NonSinglet,(2)} = \left(\frac{\hat{\alpha}_s}{4\pi}\right)^2\,
    S_\varepsilon^2\,
    \left(\frac{Q^2}{\mu^2}\right)^2\,\BigLeftHook\frac{1}{\varepsilon}\BigLeftBrace
    -2\beta_0\bar{c}_{L,q}^{(1)} + P_{qq}^{(0)}\otimes\bar{c}_{L,q}^{(1)}\BigRightBrace
    + \bar{c}_{L,q}^{\NonSinglet,{\rm nid},(2)}}\nonumber\\[2ex]
  && {}+ \bar{c}_{L,q}^{\NonSinglet,{\rm id},(2)} - 2\beta_0 a_{L,q}^{(1)}
  + P_{qq}^{(0)}\otimes a_{L,q}^{(1)}\BigRightHook,\\[2ex]
  \label{eq:4_9}
  \lefteqn{
    \hat{\cal F}_{T,q}^{\NonSinglet,(2)} = \left(\frac{\hat{\alpha}_s}{4\pi}\right)^2\,
    S_\varepsilon^2\,
    \left(\frac{Q^2}{\mu^2}\right)^\varepsilon\,\BigLeftHook\frac{1}{\varepsilon^2}\BigLeftBrace
    \frac{1}{2}P_{qq}^{(0)}\otimes P_{qq}^{(0)} - \beta_0 P_{qq}^{(0)}\BigRightBrace
    + \frac{1}{\varepsilon}\BigLeftBrace\frac{1}{2}(P_{qq}^{(1),\NonSinglet}}\nonumber\\[2ex]
  && {}+ P_{q\bar{q}}^{(1),\NonSinglet})
  - 2\beta_0\bar{c}_{T,q}^{(1)} + P_{qq}^{(0)}\otimes\bar{c}_{T,q}^{(1)}\BigRightBrace
  + \bar{c}_{T,q}^{\NonSinglet,{\rm nid},(2)} + \bar{c}_{T,q}^{\NonSinglet,{\rm id},(2)} 
  - 2\beta_0 a_{T,q}^{(1)} \nonumber\\[2ex]
  && {}+ P_{qq}^{(0)}\otimes a_{T,q}^{(1)}\BigRightHook.
\end{eqnarray}
The convolution symbol denoted by $\otimes$ is defined by
\begin{equation}
  \label{eq:4_10}
  (f\otimes g)(z) = \int_0^1\,dz_1\,\int_0^1\,dz_2\,\delta(z-z_1z_2)f(z_1)g(z_2).
\end{equation}
The second order DGLAP splitting functions denoted by $P_{ij}^{(1)}$ ($i,j=q,g$) are different
for deep inelastic structure functions (spacelike process) and fragmentation functions
(timelike process). For the latter case they have been calculated in
\cite{Cur80,*Fur80,-Fur82,Flo81}.
In order to
solve the Altarelli-Parisi equations of the fragmentation densities $D_p^H$ it is
convenient to split them into two parts which in the \MS-scheme are given by
\begin{eqnarray}
  \label{eq:4_11}
  \lefteqn{
    P_{qq}^{\NonSinglet,(1)}(z) = n_fC_FT_f\BigLeftHook -\frac{160}{9}{\cal D}_0(z) - \frac{16}{9}
    + \frac{176}{9}z - \frac{16}{3}\frac{1+z^2}{1-z}\ln z}\nonumber\\[2ex]
  && {}- \delta(1-z)(\frac{4}{3}
  + \frac{32}{3}\zeta(2))\BigRightHook\nonumber\\[2ex]
  && {}+C_F^2\BigLeftHook \frac{1+z^2}{1-z}\ln z \left(12 + 16\ln(1-z) - 16\ln z\right)
  - 40 (1-z) \nonumber\\[2ex]
  && {}- (28+12 z)\ln z + 4(1+z)\ln^2 z + \delta(1-z)(3 - 24\zeta(2) + 48\zeta(3))
  \BigRightHook\nonumber\\[2ex]
  && {}+C_AC_F\BigLeftHook (\frac{536}{9} - 16\zeta(2)){\cal D}_0(z)+8(1+z)\zeta(2)
  + 8(1+z)\ln z + \frac{212}{9}\nonumber\\[2ex]
  && {}- \frac{748}{9}z + \frac{1+z^2}{1-z}\left(4\ln^2 z
  + \frac{44}{3}\ln z\right) + \delta(1-z)(\frac{17}{3}+\frac{88}{3}\zeta(2)-24\zeta(3))
  \BigRightHook,\nonumber\\[2ex]
  && \\[2ex]
  \label{eq:4_12}
  \lefteqn{
    P_{q\bar{q}}^{\NonSinglet,(1)}(z) = (C_F^2-\frac{1}{2}C_AC_F)\BigLeftHook \frac{1+z^2}{1+z}
    \BigLeftParen 8\ln^2 z - 32\ln z\ln(1+z) - 32\Li_2(-z)}\nonumber\\[2ex]
  && {}- 16\zeta(2)\BigRightParen + 32(1-z) + 16(1+z)\ln z \BigRightHook,
\end{eqnarray}
where $\Li_n(x)$ denote the polylogarithmic functions which can be found
in \cite{Bar72,*Lew83,*Dev84}. The splitting function $P_{q\bar{q}}^{\NonSinglet,(1)}$
\eref{eq:4_12} arises when the
(anti) quarks $p_1$ and $p_2$ in reaction \eref{eq:3_1} become identical
and it is only determined by the interference terms $AB$ and $CD$ in
fig.~\ref{fig:f8}. Like the splitting functions we have also decomposed
the second order coefficients $\bar{c}_{k,q}^{\NonSinglet,(2)}$
($k=T, L$) into two parts i.e. $\bar{c}_{k,q}^{\NonSinglet,{\rm nid},(2)}$
and $\bar{c}_{k,q}^{\NonSinglet,{\rm id},(2)}$. The latter is due to
identical (anti) quark contributions and like $P_{q\bar{q}}$ it
originates from combinations $AB$ and $CD$ in \fref{fig:f8}. All
coefficients $\bar{c}_{k,q}^{(i)}$ ($i=0,1$) are computed in the
\MS-scheme indicated by a bar and they show up in the perturbation series of the
coefficient functions as we will see below. The coefficients $a_{k,q}^{(1)}$
are presented
in \eref{eq:3_17},\eref{eq:3_18} and $\beta_0$ is the lowest order coefficient in the
beta-function defined by
\begin{equation}
  \label{eq:4_13}
  \beta(\alpha_s) = -2\alpha_s\left[\beta_0\frac{\alpha_s}{4\pi} + \beta_1
  \frac{\alpha_s}{4\pi}+\cdots\right],\hspace*{4mm} \beta_0 = \frac{11}{3}C_A
  - \frac{4}{3}T_f n_f,
\end{equation}
where \alphas~now stands for the renormalized coupling (see below). The purely
singlet contributions (see \eref{eq:4_7}) are given by
\begin{eqnarray}
  \label{eq:4_14}
  \lefteqn{
    \hat{\cal F}_{L,q}^{\PureSinglet,(2)} = n_f\,\left(\frac{\hat{\alpha}_s}{4\pi}\right)^2\,
    S_\varepsilon^2\,
    \left(\frac{Q^2}{\mu^2}\right)^\varepsilon\,\BigLeftHook\frac{1}{\varepsilon}
    \BigLeftBrace
    \frac{1}{2}P_{qg}^{(0)}\otimes\bar{c}_{L,g}^{(1)}\BigRightBrace
    + \bar{c}_{L,q}^{\PureSinglet,(2)}
    }\nonumber\\[2ex]
  && {}+ \frac{1}{2}P_{qg}^{(0)}\otimes a_{L,g}^{(1)}\BigRightHook,\\[2ex]
  \label{eq:4_15}
  \lefteqn{
    \hat{\cal F}_{T,q}^{\PureSinglet,(2)} = n_f\,\left(\frac{\hat{\alpha}_s}{4\pi}\right)^2\,
    S_\varepsilon^2\,
    \left(\frac{Q^2}{\mu^2}\right)^\varepsilon\,\BigLeftHook\frac{1}{\varepsilon^2}
    \BigLeftBrace\frac{1}{2}P_{gq}^{(0)}\otimes P_{qg}^{(0)}\BigRightBrace
    + \frac{1}{\varepsilon}\BigLeftBrace\frac{1}{2}P_{qq}^{\PureSinglet,(1)} 
    }\nonumber\\[2ex]
  && {}+\frac{1}{2}P_{qg}^{(0)}\otimes\bar{c}_{T,g}^{(1)}\BigRightBrace +
    \bar{c}_{T,q}^{\PureSinglet,(2)} + \frac{1}{2}P_{qg}^{(0)}\otimes a_{T,g}^{(1)}\BigRightHook,
\end{eqnarray}
where $a_{L,g}^{(1)}$ and $a_{T,g}^{(1)}$ are presented in \eref{eq:3_19} and
\eref{eq:3_20} respectively.
The above expressions are determined by the combinations $C^2$ (non-identical
(anti-) quarks) or $C^2$ and $D^2$ (identical (anti-) quarks) in fig.~\ref{fig:f8}.
The timelike splitting function $P_{qq}^{\PureSinglet,(1)}$ can be inferred from
\cite{Cur80,*Fur80,-Fur82,Flo81}
and it reads (\MS-scheme)
\begin{eqnarray}
  \label{eq:4_16}
  \lefteqn{
    P_{qq}^{\PureSinglet,(1)}(z) = C_F T_f\BigLeftHook -\frac{320}{9z} - 128 + 64 z
    + \frac{896}{9} z^2 + 16(1+z)\ln^2 z - (80}\nonumber\\[2ex]
  && {}+144 z+\frac{128}{3}z^2)\ln z
  \BigRightHook.
\end{eqnarray}
From \eref{eq:4_7} we can now also obtain the singlet parton fragmentation function
$\hat{\cal F}_{k,q}^{\Singlet,(2)}$ ($k=T,L$). Adding eqs. \eref{eq:4_8} and
\eref{eq:4_14} provides us with $\hat{\cal F}_{L,q}^{\Singlet,(2)}$ whereas the sum of eqs.
\eref{eq:4_9} and \eref{eq:4_15} leads to $\hat{\cal F}_{T,q}^{\Singlet,(2)}$. In the same
way we obtain from \eref{eq:4_11}, \eref{eq:4_12} and \eref{eq:4_16} the singlet
splitting function
\begin{equation}
  \label{eq:4_17}
  P_{qq}^{\Singlet,(1)} = P_{qq}^{\NonSinglet,(1)} + P_{q\bar{q}}^{\NonSinglet,(1)}
  + P_{qq}^{\PureSinglet,(1)}.
\end{equation}
Finally the order \alphastwo~contributions to $\hat{\cal F}_{k,g}$ become
\begin{eqnarray}
  \label{eq_4_18}
  \lefteqn{
    \hat{\cal F}_{L,g}^{(2)} = n_f\,\left(\frac{\hat{\alpha}_s}{4\pi}\right)^2\,
    S_\varepsilon^2\,
    \left(\frac{Q^2}{\mu^2}\right)^\varepsilon\,\BigLeftHook\frac{1}{\varepsilon}
    \BigLeftBrace -2\beta_0\bar{c}_{L,g}^{(1)} + P_{gg}^{(0)}\otimes\bar{c}_{L,g}^{(1)}
    + 2P_{gq}^{(0)}\otimes \bar{c}_{L,q}^{(1)}\BigRightBrace}\nonumber\\[2ex]
  && {}+ \bar{c}_{L,g}^{(2)}
  - 2\beta_0 a_{L,g}^{(1)} + P_{gg}^{(0)}\otimes a_{L,g}^{(1)} + 2P_{gq}^{(0)}\otimes
  a_{L,q}^{(1)}\BigRightHook,\\[2ex]
  \label{eq:4_19}
  \lefteqn{
    \hat{\cal F}_{T,g}^{(2)} = n_f\,\left(\frac{\hat{\alpha}_s}{4\pi}\right)^2\,
    S_\varepsilon^2\,
    \left(\frac{Q^2}{\mu^2}\right)^\varepsilon\,\BigLeftHook\frac{1}{\varepsilon^2}
    \BigLeftBrace P_{gq}^{(0)}\otimes(P_{gg}^{(0)} + P_{qq}^{(0)}) - 2\beta_0 P_{gq}^{(0)}
    \BigRightBrace}\nonumber\\[2ex]
  && {}+ \frac{1}{\varepsilon}\BigLeftBrace P_{gq}^{(1)} - 2\beta_0\bar{c}_{T,g}^{(1)}
  + P_{gg}^{(0)}\otimes\bar{c}_{T,g}^{(1)} + 2P_{gq}^{(0)}\otimes\bar{c}_{T,q}^{(1)}
  \BigRightBrace + \bar{c}_{T,g}^{(2)} - 2\beta_0 a_{T,g}^{(1)} \nonumber\\[2ex]
  && {}+ P_{gg}^{(0)}\otimes
  a_{T,g}^{(1)} + 2P_{gq}^{(0)}\otimes a_{T,g}^{(1)}\BigRightHook,
\end{eqnarray}
where the timelike splitting function $P_{gq}^{(1)}$ in the \MS-scheme can be
found in \cite{Cur80,*Fur80,-Fur82,Flo81}. It is given by
\begin{eqnarray}
  \label{eq:4_20}
  \lefteqn{
    P_{gq}^{(1)} = C_F^2\BigLeftHook -4 + 36 z + (-64 + 4 z)\ln z + 16 z\ln(1-z)
    + (8-4z)\ln^2 z}\nonumber\\[2ex]
  && {}+ (\frac{16}{z} - 16 + 8 z)\ln^2(1-z) + (\frac{64}{z} - 64
  + 32 z)\ln z\ln(1-z) \nonumber\\[2ex]
  && {}+ (\frac{128}{z} - 128 + 64 z)\Li_2(1-z) +
  (-\frac{128}{z} + 128 - 64 z)\zeta(2)\BigRightHook\nonumber\\[2ex]
  && {}+C_AC_F\BigLeftHook \frac{136}{9z} + 40 - 8z - \frac{352}{9}z^2 +
  (-\frac{48}{z} + 64 + 72 z + \frac{64}{3}z^2)\ln z\nonumber\\[2ex]
  && {}- 16 z\ln(1-z) -
  (\frac{32}{z} + 16 + 24 z)\ln^2 z + (-\frac{16}{z} + 16 - 8z)\ln^2(1-z)
  \nonumber\\[2ex]
  && {}+
  (-\frac{32}{z} + 32 - 16 z)\ln z\ln(1-z) + (-\frac{128}{z} + 128 - 64 z)\Li_2(1-z)
  \nonumber\\[2ex]
  && {}+ (\frac{32}{z} + 32 + 16 z)\Li_2(-z) + (\frac{32}{z} + 32 + 16 z)\ln z\ln(1+z)
  + (\frac{128}{z} \nonumber\\[2ex]
  && {}- 96 + 64 z)\zeta(2)\BigRightHook.
\end{eqnarray}
The pole terms $(1/\varepsilon)^m$ showing up in the parton fragmentation functions
$\hat{\cal F}_{k,p}$ ($k=T,L$; $p=q,g$) are due to UV and C-divergences. In order
to get the coefficient functions corresponding to the fragmentation process
\eref{eq:2_1} these singularities have to be removed via coupling constant
renormalization and mass factorization.\\
The coupling constant renormalization can be achieved by replacing the bare
(unrenormalized) coupling constant $\hat{\alpha}_s$ by
\begin{equation}
  \label{eq:4_21}
  \frac{\hat{\alpha}_s}{4\pi} = \frac{\alpha_s(R^2)}{4\pi}\left(1
    + \frac{\alpha_s(R^2)}{4\pi}
    \frac{2\beta_0}{\varepsilon}S_\varepsilon\left(\frac{R^2}{\mu^2}\right)^{\varepsilon/2}
  \right),
\end{equation}
where $R$ represents the renormalization scale. After having removed the UV
singularities the remaining pole terms can be attributed to final state collinear
divergence only because $\hat{\cal F}_{k,p}$ is a semi-inclusive quantity. The latter
singularities are removed by mass factorization which proceeds in the following way
\begin{eqnarray}
  \label{eq:4_22}
  \hat{\cal F}_{k,q}^\NonSinglet &=& \Gamma_{qq}^\NonSinglet\otimes
  \mathbb{C}_{k,q}^\NonSinglet, \\[2ex]
  \label{eq:4_23}
  \hat{\cal F}_{k,q}^\Singlet    &=& \Gamma_{qq}^\Singlet\otimes
  \mathbb{C}_{k,q}^\Singlet +
  n_f\,\Gamma_{qg}\otimes \mathbb{C}_{k,g},\\[2ex]
  \label{eq:4_24}
  \hat{\cal F}_{k,g}      &=& 2\Gamma_{gq}\otimes
  \mathbb{C}_{k,q}^\Singlet + \Gamma_{gg}\otimes \mathbb{C}_{k,g},
\end{eqnarray}
with $\Gamma_{gq} = \Gamma_{g\bar{q}}$, $\Gamma_{qg} = \Gamma_{\bar{q}g}$. The
quantities $\Gamma_{ij}$ are called transition functions in which all C-divergences are
absorbed so that the fragmentation coefficient function $\mathbb{C}_{k,p}$ are
finite. Both functions are expanded in the renormalized coupling constant
$\alpha_s(R^2)$ and depend explicitly on the renormalization scale $R$ and the
factorization scale $M$ which implies that they are scheme dependent. If we
expand $\Gamma_{ij}$ in the unrenormalized coupling constant $\hat{\alpha_s}$
the expressions become very simple. Choosing the \MS-scheme they take the
following form
\begin{eqnarray}
  \label{eq:4_25}
  \lefteqn{
    \overline{\Gamma}_{qq}^\NonSinglet = \unit + \frac{\hat{\alpha}_s}{4\pi}\,S_\varepsilon\,
    \left(\frac{M^2}{\mu^2}\right)
    ^{\varepsilon/2}\BigLeftHook\frac{1}{\varepsilon}P_{qq}^{(0)}\BigRightHook
    + \left(\frac{\hat{\alpha}_s}{4\pi}\right)^2\,S_\varepsilon^2\,\left(
      \frac{M^2}{\mu^2}\right)
    ^\varepsilon\BigLeftHook\frac{1}{\varepsilon^2}}\nonumber\\[2ex]
  && \BigLeftBrace\frac{1}{2}P_{qq}^{(0)}
  \otimes P_{qq}^{(0)}-\beta_0 P_{qq}^{(0)}\BigRightBrace + \frac{1}{\varepsilon}
  \BigLeftBrace
  \frac{1}{2}P_{qq}^{\NonSinglet,(1)}+\frac{1}{2}P_{q\bar{q}}^{\NonSinglet,(1)}
  \BigRightBrace\BigRightHook,  \\[2ex]
  \label{eq:4_26}
  \lefteqn{
    \overline{\Gamma}_{qq}^\Singlet = \overline{\Gamma}_{qq}^\NonSinglet
    + 2n_f\,\overline{\Gamma}_{qq}^\PureSinglet,}\\[2ex]
  \label{eq:4_27}
  \lefteqn{
    \overline{\Gamma}_{qq}^\PureSinglet = \left(\frac{\hat{\alpha}_s}{4\pi}\right)^2\,
    S_\varepsilon^2\,
    \left(\frac{M^2}{\mu^2}\right)^\varepsilon\,\BigLeftHook\frac{1}{\varepsilon^2}
    \BigLeftBrace
    \frac{1}{4}P_{gq}^{(0)}\otimes P_{qg}^{(0)}\BigRightBrace + \frac{1}{\varepsilon}
    \BigLeftBrace\frac{1}{4}P_{qq}^{\PureSinglet,(1)}\BigRightBrace\BigRightHook,}\\[2ex]
  \label{eq:4_28}
  \lefteqn{
    \overline{\Gamma}_{gq} = \frac{\hat{\alpha}_s}{4\pi}\,S_\varepsilon\,\left(
      \frac{M^2}{\mu^2}\right)
    ^{\varepsilon/2}\,\BigLeftHook\frac{1}{\varepsilon}P_{gq}^{(0)}\BigRightHook
    + \left(\frac{\hat{\alpha}_s}{4\pi}\right)^2\,S_\varepsilon^2\,\left(\frac{M^2}{\mu^2}
    \right)
    ^\varepsilon\,\BigLeftHook\frac{1}{\varepsilon}\BigLeftBrace\frac{1}{2}P_{gq}^{(0)}
    \otimes}\nonumber\\[2ex]
  && (P_{gg}^{(0)}+P_{qq}^{(0)})-\beta_0 P_{gq}^{(0)} + \frac{1}{\varepsilon}
  \BigLeftBrace\frac{1}{2}P_{gq}^{(1)}\BigRightBrace\BigRightHook,\\[2ex]
  \label{eq:4_29}
  \lefteqn{
    \overline{\Gamma}_{qg} = \frac{\hat{\alpha}_s}{4\pi}\,S_\varepsilon\,\left(
      \frac{M^2}{\mu^2}\right)
    ^{\varepsilon/2}\,\BigLeftHook\frac{1}{2\varepsilon}P_{qg}^{(0)}\BigRightHook,}\\[2ex]
  \label{eq:4_30}
  \lefteqn{
    \overline{\Gamma}_{gg} = \unit + \frac{\hat{\alpha}_s}{4\pi}\,S_\varepsilon\,
    \left(\frac{M^2}{\mu^2}\right)^{\varepsilon/2}\,
    \BigLeftHook\frac{1}{\varepsilon}P_{gg}^{(0)}\BigRightHook,}
\end{eqnarray}
where the $\unit$ in \eref{eq:4_25} and \eref{eq:4_30} is a shorthand notation for
$\delta(1-z)$. Notice that we have expanded the $\Gamma_{ij}$ above in
sufficiently higher order of \alphas~in order to get the coefficient functions
finite. Therefore the computation of $\hat{\cal F}_{k,p}$ allows us to determine
the DGLAP-splitting functions $P_{qq}^{\NonSinglet,(1)}$, $P_{q\bar{q}}^{\NonSinglet,(1)}$,
$P_{qq}^{\PureSinglet,(1)}$, and $P_{gq}^{(1)}$ in an alternative way which is different from the
method used in \cite{Cur80,*Fur80,-Fur82,Flo81}. Further the
transition functions satisfy the following relations which originate from energy
momentum conservation
\begin{eqnarray}
  \label{eq:4_31}
  \lefteqn{
    \int_0^1dz\,z\,(\Gamma_{qq}^\Singlet(z) + \Gamma_{gq}(z)) = 1,}\\[2ex]
  \label{eq:4_32}
  \lefteqn{
    \int_0^1dz\,z\,(\Gamma_{gg}(z) + 2n_f\,\Gamma_{qg}(z)) = 1.}
\end{eqnarray}
If we substitute $\hat{\cal F}_{k,p}$ \eref{eq:4_22} - \eref{eq:4_24} into eq.
\eref{eq:4_3} the C-singularities are absorbed by the bare fragmentation
densities $\hat{D}_p^H$ as follows
\begin{eqnarray}
  \label{eq:4_33} 
  D_{\NonSinglet,p}^H &=& \Gamma_{qq}^\NonSinglet\otimes \hat{D}_{\NonSinglet,p}^H, \\[2ex]
  \label{eq:4_34}
  D^H_\Singlet &=& \Gamma_{qq}^\Singlet\otimes \hat{D}_\Singlet^H + 2\Gamma_{gq}\otimes\hat{D}_g^H,
  \\[2ex]
  \label{eq:4_35}
  D_g^H &=& n_f\,\Gamma_{qg}\otimes\hat{D}_\Singlet^H + \Gamma_{gg}\otimes\hat{D}_g^H,
\end{eqnarray}
Here $D^H_{\NonSinglet,p}$ and $D^H_\Singlet$ denote the non-singlet and singlet combinations
of the parton
fragmentation densities as defined in \eref{eq:2_5_2}, \eref{eq:2_5_1}. The same definition
holds for the bare densities $\hat{D}_{\NonSinglet,p}^H$ and $\hat{D}_\Singlet^H$. The densities
$D_{\NonSinglet,p}^H$,
$D_\Singlet^H$, and $D_g^H$ depend on the renormalization scale $R$ and the mass
factorization scale $M$ which are usually set to be equal.\\
Substituting eqs. \eref{eq:4_22}-\eref{eq:4_23}, \eref{eq:4_33}-\eref{eq:4_35}
in \eref{eq:4_3} and using \eref{eq:4_7} we obtain after rearranging terms the
structure function $F_k^{(V,V')}(x,Q^2)$ expressed into the renormalized parton
fragmentation densities $D_p^H$ and the fragmentation coefficient functions
$\mathbb{C}_{k,p}$ ($p=q,g$).
\begin{eqnarray}
  \label{eq:4_39}
  \lefteqn{
    F_k^{(V,V')}(x,Q^2) = \int_x^1\,\frac{dz}{z}\,\BigLeftHook
    \sum_{p=1}^{n_f}\,(v_p^{(V)}v_p^{(V')} + a_p^{(V)}a_p^{(V')})\BigLeftBrace
    D^H_\Singlet\left(\frac{x}{z},M^2\right)\cdot}\nonumber\\[2ex]
  && \cdot\mathbb{C}_{k,q}^\Singlet(z,Q^2/M^2)
  + D_g^H\left(\frac{x}{z},M^2\right)\,\mathbb{C}_g(z,Q^2/M^2)\BigRightBrace\nonumber\\[2ex]
  &&+ \sum_{p=1}^{n_f}\,(v_p^{(V)}v_p^{(V')} + a_p^{(V)}a_p^{(V')})\,
  D_{\NonSinglet,p}^H\left(\frac{x}{z},M^2\right)
  \mathbb{C}_{k,q}^\NonSinglet(z,Q^2/M^2)\BigRightHook,
\end{eqnarray}
where we have chosen $R=M$.\\
Like the parton fragmentation functions $\hat{\cal F}_{k,p}$ in eqs.
\eref{eq:4_8}-\eref{eq:4_19} we can express the coefficient functions $\mathbb{C}_{k,p}$
($p=q,g$) into the renormalization group coefficients. In the \MS-scheme they take
the following form. The non-singlet coefficient functions become
\begin{eqnarray}
  \label{eq:4_40}
  \lefteqn{
    \overline{\mathbb{C}}_{L,q}^\NonSinglet = \frac{\alpha_s}{4\pi}\,\BigLeftHook
    \bar{c}_{L,q}^{(1)} \BigRightHook +
    \left(\frac{\alpha_s}{4\pi}\right)^2\,\BigLeftHook\BigLeftBrace-\beta_0
    \bar{c}_{L,q}^{(1)}
    + \frac{1}{2}P_{qq}^{(0)}\otimes\bar{c}_{L,q}^{(1)}\BigRightBrace L_M
    + \bar{c}_{L,q}^{\NonSinglet,(2),{\rm nid}}} \nonumber\\[2ex]
  && {}+ \bar{c}_{L,q}^{\NonSinglet,(2),{\rm id}}
  \BigRightHook,\\[2ex]
  \label{eq:4_41}
  \lefteqn{
    \overline{\mathbb{C}}_{T,q}^\NonSinglet = \unit + \frac{\alpha_s}{4\pi}\,
    \BigLeftHook\frac{1}{2}P_{qq}^{(0)} L_M
    + \bar{c}_{T,q}^{(1)}\BigRightHook + \left(\frac{\alpha_s}{4\pi}\right)^2\,\BigLeftHook
    \BigLeftBrace\frac{1}{8}P_{qq}^{(0)}\otimes P_{qq}^{(0)}}\nonumber\\[2ex]
  && {}- \frac{1}{4}\beta_0 P_{qq}^{(0)}
  \BigRightBrace L_M^2 + \BigLeftBrace\frac{1}{2}(P_{qq}^{(1),\NonSinglet}
  +P_{q\bar{q}}^{(1),\NonSinglet})
  -\beta_0\bar{c}_{T,q}^{(1)} + \frac{1}{2} P_{qq}^{(0)}\otimes\bar{c}_{T,q}^{(1)}
  \BigRightBrace L_M \nonumber\\[2ex]
  && {}+ \bar{c}_{T,q}^{\NonSinglet,(2),{\rm nid}}
  + \bar{c}_{T,q}^{\NonSinglet,(2),{\rm id}}\BigRightHook.
\end{eqnarray}
The singlet coefficient functions are given by
\begin{eqnarray}
  \label{eq:4_42}
  \lefteqn{
    \overline{\mathbb{C}}_{k,q}^\Singlet = \overline{\mathbb{C}}_{k,q}^\NonSinglet
    + \overline{\mathbb{C}}_{k,q}^\PureSinglet,\hspace*{4mm}(k=T,L),}\\[2ex]
  \label{eq:4_43}
  \lefteqn{
    \overline{\mathbb{C}}_{L,q}^\PureSinglet = n_f\,\left(\frac{\alpha_s}{4\pi}\right)^2\,
    \BigLeftHook\BigLeftBrace
    \frac{1}{4}P_{qg}^{(0)}\otimes\bar{c}_{L,g}^{(1)}\BigRightBrace L_M
    + \bar{c}_{L,q}^{\PureSinglet,(2)}\BigRightHook,}\\[2ex]
  \label{eq:4_44}
  \lefteqn{
    \overline{\mathbb{C}}_{T,q}^\PureSinglet = n_f\,\left(\frac{\alpha_s}{4\pi}\right)^2\,
    \BigLeftHook\BigLeftBrace
    \frac{1}{8}P_{gq}^{(0)}\otimes P_{qg}^{(0)}\BigRightBrace L_M^2 + \BigLeftBrace
    \frac{1}{2}P_{qq}^{\PureSinglet,(1)} + \frac{1}{4}P_{qg}^{(0)}\otimes\bar{c}_{T,g}^{(1)}
    \BigRightBrace L_M}\nonumber\\[2ex]
  && {}+ \bar{c}_{T,q}^{\PureSinglet,(2)}\BigRightHook.
\end{eqnarray}
The gluon coefficient functions become
\begin{eqnarray}
  \label{eq:4_45}
  \lefteqn{
    \overline{\mathbb{C}}_{L,g} = \frac{\alpha_s}{4\pi}\,\BigLeftHook\bar{c}_{L,g}^{(1)}
    \BigRightHook
    + \left(\frac{\alpha_s}{4\pi}\right)^2\,\BigLeftHook\BigLeftBrace
    -\beta_0\bar{c}_{L,g}^{(1)}
    + \frac{1}{2} P_{gg}^{(0)}\otimes\bar{c}_{L,g}^{(1)}}\nonumber\\[2ex]
  && {}+ P_{gq}^{(0)}\otimes\bar{c}_{L,q}^{(1)}
  \BigRightBrace L_M + \bar{c}_{L,g}^{(2)}\BigRightHook,\\[2ex]
  \label{eq:4_46}
  \lefteqn{
    \overline{\mathbb{C}}_{T,g} = \frac{\alpha_s}{4\pi}\,\BigLeftHook P_{gq}^{(0)} L_M
    + \bar{c}_{T,g}^{(1)}\BigRightHook + \left(\frac{\alpha_s}{4\pi}\right)^2\,
    \BigLeftHook\BigLeftBrace\frac{1}{4}P_{gq}^{(0)}\otimes(P_{gg}^{(0)}+P_{qq}^{(0)})}
  \nonumber\\[2ex]
  && {}- \frac{1}{2}\beta_0 P_{gq}^{(0)}\BigRightBrace L_M^2 + \BigLeftBrace P_{gq}^{(1)}
  - \beta_0\bar{c}_{T,g}^{(1)} + \frac{1}{2} P_{gg}^{(0)}\otimes\bar{c}_{T,g}^{(1)}
  + P_{gq}^{(0)}\otimes\bar{c}_{T,q}^{(1)}\BigRightBrace L_M \nonumber\\[2ex]
  && {}+ \bar{c}_{T,g}^{(2)}\BigRightHook.
\end{eqnarray}
Further we have defined
\begin{equation}
  \label{eq:4_47}
  \unit \equiv \delta(1-z),\hspace*{4mm} L_M = \ln\frac{Q^2}{M^2},\hspace*{4mm}
  \alpha_s \equiv \alpha_s(M^2).
\end{equation}
In the case $M\neq R$ the resulting coefficient functions can be very easily derived
from the above expressions \eref{eq:4_40}-\eref{eq:4_46} by replacing
\begin{equation}
  \label{eq:4_50}
  \alpha_s(M^2) = \alpha_s(R^2)\left[ 1+\frac{\alpha_s(R^2)}{4\pi}\,\beta_0\,
    \ln\frac{R^2}{M^2}\right].
\end{equation}
The explicit expressions for the coefficient functions \eref{eq:4_40}-\eref{eq:4_46}
are listed in appendix A.\\
Besides the \MS-scheme one also can compute the coefficient functions in the so called
annihilation
scheme (A-scheme) \cite{Nas94}. It is defined in such a way that $F^H/R_{ee}$
(see \eref{eq:2_16}, \eref{eq:2_17_1}) does not get any \alphas~corrections at
$M^2=R^2=Q^2$. In the
A-scheme the transition functions $\Gamma_{ij}$ are related to the ones in the \MS-scheme
denoted by $\overline{\Gamma}_{ij}$ (see \eref{eq:4_25}-\eref{eq:4_30}) as follows
\begin{equation}
  \label{eq:4_51}
  \Gamma_{qq}^\NonSinglet = Z_{qq}^\NonSinglet\,\overline{\Gamma}_{qq},\hspace*{4mm}
  \Gamma_{ij} = Z_{ik}\,\overline{\Gamma}_{kj},
\end{equation}
where $Z_{qq}^\NonSinglet$, $Z_{ik}$ are given by (see eqs. (2.61), (2.62) in \cite{Nas94})
\begin{eqnarray}
  \label{eq:4_52}
  Z_{qq}^\NonSinglet & = & R_{ee}^{-1}\,\overline{\mathbb{C}}_q^\NonSinglet,\\[2ex]
  \label{eq:4_53}
  Z &=& \left(
    \begin{array}{cc}
      R_{ee}^{-1}\,\overline{\mathbb{C}}_q^\Singlet & R_{ee}^{-1}\,
      \overline{\mathbb{C}}_g\\[2ex]
      0 & 1
    \end{array}
    \right).
\end{eqnarray}
The coefficient functions $\overline{\mathbb{C}}_\ell^{(r)}$ ($r=\NonSinglet,\Singlet$,
$\ell=q,g$)
correspond to the structure function $F^H$ defined in \eref{eq:2_17_1} and they are given by
\begin{equation}
  \label{eq:4_55}
  \overline{\mathbb{C}}_\ell^{(r)} = \overline{\mathbb{C}}_{T,\ell}^{(r)}
  + \overline{\mathbb{C}}_{L,\ell}^{(r)}.
\end{equation}
The coefficient functions in the A-scheme, denoted by $\mathbb{C}_{k,p}$,
are related to the ones presented in the \MS-scheme in the following way
($k=T, L$)
\begin{eqnarray}
  \label{eq:4_56}
  \mathbb{C}_{k,q}^\NonSinglet & = & \left(Z_{qq}^\NonSinglet\right)^{-1}\,
  \overline{\mathbb{C}}_{k,q}^\NonSinglet,\\[2ex]
  \label{eq:4_57}
  \mathbb{C}_{k,i} & = & \left(Z^{-1}\right)_{ji}\,\overline{\mathbb{C}}_{k,j}.
\end{eqnarray}
Expanding all coefficient functions and $R_{ee}$ in \alphas~the former take
the following form in the A-scheme
\begin{eqnarray}
  \label{eq:4_58}
  \lefteqn{ 
    \mathbb{C}_{L,q}^\NonSinglet = \frac{\alpha_s}{4\pi}\BigLeftHook \bar{c}_{L,q}^{(1)}
    \BigRightHook + \left(\frac{\alpha_s}{4\pi}\right)^2\BigLeftHook\BigLeftBrace
    -\beta_0\bar{c}_{L,q}^{(1)} + \frac{1}{2}P_{qq}^{(0)}\otimes\bar{c}_{L,q}^{(1)}
    \BigRightBrace L_M + \bar{c}_{L,q}^{(2),\NonSinglet,{\rm nid}}}\nonumber\\[2ex]
  && {}+ \bar{c}_{L,q}^{(2),\NonSinglet,{\rm id}} + R^{(1)}\bar{c}_{L,q}^{(1)}
    - \bar{c}_q^{(1)}\otimes\bar{c}_{L,q}^{(1)}\BigRightHook,\\[2ex]
  \label{eq:4_59}
  \lefteqn{
    \mathbb{C}_{T,q}^\NonSinglet = \unit + \frac{\alpha_s}{4\pi}\BigLeftHook\frac{1}{2}
    P_{qq}^{(0)} L_M + R^{(1)}.\unit - \bar{c}_{L,q}^{(1)}\BigRightHook
    + \left(\frac{\alpha_s}{4\pi}\right)^2\BigLeftHook\BigLeftBrace
    \frac{1}{8}P_{qq}^{(0)}\otimes P_{qq}^{(0)}}\nonumber\\[2ex]
  && {}- \frac{1}{4}\beta_0 P_{qq}^{(0)}
  \BigRightBrace L_M^2 + \BigLeftBrace \frac{1}{2}(P_{qq}^{\NonSinglet,(1)}
  + P_{q\bar{q}}^{\NonSinglet,(1)}) - \beta_0\bar{c}_{T,q}^{(1)} + \frac{1}{2}R^{(1)}
  P_{qq}^{(0)} \nonumber\\[2ex]
  && {}- \frac{1}{2}P_{qq}^{(0)}\otimes\bar{c}_{L,q}^{(1)}
  \BigRightBrace L_M + R^{(2)}.\unit - \bar{c}_{L,q}^{\NonSinglet,(2),{\rm nid}}
  - \bar{c}_{L,q}^{\NonSinglet,(2),{\rm id}} - R^{(1)}\bar{c}_{L,q}^{(1)} \nonumber\\[2ex]
  && {}+ \bar{c}_q^{(1)}\otimes\bar{c}_{L,q}^{(1)}\BigRightHook,\\[2ex]
  \label{eq:4_60}
  \lefteqn{
    \mathbb{C}_{L,q}^\PureSinglet = n_f\left(\frac{\alpha_s}{4\pi}\right)^2\BigLeftHook
    \BigLeftBrace\frac{1}{4}P_{qg}^{(1)}\otimes\bar{c}_{L,g}^{(1)}\BigRightBrace L_M
    + \bar{c}_{L,q}^{\PureSinglet,(2)}\BigRightHook = \overline{\mathbb{C}}_{L,q}^\PureSinglet,}
  \\[2ex]
  \label{eq:4_61}
  \lefteqn{
    \mathbb{C}_{T,q}^{\PureSinglet,(2)} = n_f\left(\frac{\alpha_s}{4\pi}\right)^2\BigLeftHook
    \BigLeftBrace\frac{1}{8}P_{qg}^{(0)}\otimes P_{gq}^{(0)}\BigRightBrace L_M^2
    + \BigLeftBrace\frac{1}{2}P_{qq}^{\PureSinglet,(1)} + \frac{1}{4}P_{qg}^{(0)}\otimes
    \bar{c}_{T,g}^{(1)}\BigRightBrace L_M }\nonumber\\[2ex]
  && {}- \bar{c}_{L,q}^{\PureSinglet,(2)}\BigRightHook,\\[2ex]
  \label{eq:4_62}
  \lefteqn{
    \mathbb{C}_{L,g} = \frac{\alpha_s}{4\pi}\BigLeftHook\bar{c}_{L,g}^{(1)}
    \BigRightHook + \left(\frac{\alpha_s}{4\pi}\right)^2\BigLeftHook\BigLeftBrace
    -\beta_0\bar{c}_{L,q}^{(1)} + \frac{1}{2}P_{gg}^{(0)}\otimes\bar{c}_{L,g}^{(0)}
    + P_{gq}^{(0)}\otimes\bar{c}_{L,q}^{(1)}\BigRightBrace L_M}\nonumber\\[2ex]
  && {}+ \bar{c}_{L,g}^{(2)} - \bar{c}_g^{(1)}\otimes\bar{c}_{L,q}^{(1)}\BigRightHook,
  \\[2ex]
  \label{eq:4_63}
  \lefteqn{
    \mathbb{C}_{T,g} = \frac{\alpha_s}{4\pi}\BigLeftHook P_{gq}^{(0)} L_M
    - \bar{c}_{L,g}^{(1)}\BigRightHook + \left(\frac{\alpha_s}{4\pi}\right)^2
    \BigLeftHook\BigLeftBrace\frac{1}{4}P_{gq}^{(0)}\otimes(P_{gg}^{(0)}
    + P_{qq}^{(0)}) }\nonumber\\[2ex]
  && {}- \frac{1}{2}\beta_0 P_{gq}^{(0)}\BigRightBrace L_M^2
  + \BigLeftBrace P_{gq}^{(0)} - \beta_0\bar{c}_{T,g}^{(1)} + \frac{1}{2}
  P_{gg}^{(0)}\otimes\bar{c}_{T,g}^{(1)} + P_{gq}^{(0)}\otimes\bar{c}_{T,g}^{(1)}
  \nonumber\\[2ex]
  && {}- \frac{1}{2} P_{qq}^{(0)}\otimes\bar{c}_g^{(1)}\BigRightBrace L_M
  - \bar{c}_{L,g}^{(2)} + \bar{c}_g^{(1)}\otimes\bar{c}_{L,q}^{(1)}
  \BigRightHook,
\end{eqnarray}
where $\unit$ is given by \eref{eq:4_47} and the coefficients $R^{(i)}$ show up in the
perturbation series for $R_{ee}$ \eref{eq:2_16}:
\begin{equation}
  \label{eq:4_64}
  R_{ee} = 1 + \frac{\alpha_s}{4\pi}R^{(1)} +
  \left(\frac{\alpha_s}{4\pi}\right)^2R^{(2)}.
\end{equation}
Notice that we have expressed the above coefficient functions into the renormalization
group coefficients $P_{ij}^{(1)}$, $\bar{c}_{k,p}^{(i)}$ presented in the
\MS-scheme. From \eref{eq:4_58}-\eref{eq:4_63} we infer that at $Q^2=M^2$
($L_M=0$) the coefficient functions in \eref{eq:4_55} become
\begin{equation}
  \label{eq:4_65}
  \mathbb{C}_q^\NonSinglet = \unit.R_{ee},\hspace*{4mm}\mathbb{C}_q^\PureSinglet = 0,
  \hspace*{4mm} \mathbb{C}_g=0.
\end{equation}
In any scheme the coefficient functions satisfy the renormalization group equations
\begin{equation}
  \label{eq:4_66}
  \BigLeftHook\BigLeftBrace M\frac{\partial}{\partial M} + \beta(\alpha_s)
  \frac{\partial}{\partial\alpha_s}\BigRightBrace \delta_{ij} -
  \gamma_{ij}^{(m)}\BigRightHook\,\tilde{\mathbb{C}}_{k,i}^{(m)} = 0,
\end{equation}
with $k=T,L$ and $i,j=q,g$. Further we have defined the Mellin transforms
\begin{eqnarray}
  \label{eq:4_67}
  \lefteqn{
    \mathbb{C}_{k,i}^{(m)}(Q^2/M^2) = \int_0^1dz\,z^{m-1}\,\mathbb{C}_{k,i}(z,Q^2/M^2),
    }\\[2ex]
  \label{eq:4_68}
  \lefteqn{
    \gamma_{ij}^{(m)} = -\int_0^1dz\,z^{m-1}\,P_{ij}(z),}
\end{eqnarray}
and introduced the following the notations
\begin{equation}
  \label{eq:4_69}
  \tilde{\mathbb{C}}_{k,q}^{(m)} = \mathbb{C}_{k,q}^{(m)},\hspace*{4mm}
  \tilde{\mathbb{C}}_{k,g}^{(m)} = \frac{1}{2}\mathbb{C}_{k,g}^{(m)}.
\end{equation}
The quantities $\gamma_{ij}^{(m)}$ are the anomalous dimensions corresponding with
the timelike cut vertex operators of spin $m$. Like the timelike splitting
functions $P_{ij}$ they are scheme dependent. The relations between the anomalous
dimensions obtained from different schemes can e.g. be found in eqs.
(3.82)-(3.86) in \cite{Mer96}.
%

\newpage
\section{Results}
In this section we will discuss the order \alphastwo~contributions to the
longitudinal and transverse cross sections and their corresponding fragmentation
functions. In particular we investigate how the leading order (LO) longitudinal
quantities, which already exist in the literature \cite{Nas94}, \cite{Lar93},
\cite{Bai79}, are modified by including the order \alphastwo~contributions. We
will do the same for the transverse quantities for which a next-to-leading order
(NLO) result already exists. Further we study the dependence of the above quantities
on the mass factorization scale $M$ and the renormalization scale $R$ and show that the
sensitivity to these scales becomes less when higher order corrections are included.
Before we proceed we want to emphasize that with all higher order QCD corrections at
hand it is only possible to perform a full NLO analysis on the cross sections and the
fragmentation functions. The order \alphastwo~contributions also allow for a
next-to-next-to-leading order (NNLO) analysis of the transverse cross section
$\sigma_T(Q^2)$ but not for the transverse fragmentation function $F_T(x,Q^2)$. For the
latter one also needs the three-loop timelike splitting functions which have not been
calculated yet. Therefore the order \alphastwo~contributions to $F_T(x,Q^2)$
have to be considered as an estimate of the NNLO result. Nevertheless we will use
the notation $F_T^{NNLO}$ to indicate the order \alphastwo~corrected transverse
structure function.\\
The longitudinal and transverse cross section $\sigma_k(Q^2)$ ($k=T,~L$) defined in
\eref{eq:2_14} are obtained from the coefficient functions calculated in the previous
sections as follows
\begin{equation}
  \label{eq:5_1}
  \sigma_k(Q^2) = \sigma^{(0)}_{\rm tot}(Q^2)\,\int_0^1dz\,z\BigLeftHook
  \mathbb{C}^\Singlet_{k,q}(z,Q^2/M^2) + \frac{1}{2}\mathbb{C}_{k,g}(z,Q^2/M^2)
  \BigRightHook.
\end{equation}
The results are
\begin{eqnarray}
  \lefteqn{
    \sigma_L(Q^2) = \sigma^{(0)}_{\rm tot}(Q^2)\BigLeftHook\frac{\alpha_s(R^2)}{4\pi}
    C_F[3] + \left(\frac{\alpha_s(R^2)}{4\pi}\right)^2\BigLeftHook
    C_F^2\left\{-\frac{15}{2}\right\} + C_AC_F\BigLeftBrace} \nonumber\\[2ex]
  \label{eq:5_2}
  && -11 \ln\frac{Q^2}{R^2}
  -\frac{24}{5}\zeta(3) + \frac{2023}{30}\BigRightBrace + n_fC_FT_f\BigLeftBrace
  4\ln\frac{Q^2}{R^2} - \frac{74}{3}\BigRightBrace\BigRightHook,\\[2ex]
  \lefteqn{
    \sigma_T(Q^2) = \sigma^{(0)}_{\rm tot}(Q^2)\BigLeftHook 1 + \left(
    \frac{\alpha_s(R^2)}{4\pi}\right)^2\BigLeftHook C_F^2\{6\} + C_AC_F\BigLeftBrace
    -\frac{196}{5}\zeta(3) - \frac{178}{30}\BigRightBrace} \nonumber\\[2ex]
  \label{eq:5_3}
  && {}+ n_fC_FT_f\BigLeftBrace 16\zeta(3) + \frac{8}{3}\BigRightBrace\BigRightHook.
\end{eqnarray}
Addition of $\sigma_L$ and $\sigma_T$ yields the well known answer $\sigma_{\rm tot}(Q^2)$
(see \eref{eq:2_15} and \eref{eq:2_16}) which is in agreement with the literature
\cite{Che79,*Din79,*Cel80} (see also \cite{Gor91,*Sur91,*Che94,*Lar94,-Lar95}). Hence
\eref{eq:5_2} and \eref{eq:5_3} provides us with a check on our calculation of the
longitudinal and transverse coefficient functions. Notice that in lowest order
$\sigma_{\rm tot}(Q^2)$ only receives a contribution from the transverse cross
section \eref{eq:5_3} whereas the order \alphas~contribution can be only attributed
to the longitudinal part in \eref{eq:5_2}. In order \alphastwo~both $\sigma_L$
and $\sigma_T$ contribute to $\sigma_{\rm tot}$.\\
Because of the high sensitivity of expression \eref{eq:5_2} to the value of \alphas,
the longitudinal cross section provides us with an excellent tool to measure the
running coupling constant.\\
To illustrate the dependence of the cross sections on the running coupling constant
we have plotted the ratios
\begin{eqnarray}
  \lefteqn{
    R_L(Q^2) = \frac{\sigma_L(Q^2)}{\sigma_{\rm tot}(Q^2)} = \frac{\alpha_s(R^2)}{4\pi}
    C_F[3] + \left(\frac{\alpha_s(R^2)}{4\pi}\right)^2\BigLeftHook C_F^2\BigLeftBrace
    - \frac{33}{2}\BigRightBrace + C_AC_F\BigLeftBrace}\nonumber\\[2ex]
  \label{eq:5_4}
  && {}-11\ln\frac{Q^2}{R^2} - 44\zeta(3) + \frac{123}{2}\BigRightBrace + n_fC_FT_f
  \BigLeftBrace 4\ln\frac{Q^2}{R^2} - \frac{74}{3}\BigRightBrace\BigRightHook,
\end{eqnarray}
and
\begin{equation}
  \label{eq:5_5}
  R_T(Q^2) = \frac{\sigma_T(Q^2)}{\sigma_{\rm tot}(Q^2)} =
  1 - \frac{\sigma_L(Q^2)}{\sigma_{\rm tot}(Q^2)},
\end{equation}
as a function of $Q$ (CM-energy of the $e^+e^-$ system) in \fref{fig:f11} and
\fref{fig:f12} respectively. In \fref{fig:f11} we have shown $R_L$ corrected up
to order \alphas~($R_L^{LO}$) and $R_L$ corrected up to order \alphastwo~($R_L^{NLO}$).
For $R_L^{LO}$ we have used as input the leading log running coupling constant
$\alpha_s^{LL}(n_f,\Lambda_{LO}^{(n_f)},R^2)$ with $n_f=5$ and
$\Lambda_{LO}^{(5)} = 108~\MeV$ ($\alpha_s^{LL}(M_Z) = 0.122$). The input parameters
of $R_L^{NLO}$ are given by the next-to-leading log running coupling constant
$\alpha_s^{NLL}(n_f,\Lambda_{\MS}^{(nf)},R^2)$ with $n_f=5$ and
$\Lambda_{\MS}^{(5)} = 227~\MeV$ ($\alpha_s^{NLL}(M_Z) = 0.118$). Further we have shown
the variation of $R_L$ on the renormalization scale $R$ by choosing the values
$R = Q/2,~Q,~2Q$. \Fref{fig:f11} reveals that the order \alphastwo~corrections are
appreciable and they vary from 48\% ($Q=10~\GeV$) down to 28\% ($Q=200~\GeV$)
with respect to the LO result. Furthermore one observes an improvement of the
renormalization scale dependence while going from $R_L^{LO}$ to $R_L^{NLO}$.
In \fref{fig:f12} we have plotted $R_T$ \eref{eq:5_5} up to first order ($R_T^{NLO}$)
and up to second order ($R_T^{NNLO}$) in the running coupling constant. As input
we have used for $R_T^{NLO}$ and $R_T^{NNLO}$ the coupling constants $\alpha_s^{LL}$
and $\alpha_s^{NLL}$ respectively. The features of \fref{fig:f12} are the same as those
observed in \fref{fig:f11}. In particular $R_T^{NNLO}$ becomes less dependent on the
renormalization scale as is shown for $R^{NLO}$. In \frefs{fig:f11},~\ref{fig:f12} we have
also presented the values $R_L$ and $R_T$ at $Q=M_Z$ measured by the OPAL-experiment
\cite{Ake95} which are given by
\begin{eqnarray}
  \label{eq:5_6}
  \lefteqn{R_L = 0.057\pm 0.005,}\\[2ex]
  \label{eq:5_7}
  \lefteqn{R_T = 0.943\pm 0.005.}
\end{eqnarray}
One observes a considerable improvement in the ratios $R_k$ ($k=T,~L$) when the order
\alphastwo~contributions are included. However there is still a little discrepancy between
$R_L^{NLO}$ and $R_T^{NNLO}$, taken at $R=Q=M_Z$, and the
data. This can either be removed by choosing a larger $\Lambda_{\MS}$ or by
including the masses of the heavy quarks $c$ and $b$ in the calculation of the
coefficient functions. Also a contribution of higher twist effects can maybe not
neglected (see \cite{Nas94,Web94}).\\
We now want to investigate the effect of the order \alphastwo~contributions to the
longitudinal and transverse fragmentation functions $F_L(x,Q^2)$ and $F_T(x,Q^2)$ as
defined in \eref{eq:2_17}.\\
For our analysis we have chosen the fragmentation density sets in \cite{Bin95},
\cite{Bin95b} which will be called BKK1 and BKK2 respectively. The input parameters
for $\alpha_s$ and the QCD scale $\Lambda$ are the same as given below \eref{eq:5_5}
except for BKK1 \cite{Bin95} where one has chosen
$\Lambda_{LO}^{(5)} = \Lambda_{\MS}^{(5)} = 190~\MeV$. The definitions for $F_L^{LO}$
and $F_L^{NLO}$ are the same as those given above for $R_L^{LO}$ and $R_L^{NLO}$
respectively. However for both $F_T^{NLO}$ (order \alphas~corrected) and
$F_T^{NNLO}$ (order \alphastwo~corrected) we use $\alpha_s^{NLL}(5,\Lambda_{\MS}^{(5)},R^2)$.
In \cite{Bin95,Bin95b} the
fragmentation densities $D_p^H(z,M^2)$ have been determined for $H=\pi^++\pi^-$,
$K^++K^-$ by fitting the total fragmentation function
$F(x,Q^2) = \sum_H\,F^H(x,Q^2)$ \eref{eq:2_17_1} with $H=\pi^\pm,~K^\pm,~P,~\bar{P}$
to the $e^+e^-$ data in the range $5.2 < Q < 91.2~\GeV$. Here the proton and anti-proton
contributions to the fragmentation functions have been estimated like
\begin{equation}
  \label{eq:5_8}
  F_k^{P+\bar{P}}(x,Q^2) = (1+f(x))\,F_k^{\pi^++\pi^-}(x,Q^2),
\end{equation}
with
\begin{eqnarray}
  \label{eq:5_9}
  \lefteqn{f(x) = 0.16, \hspace*{3mm}\mbox{in \cite{Bin95}},}\\[2ex]
  \label{eq:5_10}
  \lefteqn{f(x) = 0.195 - 1.35\,(x-0.35)^2, \hspace*{3mm}\mbox{in \cite{Bin95b}}.}
\end{eqnarray}
Further we introduce the notation
\begin{equation}
  \label{eq:5_11}
  F_k(x,Q^2) = \sum_H\,F_k^H(x,Q^2),
\end{equation}
where we sum over $H=\pi^+,~\pi^-,~K^+,~K^-,~P,~\bar{P}$.\\
Notice that $f(x)$ in \eref{eq:5_10} becomes negative when $x > 0.73$ so that
$F_k^{P+\bar{P}}(x,Q^2)$ ceases to be valid above this $x$-value. In \cite{Bin95}
(BKK1) the fit has been only made to the TPC/$2\gamma$-data \cite{Aih88} ($Q=29~\GeV$)
whereas in \cite{Bin95b} (BKK2) one also included the data coming from the ALEPH
\cite{Bus95,-Bus95b} and OPAL \cite{Ake93,-Ake94} collaboration. Since the range
of $Q$-values covered by the BKK2 parametrization is larger than the one given by
BKK1 the scale evolution of the fragmentation densities turns out to be better when
the BKK2-set \cite{Bin95b} is chosen. However this improvement goes at the expense
of the description of the longitudinal fragmentation function $F_L(x,Q^2)$ as we
will show below. For each set there exists a leading log and a next-to-leading log
parametrization of $D_p^H(z,M^2)$ ($p=q,g$). The latter is presented in the \MS-scheme
so that we have to choose the corresponding coefficient functions in appendix A.
Further we set the factorization scale $M$ equal to the renormalization scale $R$.\\
In \fref{fig:f13} we have plotted $F_L(x,Q^2)$ in LO and NLO at $M=Q=M_Z$ and compared
the results with the ALEPH \cite{Bus95,-Bus95b} and OPAL \cite{Ake95} data. Here
we have chosen the BKK1-set because the BKK2-set leads to a much worse result. The
latter already happens in LO as was noticed in \cite{Bin95b} where one had to choose
a very small factorization scale. Here we have chosen the BKK1-set. We observe
that $F_L^{LO}$ is below the data in particular in the small $x$-region. The
agreement with the data becomes better when the order \alphas~corrections are
included although at very small $x$ $F_L^{NLO}$ is still smaller than the values
given by experiment. In the case of the BKK2-set (not shown in the figure) one
gets a result which is far below the data. This was already noticed in fig.~5 of
\cite{Bin95b} where one had to choose a very small factorization scale $M$ ($M=20~\GeV$)
to bring $F_L^{LO}$ in agreement with experiment. In NLO the discrepancy between
$F_L^{NLO}$, in the case of BKK2, and the data becomes even larger which is due to the
kaon contribution. It turns out that the convolution of $D_p^{K^++K^-}(z,M^2)$
with the order \alphastwo~contribution from the coefficient functions given in
\eref{eq:2_4}, leads to a negative $F_L^{K^++K^-}$. This example illustrates
the importance of the measurement of $F_L(x,Q^2)$ and the higher order corrections for
the determination of the fragmentation densities. We have also shown the results
for $F_T^{NLO}$ and $F_T^{NNLO}$ at $M=Q=M_Z$ in \fref{fig:f14} using the BKK1-set.
Both fragmentation functions agree with the data except that $F_T^{NNLO}$ gets a little
bit worse at very small $x$. Furthermore $F_T^{NLO}$ and $F_T^{NNLO}$ hardly
differ from each other which means that the order \alphastwo~corrections are small.
We do not expect that this will change when the three-loop splitting functions
are included. One also notices that $F_L$ constitutes the smallest part of the
total fragmentation function $F=F_T+F_L$ which can be inferred from \frefs{fig:f13},~
\ref{fig:f14}. This in particular holds at large $x$ where $F_T >> F_L$. Hence a fit
of the fragmentation densities to the data of $F_T$ is not
sufficient to give a precise prediction for $F_L$ and one
has to include the data of the latter to provide us with better fragmentation
densities. This in particular holds for $D_g^H(z,M^2)$ in the small $z$-region. The
order \alphastwo~contribution to $F_L$ will certainly change the parametrization of the
gluon fragmentation density given by ALEPH in \cite{Bus95,-Bus95b} and OPAL in \cite{Ake95}.\\
To illustrate the effect of the order \alphastwo~contributions to the coefficient
functions calculated in this paper at various $e^+e^-$ collider energies we have
studied the $K$-factors
\begin{equation}
  \label{eq:5_12}
  K_L^H = \frac{F_L^{H,NLO}(x,Q^2)}{F_L^{H,LO}(x,Q^2)},
\end{equation}
\begin{equation}
  \label{eq:5_13}
  K_T^H = \frac{F_T^{H,NNLO}(x,Q^2)}{F_T^{H,NLO}(x,Q^2)}.
\end{equation}
In \fref{fig:f15} we have plotted \eref{eq:5_12} for $H=\pi^++\pi^-$ at
$Q=5.2,~10,~29,~35,~55,~91.2~\GeV$ choosing the BKK2-set since the latter shows a
better scale evolution. From \fref{fig:f15} one infers that the corrections are large
at small $x$ where they vary between 2 ($Q=5.2~\GeV$) and 1.4 ($Q=91.1~\GeV$). The
corrections become smaller when $x$ increases. A similar plot is made for
$K_T^{\pi^++\pi^-}$ in \fref{fig:f16}. Here the order \alphastwo~corrections are much
smaller than in the longitudinal case except at large $x$ where they are of the same size.
Furthermore at low $x$ the order \alphastwo~corrections become negative
($K_T^{\pi^++\pi^-} < 1$) which is already revealed by \fref{fig:f14} for $Q=M_Z$.
Again the largest correction occurs at smallest $Q$. This
can be mainly attributed to the running coupling constant which
becomes large when $Q$ gets small. In \fref{fig:f17} we investigate the dependence of
$K_T$ \eref{eq:5_13}
on the specific set of fragmentation densities used. The same was done for $K_L$
\eref{eq:5_12} in \cite{Rij96} where we compared the BKK1-set with the one in
\cite{Nas94} which is presented in the A-scheme. It turned out that $K_L^H$ is very
sensitive to the chosen parametrization. Choosing the set in \cite{Nas94} and
$M=Q=M_Z$ it turns out that $K_L$ is
mildly dependent on $x$. Using the same input a similar
observation can be made for $K_T$ which shows a constant behaviour over the
whole $x$-region (see \fref{fig:f17}). On the other hand $K_L$ (see fig.~2 in
\cite{Rij96}) and $K_T$ steeply rise when $x$ tends to one if the BKK1 or BKK2 sets
are chosen. Hence we conclude that the $K$-factors heavily depend on the chosen
parametrization for the fragmentation densities.\\
In the next figures we study the factorization scale dependence of the fragmentation
functions and show the decrease in sensitivity on the scale choice for $M$ when
higher order corrections are included.\\
In \fref{fig:f18} we have plotted $F_L^{NLO}(x,Q^2)$ at three different scales
$M=Q/2,~Q,~2Q$ where $Q=M_Z$. Like in \fref{fig:f13} we have chosen the BKK1-set
since in this case we get agreement with the data. From \fref{fig:f18} one
infers that the scale variation of $F_L$ is small and that all scales describe
the data rather well. To show the improvement in the scale dependence more clearly
it is convenient to plot the following quantity
\begin{eqnarray}
  \label{eq:5_14}
  \lefteqn{
    \Delta_k^r(x,Q^2) = \frac{{\rm max}\left(F_k^r(\frac{1}{2}Q),F_k^r(Q),
      F_k^r(2Q)\right) - {\rm min}\left(F_k^r(\frac{1}{2}Q),F_k^r(Q),
      F_k^r(2Q)\right)}
    {{\rm average}\left(F_k^r(\frac{1}{2}Q),F_k^r(Q),F_k^r(2Q)\right)},}\nonumber\\[2ex]
  &&
\end{eqnarray}
for $r=LO,~NLO,~NNLO$ and $k=T,~L$.\\
In \fref{fig:f19} one can see that $\Delta_L^{NLO} < \Delta_L^{LO}$ as long as
$x > 0.1$ which implies that $F_L^{NLO}$ is less sensitive to the scale $M$ than
$F_L^{LO}$. The same observation is made for $F_T$ (\fref{fig:f20}) and
$\Delta_T$ (\fref{fig:f21}). In \fref{fig:f20} we have plotted $F_T^{NLO}$ at the same
scales as above. At all three scales the data are described very well. In addition
we also show $F_T^{NNLO}$ for $M=Q=M_Z$. At very small $x$ we found that
$F_T^{NNLO} < F_T^{NLO}$
for all three scales and $F_T^{NNLO}$ is also below the data. In \fref{fig:f21}
we see again an improvement while going from LO to NNLO except when $x < 0.1$.
The reason that for $x < 0.1$ the scale dependence of $F_k^{NLO}$ ($k=T,~L$) is
much larger than the one found in $F_k^{LO}$ can be found in \cite{Bin95} where is
stated that the scale evolution of the fragmentation densities is only reliable
in the region $0.1 < x < 0.9$.\\
Finally we also study the scale dependence of the fragmentation functions at a lower
energy. As an example we take the total fragmentation function $F^H$ with
$H=\pi^++\pi^-$ \eref{eq:2_17_1} and investigate its behaviour for different choices of
the factorization scale $M$ where again $M=Q/2,~Q,~2Q$. Contrary to the previous plots
we have chosen
the BKK2-set which range of validity is bounded by $Q\leq 100~\GeV$ and $0.1 < x < 0.8$.
Further we take $Q=29~\GeV$ and compare the theoretical result with the TPC/$2\gamma$-data
\cite{Aih88}. In \fref{fig:f22} we show $F^{H,NLO}$ at three different scales. The
scale variation is small and only noticeable at large $x$. The data are very well
described by $F^{H,NLO}$ at the three different scales except at very small $x$ where
the BKK2 parametrization is not reliable anymore. The same holds for $F^{H,NNLO}$
which hardly differs from $F^{H,NLO}$ so that even at lower energies the order
\alphastwo~corrections are very small. To show
the improvement of the scale dependence in a better way we have plotted
$\Delta^{H,LO}$, $\Delta^{H,NLO}$ and $\Delta^{H,NNLO}$ in \fref{fig:f23}. A
comparison with \fref{fig:f21} shows that there is essentially no difference between
the $\Delta_k^r$ ($r=LO,~NLO,~NNLO$, $k=T,~L$) taken at low ($Q=29~\GeV$) and high
energies ($Q=M_Z=91.2~\GeV$).\\
Summarizing our paper we have computed the order \alphastwo~contributions to the
longitudinal and transverse coefficient functions. The effect of these
contributions to the longitudinal and transverse cross sections are large which
allow us for a better determination of the strong coupling constant \alphas. The
corrections to the longitudinal fragmentation function $F_L^H$ are appreciable
too which has important consequences for the determination of the gluon fragmentation
density $D_g^H(z,M^2)$.\\
Furthermore one can now make a full NLO analysis of $F_L^H$. A NNLO description
of the transverse structure function is still not possible because of the missing
three-loop DGLAP splitting functions. However the order \alphastwo~contributions from
the coefficient functions indicate that probably the NNLO corrections are very small.
%

\newpage
\appendix
\section{The coefficient functions in the \MS-scheme}
In this appendix we will present the explicit expressions for the coefficient functions of the
fragmentation process in \eref{eq:2_1} which are calculated in section 4 in the \MS-scheme.
In order to make the presentation self contained we also give the order \alphas~contributions
$\bar{c}_{k,p}^{(1)}$ \eref{eq:4_40}-\eref{eq:4_46} which have already been presented in the
literature \cite{Nas94,Bai79,Alt79b}. The coefficient functions $\overline{\mathbb{C}}_{k,p}$
($p=q,g$)
will be expanded in the renormalized coupling constant $\alpha_s\equiv\alpha_s(M^2)$
where we have chosen the renormalization scale $R$ to be equal to the factorization scale $M$.
If one wants to chose $R$ different from $M$, $\alpha_s(M^2)$ has to be expressed into
$\alpha_s(R^2)$ following the prescription in \eref{eq:4_50}. In this paper we will only
present
the expressions for the transverse coefficient functions since the longitudinal ones are
already shown in \cite{Rij96}. However we will make an exception for the order
\alphastwo~corrections
to the non-singlet part $\mathbb{C}_{L,q}^\NonSinglet$. For future purposes we want to split it
into
a part due to identical quark contributions (AB and CD in fig.~\ref{fig:f8}) represented
by $\bar{c}_{L,q}^{\NonSinglet,(2),{\rm id}}$ and a remaining part given by
$\bar{c}_{L,q}^{\NonSinglet,(2),{\rm nid}}$ (see \eref{eq:4_40} and the discussion below
\eref{eq:4_12}). The same will be done for the transverse
coefficient $\bar{c}_{T,q}^{\NonSinglet,(2)}$ in \eref{eq:4_41}. The expression for the non-singlet
coefficient is very long and we will split it up to the various contributions. First we have
the soft plus virtual gluon contributions which are represented by the distributions
$\delta(1-z)$ and ${\cal D}_0(z)$ \eref{eq:3_35_1}. They are indicated by
$\left.\mathbb{C}_{T,q}^\NonSinglet\right|_{S+V}$. The remaining part which is integrable at $z=1$
will be called $\left.\mathbb{C}_{T,q}^\NonSinglet\right|_H$ where $H$ refers to hard gluon
contributions although $\left.\mathbb{C}_{T,q}^\NonSinglet\right|_H$ also originates from
subprocesses with (anti) quarks in the final state
(see fig.~\ref{fig:f8} except for $C^2$ and $D^2$). Following
this prescription the non-singlet coefficient function is constituted by the following parts
\begin{eqnarray}
  \lefteqn{
    \overline{\mathbb{C}}_{T,q}^\NonSinglet =
    \overline{\mathbb{C}}_{T,q}^{\NonSinglet,{\rm nid}} +
    \overline{\mathbb{C}}_{T,q}^{\NonSinglet,{\rm id}},}\\[2ex]
  \lefteqn{
    \overline{\mathbb{C}}_{T,q}^{\NonSinglet,{\rm nid}} =
    \left.\overline{\mathbb{C}}_{T,q}^{\NonSinglet,{\rm nid}}\right|_{S+V} +
    \left.\overline{\mathbb{C}}_{T,q}^{\NonSinglet,{\rm nid}}\right|_H,}
\end{eqnarray}
\begin{eqnarray}
  \lefteqn{
    \left.\overline{\mathbb{C}}_{T,q}^{\NonSinglet,{\rm nid}}\right|_{S+V} = \delta(1-z) + C_F\,
    \frac{\alpha_s}{4\pi}
    \BigLeftHook (4{\cal D}_0(z) + 3\delta(1-z))L_M + 4 {\cal D}_1(z) - 3 {\cal D}_0(z)}
  \nonumber\\[2ex]
  && {}+ \delta(1-z)(- 9 + 8 \zeta(2)) \BigRightHook \nonumber\\[2ex]
  && {}+ \left(\frac{\alpha_s}{4\pi}\right)^2\BigLeftHook C_F^2 \BigLeftBrace (16 {\cal D}_1(z)
  + 12 {\cal D}_0(z)
  ) L_M^2 + (24 {\cal D}_2(z) - 12 {\cal D}_1(z) + (16 \zeta(2) \nonumber\\[2ex]
  && {}- 45) {\cal D}_0(z)) L_M + \delta(1-z)
  \BigLeftHook (\frac{9}{2} - 8\zeta(2)) L_M^2 + (40 \zeta(3) + 24\zeta(2) \nonumber\\[2ex]
  && - \frac{51}{2}) L_M\BigRightHook\BigRightBrace \nonumber \\[2ex]
  && {}+ C_A C_F \BigLeftBrace (-\frac{22}{3} {\cal D}_0(z)) L_M^2 + (-\frac{44}{3} {\cal D}_1(z) + (
  \frac{367}{9} - 8\zeta(2)) {\cal D}_0(z)) L_M \nonumber\\[2ex]
  && {}+ \delta(1-z)
  \BigLeftHook -\frac{11}{2} L_M^2 + (-12 \zeta(3) - \frac{44}{3}\zeta(2) + \frac{215}{6})
  L_M \BigRightHook\BigRightBrace \nonumber\\[2ex]
  && {}+n_f C_F T_f \BigLeftBrace (\frac{8}{3} {\cal D}_0(z)) L_M^2 + (\frac{16}{3} {\cal D}_1(z)
  - \frac{116}{9} {\cal D}_0(z)) L_M + \delta(1-z) \BigLeftHook 2 L_M^2 \nonumber\\[2ex]
  && {}+ (\frac{16}{3}\zeta(2)
  - \frac{38}{3}) L_M\BigRightHook\BigRightBrace +
  \left.\bar{c}_{T,q}^{\NonSinglet,(2),{\rm nid}}\right|_{S+V}\BigRightHook,
\end{eqnarray}
with
\begin{eqnarray}
  \lefteqn{
    \left.\bar{c}_{T,q}^{\NonSinglet,(2),{\rm nid}}\right|_{S+V} = C_F^2\BigLeftHook
      8 {\cal D}_3(z) - 18 {\cal D}_2(z)
      + (16 \zeta(2) - 27) {\cal D}_1(z) + (-8 \zeta(3)} \nonumber\\[2ex]
  && {}+ \frac{51}{2}) {\cal D}_0(z)
  + \delta(1-z)\left(30 \zeta(2)^2 - 78 \zeta(3) - 39 \zeta(2) + \frac{331}{8}\right)
  \BigRightHook\nonumber\\[2ex]
  && {}+C_AC_F\BigLeftHook - \frac{22}{3} {\cal D}_2(z) + (\frac{367}{9} - 8\zeta(2))
  {\cal D}_1(z) + (40 \zeta(3) + \frac{44}{3}\zeta(2) \nonumber\\[2ex]
  && {}- \frac{3155}{54}) {\cal D}_0(z) + \delta(1-z)\left( - \frac{49}{5} \zeta(2)^2
  + \frac{140}{3} \zeta(3) + \frac{215}{3} \zeta(2) - \frac{5465}{72}\right)
  \BigRightHook\nonumber\\[2ex]
  && {}+n_fC_FT_f\BigLeftHook\frac{8}{3} {\cal D}_2(z)
  - \frac{116}{9} {\cal D}_1(z) + (\frac{494}{27}
  - \frac{16}{3} \zeta(2)) {\cal D}_0(z)\nonumber\\[2ex]
  && {}+ \delta(1-z)\left(\frac{8}{3}\zeta(3) - \frac{76}{3}\zeta(2) + \frac{457}{18}\right)
  \BigRightHook,\\[2ex]
  \label{eq:A_4}
  \lefteqn{
    \left.\overline{\mathbb{C}}_{T,q}^{\NonSinglet,{\rm nid}}\right|_H = C_F\,\frac{\alpha_s}{4\pi}
    \BigLeftHook
    -2 (1+z) L_M -2 (1+z)\ln(1-z) + 4\frac{1+z^2}{1-z}\ln z}\nonumber\\[2ex]
  && {}+ 3(1-z)\BigRightHook
  +\left(\frac{\alpha_s}{4\pi}\right)^2 \BigLeftHook C_F^2\BigLeftBrace \BigLeftHook
  -8 (1+z) (\ln(1-z)-\frac{3}{4}\ln z) - \frac{8}{1-z}\ln z \nonumber\\[2ex]
  && {}- 10 - 2 z\BigRightHook L_M^2
  + \BigLeftHook 4 (1+z)(\Li_2(1-z) - 3\ln z\ln(1-z) - 3\ln^2(1-z) \nonumber\\[2ex]
  && {}+ \frac{11}{2}\ln^2 z
  - 2\zeta(2)) + \frac{32}{1-z}(\ln z \ln(1-z) - \ln^2 z + \frac{3}{2}\ln z) \nonumber\\[2ex]
  && {}+ 4 (1-z)\ln(1-z)
  - (52 + 20 z)\ln z + 5 + 31 z\BigRightHook L_M\BigRightBrace \nonumber\\[2ex]
  && {}+ C_A C_F\BigLeftBrace\BigLeftHook \frac{11}{3}(1+z) L_M^2 + \BigLeftHook
  (1+z)(-2\ln^2 z + 4\zeta(2) + \frac{22}{3}\ln(1-z) \nonumber\\[2ex]
  && {}+ \frac{34}{3}\ln z) + \frac{1}{1-z}
  (4\ln^2 z- \frac{44}{3}\ln z) + \frac{7}{9} - \frac{275}{9}z\BigRightHook L_M\BigRightBrace
  \nonumber\\[2ex]
  && {}+ n_f C_F T_f \BigLeftBrace -\frac{4}{3}(1+z) L_M^2 + \BigLeftHook -\frac{8}{3}(1+z)
  (\ln(1-z) + \ln z) \nonumber\\[2ex]
  && {}+ \frac{16}{3}\frac{1}{1-z}\ln z + \frac{28}{9} + \frac{52}{9} z
  \BigRightHook L_M \BigRightBrace + \left.\bar{c}_{T,q}^{\NonSinglet,(2),{\rm nid}}\right|_H
  \BigRightHook,
\end{eqnarray}
with
\begin{eqnarray}
  \lefteqn{
    \left.\bar{c}_{T,q}^{\NonSinglet,(2),{\rm nid}}\right|_H = C_F^2\BigLeftHook 16 (1+2 z)(-2
    \Li_3(-z) + \ln z\Li_2(-z)) + \frac{1}{1-z}(96 \Li_3(-z)} \nonumber\\[2ex]
  && {}+ 72\zeta(3) - 48\ln z \Li_2(-z) - 192
  S_{1,2}(1-z) - 24 \Li_3(1-z) + 8\ln(1-z)\cdot
  \nonumber\\[2ex]
  && \cdot\Li_2(1-z) - 80\ln z\Li_2(1-z) + 20\ln z\ln^2(1-z)
  - 4\ln^2 z\ln(1-z)
  \nonumber\\[2ex]
  && {}- \frac{80}{3}\ln^3 z + 120\zeta(2)\ln z + 12\Li_2(1-z)
  - 12\ln z\ln(1-z) + 33\ln^2 z
  \nonumber\\[2ex]
  && {}- 106 \ln z) + (1+z)( 8\Li_3(1-z) + 52\ln z\Li_2(1-z)
  - 8\ln z\ln^2(1-z)
  \nonumber\\[2ex]
  && {}+ 8\ln^2z \ln(1-z) - 64\zeta(2)\ln z - 32\zeta(3) + 17\ln^3 z
  - 4 \ln^3(1-z))
  \nonumber\\[2ex]
  && {}+ (100 + 116 z)S_{1,2}(1-z) + (-16 z)\zeta(2)\ln(1-z) + (-48+24 z)\cdot
  \nonumber\\[2ex]
  && \cdot\Li_2(1-z) - (20+4 z)\ln z\ln(1-z) + (8+4 z)\ln^2(1-z) + (-45 \nonumber\\[2ex]
  && {}- 23 z + 8 z^2
  + \frac{12}{5} z^3)\ln^2 z + (20 - 36 z - 16 z^2 - \frac{24}{5} z^3) \zeta(2)
  + (-\frac{24}{5z^2}
  \nonumber\\[2ex]
  && {}- \frac{16}{z} + 8 + 8 z - 16 z^2 - \frac{24}{5} z^3)(\Li_2(-z)
  + \ln z\ln(1+z)) + (-29 + 67 z)\cdot
  \nonumber\\[2ex]
  && \cdot\ln(1-z) + (\frac{24}{5z} + \frac{218}{5} + \frac{248}{5} z
  + \frac{24}{5} z^2) \ln z - \frac{24}{5z} + \frac{187}{10} - \frac{187}{10} z
  + \frac{24}{5} z^2 \BigRightHook
  \nonumber\\[2ex]
  && {}+ C_A C_F \BigLeftHook 8(1+2 z)(2\Li_3(-z)-\ln z\Li_2(-z)) + \frac{1}{1-z}(-48\Li_3(-z)
  \nonumber\\[2ex]
  && {}- 36\zeta(3) + 24\ln z\Li_2(-z) + 24\Li_3(1-z) - 8\ln(1-z)\Li_2(1-z) \nonumber\\[2ex]
  && {}- 8\ln z\Li_2(1-z)
  + 6\ln^3 z - 24\zeta(2)\ln z - \frac{44}{3}\Li_2(1-z)
  \nonumber\\[2ex]
  && - \frac{44}{3}\ln z\ln(1-z)
  -\frac{11}{3}\ln^2 z+\frac{206}{3}\ln z) + (1+z)(-12\Li_3(1-z)
  \nonumber\\[2ex]
  && {}+ 4\ln(1-z)\Li_2(1-z)
  + 4\ln z\Li_2(1-z) + 12\zeta(2)\ln z - 2\zeta(3) - 3\ln^2 z
  \nonumber\\[2ex]
  && {}+ \frac{34}{3}\Li_2(1-z)
  + \frac{34}{3}\ln z\ln(1-z) + \frac{11}{3}\ln^2(1-z)) + (4-4 z)S_{1,2}(1-z)
  \nonumber\\[2ex]
  && {}+ (8 z)\zeta(2)\ln(1-z)
  + (\frac{47}{6} + \frac{47}{6} z - 4 z^2 - \frac{6}{5} z^3)\ln^2 z + (-\frac{28}{3}
  - \frac{28}{3} z + 8 z^2
  \nonumber\\[2ex]
  && {}+ \frac{12}{5} z^3)
  \zeta(2) + (\frac{12}{5z^2} + \frac{8}{z}
  - 4 - 4 z + 8 z^2 + \frac{12}{5} z^3)(\Li_2(-z) + \ln z\ln(1+z))
  \nonumber\\[2ex]
  && {}+ (\frac{97}{9}
  - \frac{383}{9} z)\ln(1-z) - (\frac{12}{5z} + \frac{122}{15} + \frac{184}{5} z
  + \frac{12}{5} z^2)\ln z + \frac{12}{5z}
  \nonumber\\[2ex]
  && {}+ \frac{2513}{270} + \frac{3587}{270} z
  - \frac{12}{5} z^2\BigRightHook
  \nonumber\\[2ex]
  && {}+n_f C_F T_f\BigLeftHook \frac{8}{3}\frac{1+z^2}{1-z}(\Li_2(1-z) + \ln z\ln(1-z)
  + \frac{1}{4}\ln^2 z) + \frac{4}{3}(1+z)\cdot
  \nonumber\\[2ex]
  && \cdot(2 \zeta(2) - \ln^2(1-z)) + (\frac{28}{9}
  + \frac{52}{9} z)\ln(1-z) + (\frac{20}{3} + 12 z - \frac{64}{3}\frac{1}{1-z})\ln z
  \nonumber\\[2ex]
  && {}- \frac{118}{27} - \frac{34}{27} z\BigRightHook, \\[2ex]
  \label{eq:A_7}
  \lefteqn{
    \overline{\mathbb{C}}_{T,q}^{\NonSinglet,(2),{\rm id}} = \left(\frac{\alpha_s}{4\pi}\right)^2
    \BigLeftHook
    (C_F^2 - \frac{1}{2} C_A C_F)\BigLeftBrace\BigLeftHook 4\frac{1+z^2}{1+z}(
    -4 \Li_2(-z) - 4\ln z\ln(1+z)} \nonumber\\[2ex]
  && {}+ \ln^2 z
  - 2\zeta(2)) + 8(1+z)\ln(z) + 16(1-z)\BigRightHook
  L_M \BigRightBrace + \bar{c}_{T,q}^{\NonSinglet,(2),{\rm id}}\BigRightHook,\\[2ex]
  \label{eq:A_6}
  \lefteqn{
    \bar{c}_{T,q}^{\NonSinglet,(2),{\rm id}} =
    (C_F^2 - \frac{1}{2}C_A C_F)\BigLeftHook 16 \frac{1+z^2}{1+z}(
    \Li_3\left(\frac{1-z}{1+z}\right) - \Li_3\left(-\frac{1-z}{1+z}\right)}
  \nonumber\\[2ex]
  && {}+ \frac{1}{2}S_{1,2}(1-z) - \Li_3(1-z)
  - S_{1,2}(-z) + \frac{1}{2}\Li_3(-z) + \frac{1}{2}\ln z\Li_2(1-z) \nonumber\\[2ex]
  && {}- \ln(1-z)\Li_2(-z)
  - \ln(1+z)\Li_2(-z) - \ln z\Li_2(-z) - \frac{1}{2}\zeta(2)\ln(1-z) \nonumber\\[2ex]
  && {}- \frac{1}{2}\zeta(2)
  \ln(1+z) - \zeta(2)\ln z - \ln z\ln(1-z)\ln(1+z) + \frac{1}{4}\ln^2 z\ln(1-z)
  \nonumber\\[2ex]
  && {}- \frac{1}{2}\ln z\ln^2(1+z) - \frac{3}{4}\ln^2 z\ln(1+z) + \frac{3}{8}\ln^3 z
  + \frac{1}{2}\zeta(3) - \frac{1}{2}\ln z)
  \nonumber\\[2ex]
  && {}+ 16(1+z)(-S_{1,2}(-z) + \frac{1}{2}\Li_3(-z) - \ln(1+z)\Li_2(-z)
  - \frac{1}{2}\ln z\ln^2(1+z)
  \nonumber\\[2ex]
  && {}+ \frac{1}{4}\ln^2 z\ln(1+z) - \frac{1}{2}\zeta(2)\ln(1+z)
  + \frac{1}{2}\zeta(3)+ \frac{1}{2}\Li_2(1-z) \nonumber\\[2ex]
  && {}+ \frac{1}{2}\ln z\ln(1-z)) + (-\frac{24}{5z^2}
  + \frac{16}{z} + 16 z^2 - \frac{24}{5} z^3)(\Li_2(-z)
   \nonumber\\[2ex]
   && {}+ \ln z\ln(1+z)) + (-4 + 4 z + 16 z^2
   - \frac{24}{5} z^3)\zeta(2) + (12 + 8 z- 8 z^2 \nonumber\\[2ex]
   && {}+ \frac{12}{5} z^3)\ln^2 z + 16 (1-z)\ln(1-z)
   + (\frac{24}{5z^2} + \frac{118}{5} - \frac{42}{5} z + \frac{24}{5} z^2)\ln z
   \nonumber\\[2ex]
   && {}- \frac{24}{5z} + \frac{46}{5} - \frac{46}{5} z + \frac{24}{5} z^2\BigRightHook.
\end{eqnarray}
Notice that the identical quark contribution in \eref{eq:A_7} and \eref{eq:A_6} carries the
colour factor $C_F^2-\frac{1}{2}C_AC_F$.\\
The purely singlet coefficient function becomes
\begin{eqnarray}
  \lefteqn{
    \overline{\mathbb{C}}_{T,q}^\PureSinglet = \left(\frac{\alpha_s}{4\pi}\right)^2
    \BigLeftHook C_FT_f \BigLeftBrace 8(1+z)\ln z + \frac{16}{3z}
    + 4 - 4 z - \frac{16}{3} z^2
    \BigRightBrace L_M^2}\nonumber\\[2ex]
  && {}+ C_FT_f \BigLeftBrace 16(1+z)(\Li_2(1-z) + \ln z\ln(1-z)
  + \frac{3}{2}\ln^2 z + (\frac{32}{3z} + 8 - 8 z \nonumber\\[2ex]
  && {}- \frac{32}{3} z^2)\ln(1-z) + (\frac{64}{3z} - 8 - 40 z - \frac{32}{3} z^2)\ln z
  - \frac{16}{3z} - \frac{184}{3}
  + \frac{136}{3} z \nonumber\\[2ex]
  && {}+ \frac{64}{3} z^2\BigRightBrace L_M + \bar{c}_{T,q}^{\PureSinglet,(2)}\BigRightHook,\\[2ex]
  \lefteqn{
    \bar{c}_{T,q}^{\PureSinglet,(2)} = C_FT_f \BigLeftHook 8(1+z)( 6S_{1,2}(1-z)
    - 2\Li_3(1-z) + 2\ln(1-z)\Li_2(1-z)} \nonumber\\[2ex]
  && {}+ 6\ln z\Li_2(1-z) - 2\zeta(2)\ln z + \ln z\ln^2(1-z)
  + 3\ln^2 z\ln(1-z) \nonumber\\[2ex]
  && {}+ \frac{11}{6}\ln^3 z + (\frac{64}{3z} - 8 - 40 z - \frac{32}{3} z^2)
  (\Li_2(1-z) + \ln z\ln(1-z)) + (\frac{16}{3z} + 4 \nonumber\\[2ex]
  && {}- 4 z- \frac{16}{3} z^2)\ln^2(1-z)
  + (\frac{64}{3z} - 14 - 14 z + \frac{16}{3} z^2)\ln^2 z - (\frac{32}{3z} + 32
  + 32 z \nonumber\\[2ex]
  && {}+ \frac{32}{3} z^2)(\Li_2(-z) + \ln z\ln(1+z)) + (-\frac{64}{3z} - 8 - 24 z
  + \frac{32}{3} z^2)\zeta(2) + (-\frac{16}{3z} \nonumber\\[2ex]
  && {}- \frac{184}{3} + \frac{136}{3} z
  + \frac{64}{3} z^2)\ln(1-z) - (\frac{32}{3z} + \frac{400}{3} + \frac{208}{3} z
  + \frac{256}{9} z^2)\ln z \nonumber\\[2ex]
  && {}- \frac{160}{27z} - \frac{236}{3} + \frac{140}{3} z
  + \frac{1024}{27} z^2\BigRightHook.
\end{eqnarray}
The gluonic coefficient function is equal to
\begin{eqnarray}
  \lefteqn{
    \overline{\mathbb{C}}_{T,g} = C_F\,\frac{\alpha_s}{4\pi}\BigLeftHook
    \BigLeftBrace \frac{8}{z} - 8 + 4 z\BigRightBrace L_M + (\frac{8}{z}
    - 8 + 4 z)(\ln(1-z) + 2\ln z) -\frac{8}{z} + 8\BigRightHook}\nonumber\\[2ex]
  && {}+\left(\frac{\alpha_s}{4\pi}\right)^2\BigLeftHook C_F^2\BigLeftBrace\BigLeftHook
  (\frac{16}{z} - 16 + 8 z)\ln(1-z) + (8 - 4 z)\ln z + 8 - 2 z\BigRightHook L_M^2
  \nonumber\\[2ex]
  && {}+ \BigLeftHook (\frac{64}{z} - 48 + 24 z)(\Li_2(1-z) + \ln z\ln(1-z)) + (\frac{32}{z}
  - 32 + 16 z)\cdot \nonumber\\[2ex]
  && \cdot(\ln^2(1-z) - 2\zeta(2)) + (24 - 12 z)\ln^2 z + (-\frac{48}{z} + 64 - 12 z
  )\ln(1-z) \nonumber\\[2ex]
  && {}+ (-16 + 20 z)\ln z + \frac{8}{z} - 20 + 8z \BigRightHook L_M
  \nonumber\\[2ex]
  && {}+ C_A C_F\BigLeftBrace\BigLeftHook (\frac{16}{z} - 16 + 8 z)\ln(1-z)
  - (\frac{16}{z} + 16 + 16 z)\ln z - \frac{124}{3z} \nonumber\\[2ex]
  && {}+ 32 + 4 z + \frac{16}{3} z^2\BigRightHook
  L_M^2 + \BigLeftHook -(\frac{64}{z} + 48 z)\Li_2(1-z) - (64 + 16 z)\cdot\nonumber\\[2ex]
  && \cdot\ln z\ln(1-z)
  + (\frac{32}{z} + 32 + 16 z)(\Li_2(-z) + \ln z\ln(1+z)) + (\frac{96}{z} \nonumber\\[2ex]
  && {}- 64 + 48 z) \zeta(2)
  + (\frac{16}{z} - 16 + 8 z)\ln^2(1-z) - (\frac{64}{z} + 48 + 56 z)\ln^2 z
  \nonumber\\[2ex]
  && {}+ (-\frac{344}{3z}
  + 96 - 8z + \frac{32}{3} z^2) \ln(1-z) + (-\frac{400}{3z} + 64 + 40 z + \frac{32}{3} z^2)
  \ln z \nonumber\\[2ex]
  && {}+ \frac{188}{3z} - \frac{32}{3} - \frac{100}{3} z - \frac{32}{3} z^2\BigRightHook L_M
  + \bar{c}_{T,g}^{(2)} \BigRightHook,\\[2ex]
  \lefteqn{
    \bar{c}_{T,g}^{(2)} = C_F^2\BigLeftHook (\frac{16}{z} + 32 + 16 z)(4 S_{1,2}(-z)
    + 4\ln(1+z)\Li_2(-z) + 2\ln z\ln^2(1+z)} \nonumber\\[2ex]
  && {}- \ln^2 z\ln(1+z) + 2\zeta(2)\ln(1+z))
  + (\frac{32}{z} - 192 + 32 z)\Li_3(-z) + (-\frac{160}{z} \nonumber\\[2ex]
  && {}+ 240 - 120 z) S_{1,2}(1-z)
  - \frac{16}{z}\Li_3(1-z) + (\frac{48}{z} - 32 + 16 z)\ln(1-z)\cdot\nonumber\\[2ex]
  && \cdot\Li_2(1-z)
  + (48-24 z)\ln z\Li_2(1-z) + (-\frac{32}{z} + 64 - 32 z)\ln z\Li_2(-z)
  \nonumber\\[2ex]
  && {}+ (\frac{40}{3z}
  - \frac{40}{3} + \frac{20}{3} z)\ln^3(1-z) + (\frac{44}{3} - \frac{22}{3}z)\ln^3 z
  + (\frac{48}{z} - 40 + 20 z)\cdot\nonumber\\[2ex]
  && \cdot\ln z\ln^2(1-z) + (\frac{32}{z} - 8 + 4 z)\ln^2 z\ln(1-z)
  + (\frac{16}{z} - 48 + 24 z)\cdot\nonumber\\[2ex]
  && \cdot\zeta(2)\ln(1-z) + (-16 + 8 z)\zeta(2)\ln z
  + (\frac{112}{z} -208 + 40 z)\zeta(3) + (-\frac{96}{z} \nonumber\\[2ex]
  && {}+ 80 - 28 z)\Li_2(1-z)
  + (-\frac{96}{z} + 80 - 12 z)\ln z\ln(1-z) + (-\frac{64}{5z^2} \nonumber\\[2ex]
  && {}- 64 - 96 z + \frac{16}{5}z^3)
  (\Li_2(-z) + \ln z\ln(1+z)) + (-\frac{48}{z} + 56 - 14 z)\cdot\nonumber\\[2ex]
  && \cdot\ln^2(1-z) + (-32 + 83 z
  - \frac{8}{5} z^3)\ln^2 z + (\frac{96}{z} - 112 - 44 z + \frac{16}{5} z^3)\zeta(2)
  \nonumber\\[2ex]
  && {}+ (\frac{8}{z} - 24 z)\ln(1-z) + (\frac{144}{5z} + \frac{418}{5} - \frac{262}{5} z
  - \frac{16}{5} z^2)\ln z + \frac{316}{5z} - \frac{604}{5} \nonumber\\[2ex]
  && {}+ \frac{154}{5} z
  - \frac{16}{5} z^2 \BigRightHook\nonumber\\[2ex]
  && {}+C_A C_F \BigLeftHook (\frac{16}{z} + 32 + 16 z)(-2 S_{1,2}(-z) -2\ln(1+z)\Li_2(-z)
  \nonumber\\[2ex]
  && {}+ \ln z\Li_2(-z) - \ln z\ln^2(1+z) - \zeta(2)\ln(1+z)) + (\frac{32}{z} + 32 + 16 z)
  \cdot\nonumber\\[2ex]
  && \cdot(\Li_3\left(-\frac{1-z}{1+z}\right) - \Li_3\left(\frac{1-z}{1+z}\right)
  + \ln z\ln(1-z)\ln(1+z) + \ln(1-z)\cdot \nonumber\\[2ex]
  && \cdot\Li_2(-z)) + (\frac{48}{z} + 32 + 16 z)\Li_3(-z) + (-\frac{240}{z} - 160 z)
  S_{1,2}(1-z) + (\frac{48}{z} \nonumber\\[2ex]
  && {}+ 80 + 40 z)\Li_3(1-z) - (\frac{48}{z} + 16 + 40 z)
  \ln(1-z)\Li_2(1-z) - (\frac{192}{z} + 32 \nonumber\\[2ex]
  && {}+ 144 z)\ln z\Li_2(1-z) + (\frac{8}{3z}
  - \frac{8}{3} + \frac{4}{3} z)\ln^3(1-z) - (\frac{160}{3z} + \frac{88}{3}
  \nonumber\\[2ex]
  && {}+ \frac{124}{3} z)\ln^3 z - (\frac{8}{z} + 24 + 12 z)\ln z\ln^2(1-z) - (\frac{48}{z}
  + 64 + 48 z)\cdot\nonumber\\[2ex]
  && \cdot\ln^2 z\ln(1-z) + (\frac{16}{z} + 32)\zeta(2)\ln(1-z) + (\frac{224}{z}
  - 96 + 128 z)\zeta(2)\ln z \nonumber\\[2ex]
  && {}+ (\frac{40}{z} + 48 + 24 z)\ln^2 z\ln(1+z) + (\frac{8}{z}
  + 40 + 12 z)\zeta(3) + (-\frac{304}{3z} + 32 \nonumber\\[2ex]
  && {}+ 24 z+ \frac{32}{3} z^2)\Li_2(1-z)
  + (-\frac{496}{3z} + 96 + 16 z + \frac{32}{3} z^2)\ln z\ln(1-z) \nonumber\\[2ex]
  && {}+ (\frac{80}{3z} + 80
  + 56 z + \frac{32}{3} z^2)(\Li_2(-z) + \ln z \ln(1+z)) + (-\frac{172}{3z} + 48 - 8 z
  \nonumber\\[2ex]
  && {}+ \frac{16}{3} z^2)\ln^2(1-z) + (-\frac{232}{3z} - 16 + 2 z - \frac{16}{3} z^2)\ln^2 z
  + (\frac{136}{3z} + 40 z \nonumber\\[2ex]
  && {}- \frac{32}{3} z^2)\zeta(2) + (\frac{356}{3z} - \frac{236}{3}
  - \frac{4}{3} z - \frac{32}{3} z^2)\ln(1-z) + (\frac{496}{3z} + \frac{772}{3}
  \nonumber\\[2ex]
  && {}+ \frac{172}{3} z + \frac{256}{9} z^2)\ln z + \frac{4438}{27z} - 36 - 106 z
  -\frac{928}{27}z^2 \BigRightHook.
\end{eqnarray}
Finally we also present the non-singlet longitudinal coefficient function (see also
eq. (17) in \cite{Rij96})
\begin{eqnarray}
  \lefteqn{
    \overline{\mathbb{C}}_{L,q}^\NonSinglet = \frac{\alpha_s}{4\pi}\,C_F\BigLeftHook
    2\BigRightHook + \left(\frac{\alpha_s}{4\pi}\right)^2\,\BigLeftHook C_F^2
    \BigLeftBrace \left(8\ln(1-z)-4\ln z + 2 + 4 z\right) L_M\BigRightBrace}
  \nonumber\\[2ex]
  && {}+C_AC_F\BigLeftBrace -\frac{22}{3}L_M \BigRightBrace
  +n_fC_FT_f\BigLeftBrace\frac{8}{3}L_M \BigRightBrace
  + \bar{c}_{L,q}^{\NonSinglet,(2),{\rm nid}} + \bar{c}_{L,q}^{\NonSinglet,(2),{\rm id}}
  \BigRightHook,\\[2ex]
  \lefteqn{
    \bar{c}_{L,q}^{\NonSinglet,(2),{\rm nid}} = C_F^2\,\BigLeftBrace
    16 S_{1,2}(1-z) - 32\Li_3(-z) + 16\ln z\Li_2(-z) - 16 \zeta(2)\ln(1-z)} \nonumber\\[2ex]
  && {}- 12\Li_2(1-z)
  + 4\ln z\ln(1-z) + 4\ln^2(1-z) + (-10 + 8 z+ 4 z^2 + \frac{8}{5}z^3)\cdot
  \nonumber\\[2ex]
  && \cdot\ln^2 z
  + (24 - 16 z - 8 z^2- \frac{16}{5}z^3)\zeta(2) + (\frac{24}{5z^2} + \frac{16}{z}
  - 16 z - 8 z^2 - \frac{16}{5}z^3)\cdot\nonumber\\[2ex]
  && \cdot(\Li_2(-z)+\ln z\ln(1+z)) + (14 + 4z)\ln(1-z)
  + (-\frac{24}{5z} - \frac{78}{5} + \frac{32}{5}z \nonumber\\[2ex]
  && {}+ \frac{16}{5}z^2)\ln z
  + \frac{24}{5z} - \frac{211}{5} + \frac{86}{5}z + \frac{16}{5}z^2\BigRightBrace \nonumber\\[2ex]
  && {}+ C_AC_F\,\BigLeftBrace - 8 S_{1,2}(1-z) + 16\Li_3(-z)
  - 8\ln z\Li_2(-z) + 8\zeta(2)\ln(1-z) \nonumber\\[2ex]
  && {}+ (-\frac{12}{5z^2} - \frac{8}{z} + 8z
  + 4z^2 + \frac{8}{5}z^3)
  (\Li_2(-z) + \ln z\ln(1+z)) \nonumber\\[2ex]
  && {}+ (2 - 4z - 2z^2 - \frac{4}{5}z^3)
  (\ln^2 z - 2\zeta(2)) - \frac{46}{3}\ln(1-z) + (\frac{12}{5z} + \frac{22}{15}
  + \frac{4}{5}z \nonumber\\[2ex]
  && {}- \frac{8}{5}z^2)\ln z - \frac{12}{5z} + \frac{2017}{45}
  - \frac{254}{15}z - \frac{8}{5}z^2\BigRightBrace \nonumber\\[2ex]
  && +{}n_fC_FT_f\BigLeftBrace\frac{8}{3}(\ln(1-z) + \ln z)
  - \frac{100}{9} + \frac{8}{3}z\BigRightBrace,\\[2ex]
  \lefteqn{
    \bar{c}_{L,q}^{\NonSinglet,(2),{\rm id}} =
    (C_F^2 - \frac{1}{2}C_AC_F)\BigLeftHook 32 S_{1,2}(1-z) - 16\Li_3(-z)
    + 32\ln(1+z)\Li_2(-z)}\nonumber\\[2ex]
  && {}+ 16\zeta(2)\ln(1+z) + 16\ln z\ln^2(1+z)
  - 8\ln^2z\ln(1+z) - 16\zeta(3) + (\frac{24}{5z^2} \nonumber\\[2ex]
  && {}- \frac{16}{z} - 16 - 16 z
  + 8 z^2 - \frac{16}{5}z^3)(\Li_2(-z) + \ln z\ln(1+z)) + (4 + 8z - 4z^2
  \nonumber\\[2ex]
  && {}+ \frac{8}{5}z^3)
  (\ln^2z - 2\zeta(2)) + (-\frac{24}{5z} + \frac{112}{5} - \frac{8}{5}z + \frac{16}{5}z^2)
  \ln z + \frac{24}{5z} + \frac{64}{5} - \frac{104}{5}z \nonumber\\[2ex]
  && {}+ \frac{16}{5}z^2\BigRightHook.
\end{eqnarray}
The remaining coefficient functions $\overline{\mathbb{C}}_{L,q}^\PureSinglet$ and
$\overline{\mathbb{C}}_{L,g}$ can be found in eqs. (19) and (20) respectively of \cite{Rij96}.
\newpage
\section{The coefficient functions in the annihilation scheme (A-scheme)}
A glance at the coefficient functions in the A-scheme as presented in \eref{eq:4_58}-\eref{eq:4_63}
reveals that they are all expressed into the \MS-representation of the DGLAP splitting functions
$P_{ij}^{(k)}$ and the longitudinal coefficients $\bar{c}_{L,p}^{(k)}$ ($p=q,g$). The
longitudinal coefficient functions in the A-scheme are already given in eqs. (21)-(23)
of \cite{Rij96} and we only add those formulae needed to construct the transverse coefficient
functions $\mathbb{C}_{T,p}$ in the latter scheme. For $\mathbb{C}_{T,q}^\NonSinglet$ \eref{eq:4_59}
we have to compute the following convolutions
\begin{eqnarray}
  \lefteqn{
    \frac{1}{2}(R^{(1)}\,\delta(1-z) - \bar{c}_{L,q}^{(1)})\otimes P_{qq}^{(0)} =
    C_F^2\,\BigLeftHook 12{\cal D}_0(z) + 9\delta(1-z) - 16\ln(1-z)}\nonumber\\[2ex]
  && {}+ 8\ln z - 10 - 14 z\BigRightHook,
\end{eqnarray}
and
\begin{eqnarray}
  \lefteqn{
    (R^{(1)}\,\delta(1-z) - \bar{c}_q^{(1)})\otimes\bar{c}_{L,q}^{(1)} =
    C_F^2\,\BigLeftHook 20 \Li_2(1-z) - 4\ln^2(1-z) + 4\ln z\ln(1-z)}\nonumber\\[2ex]
  && {}+ 4\ln^2 z
  - 16 \zeta(2) + (10 - 4 z)\ln(1-z) + (4 - 8 z)\ln z + 18 + 6 z\BigRightHook.
\end{eqnarray}
Note that apart from $R^{(1)}$, $R^{(2)}$ the non logarithmic ($L_M$) contributions to
$\mathbb{C}_{L,q}^\NonSinglet$ \eref{eq:4_58} and $\mathbb{C}_{T,q}^\NonSinglet$ \eref{eq:4_59}
have an opposite sign. The same observation holds for the coefficient
functions $\mathbb{C}_{k,q}^\PureSinglet$ \eref{eq:4_60}, \eref{eq:4_61} and $\mathbb{C}_{k,g}$
\eref{eq:4_62}, \eref{eq:4_63}. For $\mathbb{C}_{k,g}$ we need the convolutions
\begin{eqnarray}
  \lefteqn{
    P_{qq}^{(0)}\otimes\bar{c}_g^{(1)} = C_F^2\,\BigLeftHook\left(\frac{192}{z}-160+80z\right)\Li_2(1-z)
    + \left(\frac{64}{z} - 64 + 32 z\right)\cdot}\nonumber\\[2ex]
  && \cdot\ln^2(1-z) + \left( \frac{128}{z} - 96 + 48 z\right)
  \ln z\ln(1-z) + (32 - 16z)\ln^2 z \nonumber\\[2ex]
  && {}+ \left(-\frac{64}{z} + 64 - 32 z\right)\zeta(2)
  + (32 - 8z)\ln(1-z) + (-64 + 32 z)\ln z - \frac{80}{z} \nonumber\\[2ex]
  && {}+ 96 - 16 z\BigRightHook,
\end{eqnarray}
and
\begin{eqnarray}
  \lefteqn{
    \bar{c}_g^{(1)}\otimes\bar{c}_{L,q}^{(1)} = C_F^2\,\BigLeftHook 16 \Li_2(1-z)
    + 16 \ln^2 z + 16 \ln z\ln(1-z) + \left(\frac{16}{z} - 8 - 8 z\right)\cdot}\nonumber\\[2ex]
  && \cdot\ln(1-z) + \left(\frac{32}{z} + 16 - 16 z\right)\ln z + \frac{32}{z} - 56 + 24 z
  \BigRightHook.
\end{eqnarray}
%

%
\renewcommand{\thesection}{}
\bibliographystyle{/home/rulil0/pieter/tex/style/mybib}
\bibliography{/home/rulil0/pieter/tex/physics}

\newpage
\begin{mcbibliography}{10}
\bibitem{Zer92}{}{{``}QCD 20 years later", vol. 1, Editors: P.M. Zerwas and
  H.A. Kastrup, World Scientific, Singapore 1992}\bibitem{Mni96}{}{J. Mnich,
  Phys. Rep {\bf 271} (1996) 181}\bibitem{Mat90}{}{T. Matsuura, R. Hamberg, and
  W.L. van Neerven, Nucl. Phys. {\bf B345} (1990) 331}\bibitem{Ham91}{R.
  Hamberg, W.L. van Neerven, and T. Matsuura,}{Nucl. Phys. {\bf B359} (1991)
  343}\bibitem{Nee92}{}{W.L. van Neerven, and E.B. Zijlstra, Nucl. Phys. {\bf
  B382} (1992) 11}\bibitem{Zij92}{}{E.B. Zijlstra, and W.L. van Neerven, Nucl.
  Phys. {\bf B383} (1992) 525}\bibitem{Zij92b}{E.B. Zijlstra, and W.L. van
  Neerven,}{Phys. Lett. {\bf 297B} (1992) 377}\bibitem{Lar91}{}{S.A. Larin and
  J.A.M. Vermaseren, Phys. Lett. {\bf 256B} (1991) 345}\bibitem{Lar91b}{}{S.A.
  Larin, F.V. Tkachov and J.A.M. Vermaseren, Phys. Rev. Lett. {\bf 66} (1991)
  862}\bibitem{Lar94b}{S.A. Larin, T. van Ritbergen, and J.A.M.
  Vermaseren,}{Nucl. Phys. {\bf B427} (1994) 41}\bibitem{Gor91}{}{S.G.
  Gorishni, A.L. Kataev, and S.A. Larin, Phys. Lett. {\bf 159B} (1991)
  144}\bibitem{Sur91}{}{L.R. Surguladze, and M.A. Samuel, Phys. Rev. Lett. {\bf
  66} (1991) 560, Erratum Phys. Rev. Lett. {\bf 66} (1991)
  2416}\bibitem{Che94}{K.G. Chetyrkin and O.V. Tarasov,}{Phys. Lett. {\bf B327}
  (1994) 114}\bibitem{Lar94}{S.A. Larin, T. van Ritbergen, and J.A.M.
  Vermaseren,}{Phys. Lett. {\bf B320} (1994) 159}\bibitem{Lar95}{S.A. Larin, T.
  van Ritbergen, and J.A.M. Vermaseren,}{Nucl. Phys. {\bf B438} (1995)
  278}\bibitem{Che81}{}{K.G. Chetyrkin and F.V. Tkachov, Nucl. Phys. {\bf B192}
  (1981) 159}\bibitem{Kaz88}{}{D.I. Kazakov and A.B. Kotikov, Nucl. Phys. {\bf
  B307} (1988) 721}\bibitem{Bra79}{}{Brandelik \etal~(DASP Collaboration),
  Nucl. Phys. {\bf B148} (1979) 189}\bibitem{Alb89}{}{H. Albrecht
  \etal~(ARGUS), Z. Phys. {\bf C44} (1989) 547}\bibitem{Bra89}{}{W.
  Braunschweig \etal~(TASSO), Z. Phys. {\bf C42} (1989) 189}\bibitem{Bra90}{W.
  Braunschweig \etal~(TASSO),}{Z. Phys. {\bf C47} (1990)
  187}\bibitem{Pet88}{}{A. Peterson \etal~(Mark II), Phys. Rev. {\bf D37}
  (1988) 1}\bibitem{Aih88}{}{H. Aihara \etal~(TPC/2$\gamma$ Collaboration),
  Phys. Rev. Lett. {\bf 61} (1988) 1263}\bibitem{Pod}{}{O. Podobrin (CELLO),
  Ph. D. thesis, Universit\"at Hamburg}\bibitem{Kum90}{}{T. Kumita \etal~(AMY),
  Phys. Rev. {\bf D42} (1990) 1339}\bibitem{Li90}{}{Y.K. Li \etal~(AMY), Phys.
  Rev. {\bf D41} (1990) 2675}\bibitem{Abr93}{}{P. Abreu \etal~(DELPHI), Phys.
  Lett. {\bf B311} (1993) 408}\bibitem{Bus95}{}{D. Buskulic \etal~(ALEPH),
  Phys. Lett. {\bf B357} (1995) 487}\bibitem{Bus95b}{D. Buskulic
  \etal~(ALEPH),}{Z. Phys. {\bf C66} (1995) 355}\bibitem{Ake93}{R. Akers
  \etal~(OPAL),}{Z. Phys. {\bf C58} (1993) 387}\bibitem{Ake94}{R. Akers
  \etal~(OPAL),}{Z. Phys. {\bf C63} (1994) 181}\bibitem{Ake95}{}{R. Akers
  \etal~(OPAL), Z. Phys. {\bf C68} (1995) 203}\bibitem{Rij96}{}{P.J. Rijken and
  W.L. van Neerven, INLO-PUB-4/96, hep-ph/9604436, to be published in Phys.
  Lett. B}\bibitem{Gri72}{}{V.N. Gribov and L.N. Lipatov, Sov. J. Nucl. Phys.
  {\bf 15} (1972) 675}\bibitem{Alt77}{}{G. Altarelli and G. Parisi, Nucl. Phys.
  {\bf B126} (1977) 298}\bibitem{Dok77}{}{Yu.L. Dokshitzer, Sov. Phys. JETP
  {\bf 46} (1977) 641}\bibitem{Che79}{}{K.E. Chetyrkin, A.L. Kataev, and F.V.
  Tkachov, Phys. Lett. {\bf 85B} (1979) 277}\bibitem{Din79}{}{M. Dine, and J.
  Sapirstein, Phys. Rev. Lett. {\bf 43} (1979) 668}\bibitem{Cel80}{}{W.
  Celmaster, and R.J. Gonsalves, Phys. Rev. Lett. {\bf 44} (1980)
  560}\bibitem{Nas94}{}{P. Nason, and B.R. Webber, Nucl. Phys. {\bf B421}
  (1994) 473}\bibitem{Web94}{}{B.R. Webber, Cavendish Laboratory Report Nos.
  Cavendish-HEP-94/17, hep-ph/9411384, 1994
  (unpublished)}\bibitem{Ver}{}{J.A.M. Ver\-ma\-se\-ren, FORM2, published by
  Com\-pu\-ter Al\-ge\-bra Ne\-der\-land (CAN), Kruis\-laan 413, 1089 SJ
  Am\-ster\-dam, The Netherlands}\bibitem{Lar93}{S.A. Larin,}{Phys. Lett. {\bf
  303B} (1993) 113}\bibitem{Bai79}{}{R. Baier, and K. Fey, Z. Phys. {\bf C2}
  (1979) 339}\bibitem{Alt79b}{}{G. Altarelli, R.K. Ellis, G. Martinelli, and
  S.-Y. Pi, Nucl. Phys. {\bf B160} (1979) 301}\bibitem{Mat88a}{}{T. Matsuura,
  and W.L. van Neerven, Z. Phys. {\bf C38} (1988) 623}\bibitem{Mat89}{}{T.
  Matsuura, S.C. van der Marck, and W.L. van Neerven, Nucl. Phys. {\bf B319}
  (1989) 570}\bibitem{Kra87}{}{G. Kramer and B. Lampe, Z. Phys. {\bf C34}
  (1987); Erratum {\bf C42} (1989) 504}\bibitem{Pas79}{}{G. Pasarino and M.
  Veltman, Nucl. Phys. {\bf B160} (1979) 151}\bibitem{Bee89b}{}{W. Beenakker,
  Ph. D. thesis University of Leiden, The Netherlands
  (1989)}\bibitem{Bee89}{}{W. Beenakker, H. Kuijf, W.L. van Neerven, and J.
  Smith, Phys. Rev. {\bf D40} (1989) 54}\bibitem{Erd53}{}{Erd\'ely\'\i, Magnus,
  Oberhettinger, Tricomi, ``Higher Transcendental functions", Bateman
  Manuscript Project. Vol 1, McGraw-Hill 1953}\bibitem{Alt79}{}{G. Altarelli,
  R.K. Ellis, and G. Martinelli, Nucl. Phys. {\bf B157} (1979)
  461}\bibitem{Hum81}{}{B. Humpert, and W.L. Van Neerven, Nucl. Phys. {\bf
  B184} (1981) 225}\bibitem{Cur80}{}{G. Curci, W. Furmanski, and R. Petronzio,
  Nucl. Phys. {\bf B175} (1980) 27}\bibitem{Fur80}{}{W. Furmanski, and R.
  Petronzio, Phys. Lett. {\bf 97B} (1980) 437}\bibitem{Fur82}{W. Furmanski, and
  R. Petronzio,}{Z. Phys. {\bf C11} (1982) 293}\bibitem{Flo81}{}{E.G. Floratos,
  C. Kounnas, and R. Lacaze, Nucl. Phys. {\bf B192} (1981)
  417}\bibitem{Bar72}{}{R. Barbieri, J.A. Mignaco, and E. Remiddi, Nuovo Cim.
  {\bf 11A} (1972) 824}\bibitem{Lew83}{}{L. Lewin, ``Polylogarithms and
  Associated Functions" (North Holland, Amsterdam 1983)}\bibitem{Dev84}{}{A.
  Devoto, and D.W. Duke, Riv. Nuovo Cim. {\bf 7-6} (1984)
  1}\bibitem{Mer96}{}{R. Mertig and W.L. van Neerven, Z. Phys. {\bf C70} (1996)
  637}\bibitem{Bin95}{}{J. Binnewies, B.A. Kniehl, and G. Kramer, Z. Phys. {\bf
  C65} (1995) 471}\bibitem{Bin95b}{}{J. Binnewies, B.A. Kniehl, and G. Kramer,
  Phys. Rev. {\bf D52} (1995) 4947}\end{mcbibliography}
\newpage
\section{Figure captions}
{\bf Fig. 1} Kinematics of the process $e^+ e^- \rightarrow H+ ``X"$.\\[2ex]
{\bf Fig. 2} Born contribution given by the subprocess $V \rightarrow ``q" + \bar{q}$.
             \\[2ex]
{\bf Fig. 3} One-loop correction to the subprocess $V \rightarrow ``q" + \bar{q}$. Graphs
             with external self energies are omitted since they do not contribute in the
             case of massless quarks.\\[2ex]
{\bf Fig. 4} Graphs contributing to the subprocess $V \rightarrow ``q" + \bar{q} + g$ and
             $V \rightarrow ``g" + q + \bar{q}$.\\[2ex]
{\bf Fig. 5} Two-loop corrections to the subprocess $V \rightarrow ``q" + \bar{q}$.
             Graphs with external self energies are omitted since they do not contribute
             in the case of massless quarks.\\[2ex]
{\bf Fig. 6} One-loop corrections to the subprocesses $V \rightarrow ``q" + \bar{q} + g$
             and
             $V \rightarrow ``g" + q + \bar{q}$. Graphs with external self energies are
             omitted since they do not contribute in the case of massless quarks and
             gluons.\\[2ex]
{\bf Fig. 7} Graphs contributing to the subprocesses
             $V \rightarrow ``q" + \bar{q} + g + g$
             and $V \rightarrow ``g" + q + \bar{q} + g$.\\[2ex]
{\bf Fig. 8} Graphs contributing to the subprocess
             $V \rightarrow ``q" + \bar{q}(1) + q + \bar{q}(2)$. The cross ($\times$)
             indicates
             that the process is exclusive with respect to the quark denoted by $``q"$.
             If
             $\bar{q}(1)\neq\bar{q}(2)$ only combinations A and C have to be considered.
             When
             $\bar{q}(1)=\bar{q}(2)$ combinations B and D have to be added to A and C.
             \\[2ex]
{\bf Fig. 9} Cut diagrams obtained from the groups $C$ and $D$ in fig.~\ref{fig:f8}
             contributing to the process
             $V\rightarrow ``q"+\bar{q}(1) + q + \bar{q}(2)$.\\[2ex]
{\bf Fig. 10} Cut diagrams resulting from the combinations $AD$ and $BC$ in
              fig.~\ref{fig:f8} contributing to
              $V\rightarrow ``q"+\bar{q}(1) + q + \bar{q}(2)$ in the case that
              $\bar{q}(1)=\bar{q}(2)$.\\[2ex]
{\bf Fig. 11} The ratio $R_L=\sigma_T/\sigma_{\rm tot}$. Dotted lines:
              $R_L^{LO}$; solid lines: $R_L^{NLO}$. Lower curve: $R=2Q$;
              middle curve: $R=Q$; upper curve: $R=Q/2$. The data point at $Q=M_Z$
              is from OPAL \cite{Ake95}.\\[2ex]
{\bf Fig. 12} The ratio $R_L=\sigma_T/\sigma_{\rm tot}$. Dotted lines:
              $R_T^{NLO}$; $R_T^{NNLO}$. Lower curve: $R=2Q$;
              middle curve: $R=Q$; upper curve: $R=Q/2$. The data point at $Q=M_Z$
              is from OPAL \cite{Ake95}.\\[2ex]
{\bf Fig. 13} The longitudinal fragmentation function $F_L(x,Q^2)$ at $M=Q=M_Z$.
              Dotted line: $F_L^{LO}$; solid line: $F_L^{NLO}$. The data are from
              ALEPH \cite{Bus95,-Bus95b} and OPAL \cite{Ake95}. The fragmentation
              density set is BKK1 \cite{Bin95}.\\[2ex]
{\bf Fig. 14} The transverse fragmentation function $F_T(x,Q^2)$ at $M=Q=M_Z$.
              Dotted line: $F_T^{NLO}$; solid line: $F_T^{NNLO}$. The data are from
              ALEPH \cite{Bus95,-Bus95b} and OPAL \cite{Ake95}. The fragmentation
              density set is BKK1 \cite{Bin95}.\\[2ex]
{\bf Fig. 15} The ratio $K_L^H = F_L^{H,NLO}/F_L^{H,LO}$ with $H=\pi^++\pi^-$
              at $M=Q$ for different values of $Q$. Upper dotted line: $Q=5.2~\GeV$;
              solid line $Q=10~\GeV$; long dashed line: $Q=29~\GeV$; short dashed
              line: $Q=35~\GeV$; lower dotted line: $Q=55~\GeV$; dashed-dotted line:
              $Q=91.2~\GeV$. The fragmentation density set is BKK2 \cite{Bin95b}.\\[2ex]
{\bf Fig. 16} The same as in \fref{fig:f19} but now for the ratio
              $K_T^H = F_T^{H,NNLO}/F_T^{H,NLO}$ with $H=\pi^++\pi^-$.\\[2ex]
{\bf Fig. 17} The dependence of $K_T$ on the fragmentation density sets at
              $M=Q=M_Z$. Dotted curve: set from \cite{Nas94}; solid curve:
              BKK1 \cite{Bin95}; dashed curve: BKK2 \cite{Bin95b}.\\[2ex]
{\bf Fig. 18} The mass factorization scale dependence of $F_L^{NLO}$ at $Q=M_Z$.
              Lower dotted curve: $M=2Q$; middle dotted curve: $M=Q$; upper
              dotted curve: $M=Q/2$. The data are from ALEPH \cite{Bus95,-Bus95b}
              and OPAL \cite{Ake95}. The fragmentation density set is BKK1
              \cite{Bin95}\\[2ex]
{\bf Fig. 19} Sensitivity of $F_L^r$ ($r=LO,~NLO$) to the mass factorization
              scale represented by $\Delta_L^r$ \eref{eq:5_14} at $Q=M_Z$. Dotted
              line: $\Delta_L^{LO}$; solid line: $\Delta_L^{NLO}$. The
              fragmentation density set is BKK1 \cite{Bin95}.\\[2ex]
{\bf Fig. 20} The same as in \fref{fig:f18} but now for $F_T^{NLO}$. Also
              presented is the curve for $F_T^{NNLO}$ (solid line).\\[2ex]
{\bf Fig. 21} The same as in \fref{fig:f19} but now for $\Delta_T^r$ \eref{eq:5_14}.
              Also shown is $\Delta_T^{NNLO}$ (dashed line).\\[2ex]
{\bf Fig. 22} The mass factorization scale dependence of the total fragmentation
              function $F^{H,NLO}$ with $H=\pi^++\pi^-$ at $Q=29~\GeV$. Lower
              dotted curve: $M=2Q$; middle dotted curve: $M=Q$; upper dotted curve:
              $M=Q/2$; solid line: $F_T^{H,NNLO}$. The data are from TPC/$2\gamma$
              \cite{Aih88}. The fragmentation density set is BKK2 \cite{Bin95b}.\\[2ex]
{\bf Fig. 23} Sensitivity of $F^{H,r}$ ($r = LO,~NLO,~NNLO$) to the mass
              factorization scale represented by $\Delta^{H,r}$ with $H=\pi^++\pi^-$
              at $Q=29~\GeV$. Dotted line: $\Delta^{H,LO}$; solid line:
              $\Delta^{H,NLO}$; dashed line: $\Delta^{H,NNLO}$. The fragmentation
              density set is BKK2 \cite{Bin95b}.
%

\newpage
\begin{mytable}{|c|lclc|}{tab:t1}{List of parton subprocesses in $e^+e^-$ annihilation
    up to order \alphastwo.}
  \hline
  figure & \multicolumn{4}{c|}{Parton subprocesses} \\ \hline
               &               &                 &                            & \\[2ex]
  \ref{fig:f2} & $\alpha_s^0$: & $V \rightarrow$ & $q+\bar{q}$                & \\[4ex]
  \ref{fig:f3} & $\alpha_s^1$: & $V \rightarrow$ & $q+\bar{q}$                & (one loop correction) \\[2ex]
  \ref{fig:f4} &               & $V \rightarrow$ & $``q"+\bar{q}+g$           & \\[2ex]
  \ref{fig:f4} &               & $V \rightarrow$ & $q+\bar{q}+``g"$           & \\[4ex]
  \ref{fig:f5} & $\alpha_s^2$: & $V \rightarrow$ & $q+\bar{q}$                & (two loop correction) \\[2ex]
  \ref{fig:f6} &               & $V \rightarrow$ & $``q"+\bar{q}+g$           & (one loop correction) \\[2ex]
  \ref{fig:f7} &               & $V \rightarrow$ & $``q"+\bar{q}+g+g$         & \\[2ex]
  \ref{fig:f6} &               & $V \rightarrow$ & $q+\bar{q}+``g"$           & (one loop correction) \\[2ex]
  \ref{fig:f7} &               & $V \rightarrow$ & $q+\bar{q}+``g"+g$         & \\[2ex]
  \ref{fig:f8} &               & $V \rightarrow$ & $``q"+\bar{q}+q'+\bar{q}'$ & \\
               &               &                 &                            & \\ \hline
\end{mytable}
%
%
\clearpage

\newpage
%
%
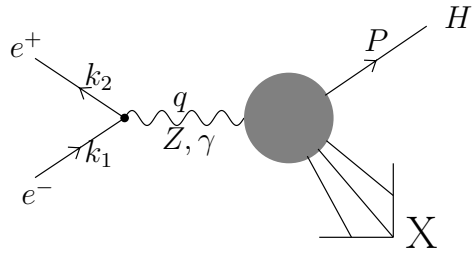
\begin{figure}
  \begin{center}
    \noindent
    \setlength{\unitlength}{25pt}
\renewcommand{\Init}{0.450 0.450 Scale Init}
\begin{picture}(7.200,4.500)
\put(0,0){\Fermion{2.000}{3.000}{5.000}{5.000}{+}}
\put(0,0){\Fermion{5.000}{5.000}{2.000}{7.000}{+}}
\put(0,0){\VectorBoson{5.000}{5.000}{9.000}{5.000}{}}
\put(0,0){\Bubble{180.000}{1.500}{1}{1.013}{10.500}{5.000}}
\put(0,0){\Bullet{5.000}{5.000}}
\put(0,0){\Fermion{11.720}{5.760}{15.120}{8.080}{+}}
\put(0,0){\Fermion{11.480}{3.960}{14.000}{1.000}{}}
\put(0,0){\Fermion{14.000}{1.000}{14.000}{3.480}{}}
\put(0,0){\Fermion{14.000}{1.000}{11.520}{1.000}{}}
\put(0,0){\Fermion{11.760}{4.240}{14.000}{2.400}{}}
\put(0,0){\Fermion{11.120}{3.720}{12.600}{1.000}{}}
\put(0.702,0.954){$e^-$}
\put(0.522,3.186){$e^+$}
\put(2.988,2.448){$q$}
\put(2.826,1.782){$Z,\gamma$}
\put(5.886,3.276){$P$}
\put(7.092,3.618){$H$}
\put(6.516,0.288){\Large{X}}
\put(1.656,1.584){$k_1$}
\put(1.656,2.754){$k_2$}
\end{picture}
    \caption{\label{fig:fl}}
  \end{center}
\end{figure}
%
%
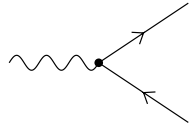
\begin{figure}
  \begin{center}
    \noindent
    \setlength{\unitlength}{25pt}
\renewcommand{\Init}{0.450 0.450 Scale Init}
\begin{picture}(4.050,2.700)
\put(0,0){\VectorBoson{2.000}{3.000}{5.000}{3.000}{}}
\put(0,0){\Fermion{8.000}{1.000}{5.000}{3.000}{+}}
\put(0,0){\Fermion{5.000}{3.000}{8.000}{5.000}{+}}
\put(0,0){\Bullet{5.000}{3.000}}
\end{picture}
    \caption{\label{fig:f2}}
  \end{center}
\end{figure}
%
%
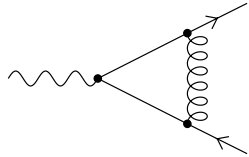
\begin{figure}
  \begin{center}
    \noindent
    \setlength{\unitlength}{25pt}
\renewcommand{\Init}{0.450 0.450 Scale Init}
\begin{picture}(4.950,4.050)
\put(0,0){\VectorBoson{2.000}{4.000}{5.000}{4.000}{}}
\put(0,0){\Gluon{8.000}{2.480}{8.000}{5.520}{}}
\put(0,0){\Fermion{5.000}{4.000}{8.000}{5.520}{}}
\put(0,0){\Fermion{8.000}{2.480}{5.000}{4.000}{}}
\put(0,0){\Fermion{8.000}{5.520}{10.000}{6.520}{+}}
\put(0,0){\Fermion{10.000}{1.480}{8.000}{2.480}{+}}
\put(0,0){\Bullet{5.000}{4.000}}
\put(0,0){\Bullet{8.000}{5.520}}
\put(0,0){\Bullet{8.000}{2.480}}
\end{picture}
    \caption{\label{fig:f3}}
  \end{center}
\end{figure}
%
%
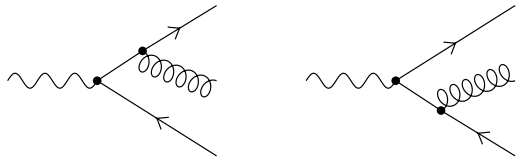
\begin{figure}
  \begin{center}
    \noindent
    \setlength{\unitlength}{25pt}
\renewcommand{\Init}{0.450 0.450 Scale Init}
\begin{picture}(8.550,4.050)
\put(0,0){\VectorBoson{1.000}{4.000}{4.000}{4.000}{}}
\put(0,0){\Fermion{4.000}{4.000}{5.520}{5.000}{}}
\put(0,0){\Gluon{5.520}{5.000}{8.000}{4.000}{}}
\put(0,0){\Fermion{5.520}{5.000}{8.000}{6.520}{+}}
\put(0,0){\Fermion{8.000}{1.480}{4.000}{4.000}{+}}
\put(0,0){\Bullet{4.000}{4.000}}
\put(0,0){\Bullet{5.520}{5.000}}
\put(0,0){\VectorBoson{11.000}{4.000}{14.000}{4.000}{}}
\put(0,0){\Fermion{14.000}{4.000}{18.000}{6.520}{+}}
\put(0,0){\Fermion{18.000}{1.480}{15.520}{3.000}{+}}
\put(0,0){\Fermion{15.520}{3.000}{14.000}{4.000}{}}
\put(0,0){\Gluon{15.520}{3.000}{18.000}{4.000}{+}}
\put(0,0){\Bullet{14.000}{4.000}}
\put(0,0){\Bullet{15.520}{3.000}}
\end{picture}
    \caption{\label{fig:f4}}
  \end{center}
\end{figure}
%
%
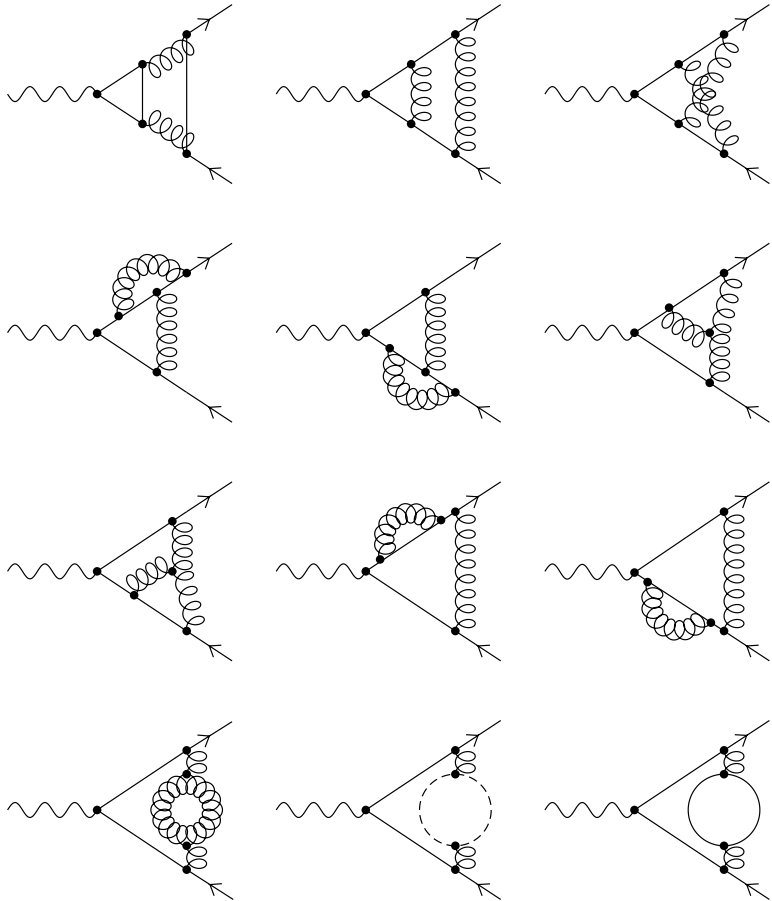
\begin{figure}
  \begin{center}
    \noindent
    \setlength{\unitlength}{25pt}
\renewcommand{\Init}{0.450 0.450 Scale Init}
\begin{picture}(12.600,14.850)
\put(0,0){\VectorBoson{1.000}{28.000}{4.000}{28.000}{+}}
\put(0,0){\Fermion{4.000}{28.000}{5.520}{29.000}{}}
\put(0,0){\Fermion{5.520}{29.000}{5.520}{27.000}{}}
\put(0,0){\Fermion{5.520}{27.000}{4.000}{28.000}{}}
\put(0,0){\Gluon{5.520}{29.000}{7.000}{30.000}{}}
\put(0,0){\Gluon{7.000}{26.000}{5.520}{27.000}{}}
\put(0,0){\Fermion{7.000}{30.000}{7.000}{26.000}{}}
\put(0,0){\Fermion{7.000}{30.000}{8.520}{31.000}{+}}
\put(0,0){\Fermion{8.520}{25.000}{7.000}{26.000}{+}}
\put(0,0){\Bullet{4.000}{28.000}}
\put(0,0){\Bullet{5.520}{29.000}}
\put(0,0){\Bullet{7.000}{30.000}}
\put(0,0){\Bullet{7.000}{26.000}}
\put(0,0){\Bullet{5.520}{27.000}}
\put(0,0){\VectorBoson{10.000}{28.000}{13.000}{28.000}{}}
\put(0,0){\Fermion{13.000}{28.000}{14.520}{29.000}{}}
\put(0,0){\Fermion{14.520}{27.000}{13.000}{28.000}{}}
\put(0,0){\Fermion{14.520}{29.000}{16.000}{30.000}{}}
\put(0,0){\Fermion{16.000}{26.000}{14.520}{27.000}{}}
\put(0,0){\Fermion{16.000}{30.000}{17.520}{31.000}{+}}
\put(0,0){\Fermion{17.520}{25.000}{16.000}{26.000}{+}}
\put(0,0){\Gluon{14.520}{27.000}{14.520}{29.000}{}}
\put(0,0){\Gluon{16.000}{26.000}{16.000}{30.000}{}}
\put(0,0){\Bullet{13.000}{28.000}}
\put(0,0){\Bullet{14.520}{29.000}}
\put(0,0){\Bullet{16.000}{30.000}}
\put(0,0){\Bullet{16.000}{26.000}}
\put(0,0){\Bullet{14.520}{27.000}}
\put(0,0){\VectorBoson{19.000}{28.000}{22.000}{28.000}{}}
\put(0,0){\Fermion{22.000}{28.000}{25.000}{30.000}{}}
\put(0,0){\Fermion{25.000}{26.000}{22.000}{28.000}{}}
\put(0,0){\Gluon{25.000}{26.000}{23.480}{29.000}{}}
\put(0,0){\Gluon{25.000}{30.000}{23.480}{27.000}{+}}
\put(0,0){\Fermion{25.000}{30.000}{26.520}{31.000}{+}}
\put(0,0){\Fermion{26.520}{25.000}{25.000}{26.000}{+}}
\put(0,0){\Bullet{22.000}{28.000}}
\put(0,0){\Bullet{23.480}{29.000}}
\put(0,0){\Bullet{25.000}{30.000}}
\put(0,0){\Bullet{25.000}{26.000}}
\put(0,0){\Bullet{23.480}{27.000}}
\put(0,0){\Fermion{8.520}{23.000}{7.000}{22.000}{-}}
\put(0,0){\Fermion{7.000}{22.000}{4.000}{20.000}{}}
\put(0,0){\VectorBoson{1.000}{20.000}{4.000}{20.000}{}}
\put(0,0){\Fermion{4.000}{20.000}{7.000}{18.000}{}}
\put(0,0){\Fermion{8.520}{17.000}{7.000}{18.000}{+}}
\put(0,0){\GluonArc{4.720}{20.560}{7.000}{22.000}{180.979}{}}
\put(0,0){\Gluon{6.000}{18.680}{6.000}{21.360}{}}
\put(0,0){\Bullet{4.000}{20.000}}
\put(0,0){\Bullet{4.720}{20.560}}
\put(0,0){\Bullet{6.000}{21.360}}
\put(0,0){\Bullet{7.000}{22.000}}
\put(0,0){\Bullet{6.000}{18.680}}
\put(0,0){\VectorBoson{10.000}{20.000}{13.000}{20.000}{}}
\put(0,0){\Fermion{13.000}{20.000}{16.000}{22.000}{}}
\put(0,0){\Fermion{13.000}{20.000}{16.000}{18.000}{}}
\put(0,0){\Fermion{16.000}{22.000}{17.520}{23.000}{+}}
\put(0,0){\Fermion{17.520}{17.000}{16.000}{18.000}{+}}
\put(0,0){\GluonArc{16.000}{18.000}{13.800}{19.480}{180.987}{}}
\put(0,0){\Gluon{15.000}{18.680}{15.000}{21.360}{}}
\put(0,0){\Bullet{13.000}{20.000}}
\put(0,0){\Bullet{13.800}{19.480}}
\put(0,0){\Bullet{15.000}{18.680}}
\put(0,0){\Bullet{16.000}{18.000}}
\put(0,0){\Bullet{15.000}{21.360}}
\put(0,0){\VectorBoson{19.000}{20.000}{22.000}{20.000}{}}
\put(0,0){\Fermion{22.000}{20.000}{25.000}{18.000}{}}
\put(0,0){\Fermion{22.000}{20.000}{25.000}{22.000}{}}
\put(0,0){\Fermion{25.000}{22.000}{26.520}{23.000}{+}}
\put(0,0){\Fermion{26.520}{17.000}{25.000}{18.000}{+}}
\put(0,0){\Bullet{22.000}{20.000}}
\put(0,0){\VectorBoson{1.000}{12.000}{4.000}{12.000}{}}
\put(0,0){\Fermion{4.000}{12.000}{7.000}{14.000}{}}
\put(0,0){\Fermion{4.000}{12.000}{7.000}{10.000}{}}
\put(0,0){\Fermion{7.000}{14.000}{8.520}{15.000}{+}}
\put(0,0){\Fermion{8.520}{9.000}{7.000}{10.000}{+}}
\put(0,0){\Bullet{4.000}{12.000}}
\put(0,0){\VectorBoson{10.000}{12.000}{13.000}{12.000}{}}
\put(0,0){\Fermion{13.000}{12.000}{16.000}{14.000}{}}
\put(0,0){\Fermion{16.000}{10.000}{13.000}{12.000}{}}
\put(0,0){\Fermion{16.000}{14.000}{17.520}{15.000}{+}}
\put(0,0){\Fermion{17.520}{9.000}{16.000}{10.000}{+}}
\put(0,0){\Gluon{16.000}{10.000}{16.000}{14.000}{}}
\put(0,0){\GluonArc{13.480}{12.400}{15.520}{13.720}{181.113}{}}
\put(0,0){\Bullet{13.000}{12.000}}
\put(0,0){\Bullet{13.480}{12.400}}
\put(0,0){\Bullet{15.520}{13.720}}
\put(0,0){\Bullet{16.000}{14.000}}
\put(0,0){\Bullet{16.000}{10.000}}
\put(0,0){\VectorBoson{19.000}{12.000}{22.000}{11.960}{}}
\put(0,0){\Fermion{22.000}{11.960}{25.000}{14.000}{}}
\put(0,0){\Fermion{25.000}{10.000}{22.000}{11.960}{}}
\put(0,0){\Fermion{25.000}{14.000}{26.520}{15.000}{+}}
\put(0,0){\Fermion{26.520}{9.000}{25.000}{10.000}{+}}
\put(0,0){\Gluon{25.000}{10.000}{25.000}{14.000}{}}
\put(0,0){\GluonArc{24.560}{10.280}{22.440}{11.640}{181.749}{}}
\put(0,0){\Bullet{22.000}{11.960}}
\put(0,0){\Bullet{22.440}{11.640}}
\put(0,0){\Bullet{24.560}{10.280}}
\put(0,0){\Bullet{25.000}{10.000}}
\put(0,0){\Bullet{25.000}{14.000}}
\put(0,0){\VectorBoson{1.000}{4.000}{4.000}{4.000}{}}
\put(0,0){\VectorBoson{10.000}{4.000}{13.000}{4.000}{}}
\put(0,0){\VectorBoson{19.000}{4.000}{22.000}{4.000}{}}
\put(0,0){\Fermion{4.000}{4.000}{7.000}{6.000}{}}
\put(0,0){\Fermion{7.000}{2.000}{4.000}{4.000}{}}
\put(0,0){\Fermion{13.000}{4.000}{16.000}{6.000}{}}
\put(0,0){\Fermion{16.000}{2.000}{13.000}{4.000}{}}
\put(0,0){\Fermion{22.000}{4.000}{25.000}{6.000}{}}
\put(0,0){\Fermion{25.000}{2.000}{22.000}{4.000}{}}
\put(0,0){\Fermion{7.000}{6.000}{8.520}{6.960}{+}}
\put(0,0){\Fermion{8.560}{1.000}{7.000}{2.000}{+}}
\put(0,0){\Fermion{16.000}{6.000}{17.520}{7.000}{+}}
\put(0,0){\Fermion{17.520}{1.000}{16.000}{2.000}{+}}
\put(0,0){\Fermion{25.000}{6.000}{26.520}{7.000}{+}}
\put(0,0){\Fermion{26.520}{1.000}{25.000}{2.000}{+}}
\put(0,0){\Bullet{4.000}{4.000}}
\put(0,0){\Bullet{13.000}{4.000}}
\put(0,0){\Bullet{22.000}{4.000}}
\put(0,0){\Gluon{7.000}{2.000}{7.000}{2.800}{}}
\put(0,0){\Gluon{7.000}{5.200}{7.000}{6.000}{}}
\put(0,0){\GluonArc{7.000}{2.800}{7.000}{5.200}{180.000}{}}
\put(0,0){\GluonArc{7.000}{5.200}{7.000}{2.800}{180.000}{}}
\put(0,0){\Bullet{7.000}{6.000}}
\put(0,0){\Bullet{7.000}{5.200}}
\put(0,0){\Bullet{7.000}{2.800}}
\put(0,0){\Bullet{7.000}{2.000}}
\put(0,0){\Gluon{16.000}{2.000}{16.000}{2.800}{}}
\put(0,0){\Gluon{16.000}{5.200}{16.000}{6.000}{}}
\put(0,0){\Gluon{25.000}{5.200}{25.000}{6.000}{}}
\put(0,0){\Gluon{25.000}{2.000}{25.000}{2.800}{}}
\put(0,0){\FermionArc{25.000}{5.200}{25.000}{2.800}{180.000}{}}
\put(0,0){\FermionArc{25.000}{2.800}{25.000}{5.200}{180.000}{}}
\put(0,0){\Bullet{16.000}{6.000}}
\put(0,0){\Bullet{16.000}{2.000}}
\put(0,0){\Bullet{16.000}{2.800}}
\put(0,0){\Bullet{16.000}{5.200}}
\put(0,0){\Bullet{25.000}{6.000}}
\put(0,0){\Bullet{25.000}{5.200}}
\put(0,0){\Bullet{25.000}{2.800}}
\put(0,0){\Bullet{25.000}{2.000}}
\put(0,0){\Gluon{24.520}{18.320}{24.520}{20.000}{}}
\put(0,0){\Gluon{24.520}{20.000}{25.000}{22.000}{}}
\put(0,0){\Gluon{24.520}{20.000}{23.160}{20.840}{+}}
\put(0,0){\Bullet{23.160}{20.840}}
\put(0,0){\Bullet{24.520}{20.000}}
\put(0,0){\Bullet{25.000}{22.000}}
\put(0,0){\Bullet{24.520}{18.320}}
\put(0,0){\Gluon{6.520}{12.000}{6.520}{13.680}{}}
\put(0,0){\Gluon{7.000}{10.000}{6.520}{12.000}{}}
\put(0,0){\Gluon{6.520}{12.000}{5.240}{11.200}{}}
\put(0,0){\Bullet{5.240}{11.200}}
\put(0,0){\Bullet{6.520}{12.000}}
\put(0,0){\Bullet{6.520}{13.680}}
\put(0,0){\Bullet{7.000}{10.000}}
\put(0,0){\ScaGhoArc{16.000}{5.200}{16.000}{2.800}{180.000}{}}
\put(0,0){\ScaGhoArc{16.000}{2.800}{16.000}{5.200}{180.000}{}}
\end{picture}
    \caption{\label{fig:f5}}
  \end{center}
\end{figure}
%
%
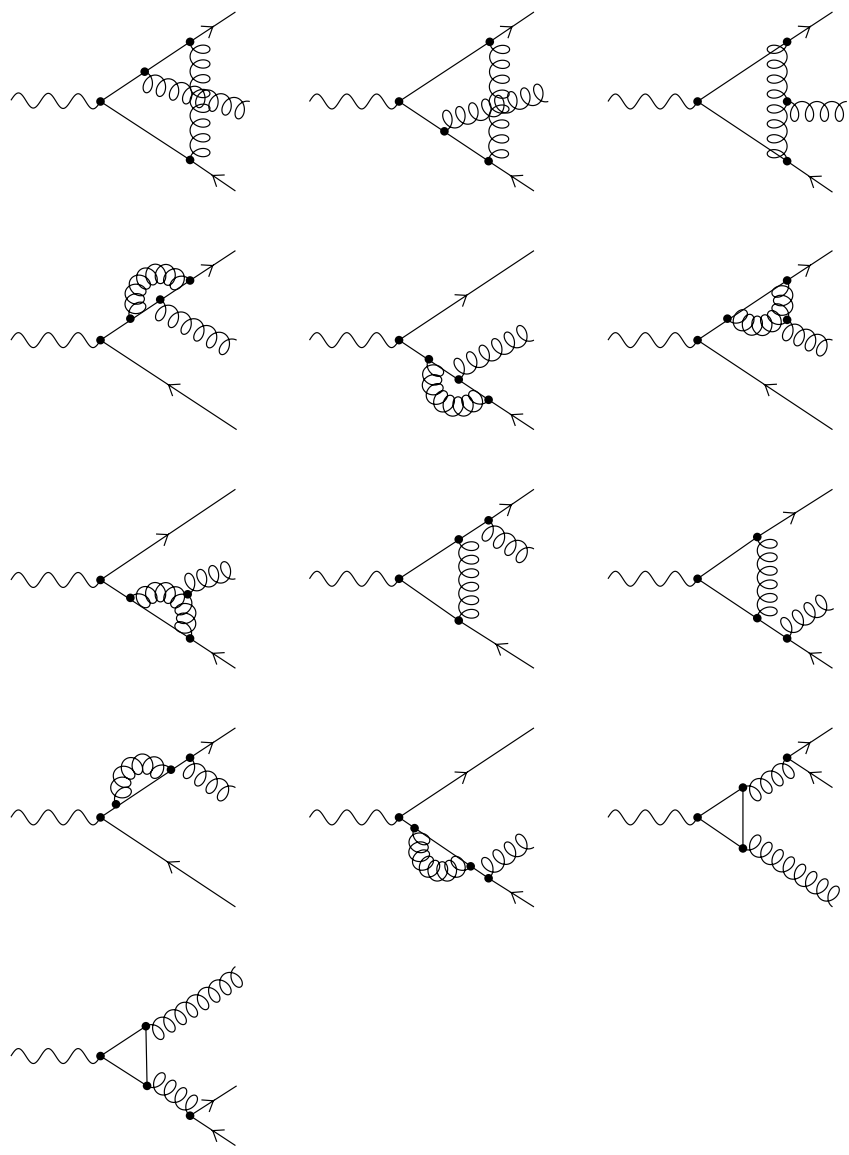
\begin{figure}
  \begin{center}
    \noindent
    \setlength{\unitlength}{25pt}
\renewcommand{\Init}{0.450 0.450 Scale Init}
\begin{picture}(13.500,18.000)
\put(0,0){\VectorBoson{1.000}{36.040}{4.000}{36.000}{}}
\put(0,0){\Fermion{4.000}{36.000}{7.000}{38.000}{}}
\put(0,0){\Fermion{7.000}{34.040}{4.000}{36.000}{}}
\put(0,0){\Fermion{7.000}{38.000}{8.520}{39.000}{+}}
\put(0,0){\Fermion{8.520}{33.000}{7.000}{34.040}{+}}
\put(0,0){\Gluon{7.000}{34.040}{7.000}{38.000}{}}
\put(0,0){\Gluon{5.480}{37.000}{9.000}{36.000}{}}
\put(0,0){\Bullet{4.000}{36.000}}
\put(0,0){\Bullet{5.480}{37.000}}
\put(0,0){\Bullet{7.000}{38.000}}
\put(0,0){\Bullet{7.000}{34.040}}
\put(0,0){\VectorBoson{11.000}{36.000}{14.000}{36.000}{}}
\put(0,0){\Fermion{14.000}{36.000}{17.040}{38.000}{}}
\put(0,0){\Fermion{17.000}{34.000}{14.000}{36.000}{}}
\put(0,0){\Fermion{17.040}{38.000}{18.520}{39.000}{+}}
\put(0,0){\Fermion{18.520}{33.000}{17.000}{34.000}{+}}
\put(0,0){\VectorBoson{21.000}{36.000}{24.000}{36.000}{}}
\put(0,0){\Fermion{24.000}{36.000}{27.000}{38.000}{}}
\put(0,0){\Fermion{27.000}{34.000}{24.000}{36.000}{}}
\put(0,0){\Fermion{27.000}{38.000}{28.520}{39.000}{+}}
\put(0,0){\Fermion{28.520}{32.960}{27.000}{34.000}{+}}
\put(0,0){\Gluon{17.000}{34.000}{17.040}{38.000}{}}
\put(0,0){\Gluon{27.000}{38.000}{27.000}{34.000}{}}
\put(0,0){\Gluon{27.000}{36.000}{29.000}{36.000}{}}
\put(0,0){\Gluon{15.520}{35.000}{19.000}{36.000}{+}}
\put(0,0){\Bullet{14.000}{36.000}}
\put(0,0){\Bullet{15.520}{35.000}}
\put(0,0){\Bullet{17.000}{34.000}}
\put(0,0){\Bullet{17.040}{38.000}}
\put(0,0){\Bullet{24.000}{36.000}}
\put(0,0){\Bullet{27.000}{38.000}}
\put(0,0){\Bullet{27.000}{36.000}}
\put(0,0){\Bullet{27.000}{34.000}}
\put(0,0){\VectorBoson{1.000}{28.000}{4.000}{28.000}{}}
\put(0,0){\VectorBoson{11.000}{28.000}{14.000}{28.000}{}}
\put(0,0){\VectorBoson{21.000}{28.000}{24.000}{28.000}{}}
\put(0,0){\VectorBoson{1.000}{19.960}{4.000}{19.960}{}}
\put(0,0){\VectorBoson{11.000}{20.000}{14.000}{20.000}{}}
\put(0,0){\VectorBoson{21.000}{20.000}{24.000}{20.000}{}}
\put(0,0){\VectorBoson{1.000}{12.000}{4.000}{12.000}{}}
\put(0,0){\VectorBoson{11.000}{12.000}{14.000}{12.000}{}}
\put(0,0){\VectorBoson{21.000}{12.000}{24.000}{12.000}{}}
\put(0,0){\Fermion{8.560}{25.000}{4.000}{28.000}{+}}
\put(0,0){\Fermion{14.000}{28.000}{18.520}{31.000}{+}}
\put(0,0){\Fermion{28.520}{25.000}{24.000}{28.000}{+}}
\put(0,0){\Fermion{4.000}{19.960}{8.520}{23.000}{+}}
\put(0,0){\Fermion{8.520}{9.000}{4.000}{12.000}{+}}
\put(0,0){\Fermion{14.000}{12.000}{18.520}{15.000}{+}}
\put(0,0){\Fermion{4.000}{28.000}{7.000}{30.000}{}}
\put(0,0){\Fermion{14.000}{28.000}{17.000}{26.000}{}}
\put(0,0){\Fermion{24.000}{28.000}{27.000}{30.000}{}}
\put(0,0){\Fermion{4.000}{19.960}{7.000}{18.000}{}}
\put(0,0){\Fermion{14.000}{20.000}{17.000}{21.960}{}}
\put(0,0){\Fermion{24.000}{20.000}{27.000}{18.000}{}}
\put(0,0){\Fermion{4.000}{12.000}{7.000}{14.000}{}}
\put(0,0){\Fermion{14.000}{12.000}{17.000}{9.960}{}}
\put(0,0){\Fermion{24.000}{12.000}{25.520}{13.000}{}}
\put(0,0){\Fermion{24.000}{12.000}{25.520}{10.960}{}}
\put(0,0){\Fermion{25.520}{10.960}{25.520}{13.000}{}}
\put(0,0){\Fermion{7.000}{30.000}{8.520}{31.000}{+}}
\put(0,0){\Fermion{18.520}{25.000}{17.000}{26.000}{+}}
\put(0,0){\Fermion{27.000}{30.000}{28.520}{31.000}{+}}
\put(0,0){\Fermion{8.520}{17.000}{7.000}{18.000}{+}}
\put(0,0){\Fermion{17.000}{21.960}{18.520}{23.000}{+}}
\put(0,0){\Fermion{18.520}{17.000}{16.000}{18.600}{+}}
\put(0,0){\Fermion{16.000}{18.600}{14.000}{20.000}{}}
\put(0,0){\Fermion{28.520}{17.000}{27.000}{18.000}{+}}
\put(0,0){\Fermion{26.000}{21.400}{28.520}{23.000}{+}}
\put(0,0){\Fermion{24.000}{20.000}{26.000}{21.400}{}}
\put(0,0){\Fermion{7.000}{14.000}{8.520}{15.000}{+}}
\put(0,0){\Fermion{18.520}{9.000}{17.000}{9.960}{+}}
\put(0,0){\GluonArc{5.000}{28.720}{7.000}{30.000}{181.550}{}}
\put(0,0){\GluonArc{17.000}{26.000}{15.000}{27.360}{184.184}{}}
\put(0,0){\GluonArc{27.000}{30.000}{25.000}{28.720}{183.711}{}}
\put(0,0){\GluonArc{5.000}{19.360}{7.000}{18.000}{180.250}{}}
\put(0,0){\GluonArc{4.520}{12.440}{6.360}{13.600}{182.416}{}}
\put(0,0){\GluonArc{16.400}{10.360}{14.520}{11.640}{181.793}{}}
\put(0,0){\Gluon{6.000}{29.360}{8.560}{28.000}{}}
\put(0,0){\Gluon{18.520}{28.000}{16.000}{26.680}{}}
\put(0,0){\Gluon{27.000}{28.680}{28.520}{28.000}{}}
\put(0,0){\Gluon{8.520}{20.000}{6.920}{19.480}{}}
\put(0,0){\Gluon{16.000}{18.600}{16.000}{21.320}{}}
\put(0,0){\Gluon{17.000}{21.960}{18.520}{21.000}{}}
\put(0,0){\Gluon{26.000}{18.680}{26.000}{21.400}{}}
\put(0,0){\Gluon{28.560}{19.000}{27.000}{18.000}{}}
\put(0,0){\Gluon{7.000}{14.000}{8.520}{13.000}{}}
\put(0,0){\Gluon{18.520}{11.000}{17.000}{9.960}{}}
\put(0,0){\Gluon{28.520}{9.000}{25.520}{10.960}{}}
\put(0,0){\Gluon{25.520}{13.000}{27.000}{14.000}{}}
\put(0,0){\Fermion{27.000}{14.000}{28.520}{15.000}{+}}
\put(0,0){\Fermion{28.520}{13.000}{27.000}{14.000}{+}}
\put(0,0){\VectorBoson{1.000}{4.000}{4.000}{4.000}{}}
\put(0,0){\Fermion{4.000}{4.000}{5.520}{5.000}{}}
\put(0,0){\Fermion{5.520}{5.000}{5.560}{3.000}{}}
\put(0,0){\Fermion{5.560}{3.000}{4.000}{4.000}{}}
\put(0,0){\Gluon{5.520}{5.000}{8.520}{7.000}{}}
\put(0,0){\Gluon{7.000}{2.000}{5.560}{3.000}{}}
\put(0,0){\Fermion{7.000}{2.000}{8.560}{3.000}{+}}
\put(0,0){\Fermion{8.520}{1.000}{7.000}{2.000}{+}}
\put(0,0){\Bullet{4.000}{28.000}}
\put(0,0){\Bullet{5.000}{28.720}}
\put(0,0){\Bullet{6.000}{29.360}}
\put(0,0){\Bullet{14.000}{28.000}}
\put(0,0){\Bullet{15.000}{27.360}}
\put(0,0){\Bullet{16.000}{26.680}}
\put(0,0){\Bullet{17.000}{26.000}}
\put(0,0){\Bullet{7.000}{30.000}}
\put(0,0){\Bullet{24.000}{28.000}}
\put(0,0){\Bullet{25.000}{28.720}}
\put(0,0){\Bullet{27.000}{30.000}}
\put(0,0){\Bullet{27.000}{28.680}}
\put(0,0){\Bullet{26.000}{21.400}}
\put(0,0){\Bullet{24.000}{20.000}}
\put(0,0){\Bullet{26.000}{18.680}}
\put(0,0){\Bullet{27.000}{18.000}}
\put(0,0){\Bullet{27.000}{14.000}}
\put(0,0){\Bullet{25.520}{13.000}}
\put(0,0){\Bullet{25.520}{10.960}}
\put(0,0){\Bullet{24.000}{12.000}}
\put(0,0){\Bullet{14.000}{12.000}}
\put(0,0){\Bullet{14.520}{11.640}}
\put(0,0){\Bullet{16.400}{10.360}}
\put(0,0){\Bullet{17.000}{9.960}}
\put(0,0){\Bullet{17.000}{21.960}}
\put(0,0){\Bullet{16.000}{21.320}}
\put(0,0){\Bullet{16.000}{18.600}}
\put(0,0){\Bullet{14.000}{20.000}}
\put(0,0){\Bullet{4.000}{19.960}}
\put(0,0){\Bullet{5.000}{19.360}}
\put(0,0){\Bullet{7.000}{18.000}}
\put(0,0){\Bullet{6.920}{19.480}}
\put(0,0){\Bullet{7.000}{14.000}}
\put(0,0){\Bullet{6.360}{13.600}}
\put(0,0){\Bullet{4.520}{12.440}}
\put(0,0){\Bullet{4.000}{12.000}}
\put(0,0){\Bullet{5.520}{5.000}}
\put(0,0){\Bullet{4.000}{4.000}}
\put(0,0){\Bullet{5.560}{3.000}}
\put(0,0){\Bullet{7.000}{2.000}}
\end{picture}
    \caption{\label{fig:f6}}
  \end{center}
\end{figure}
%
%
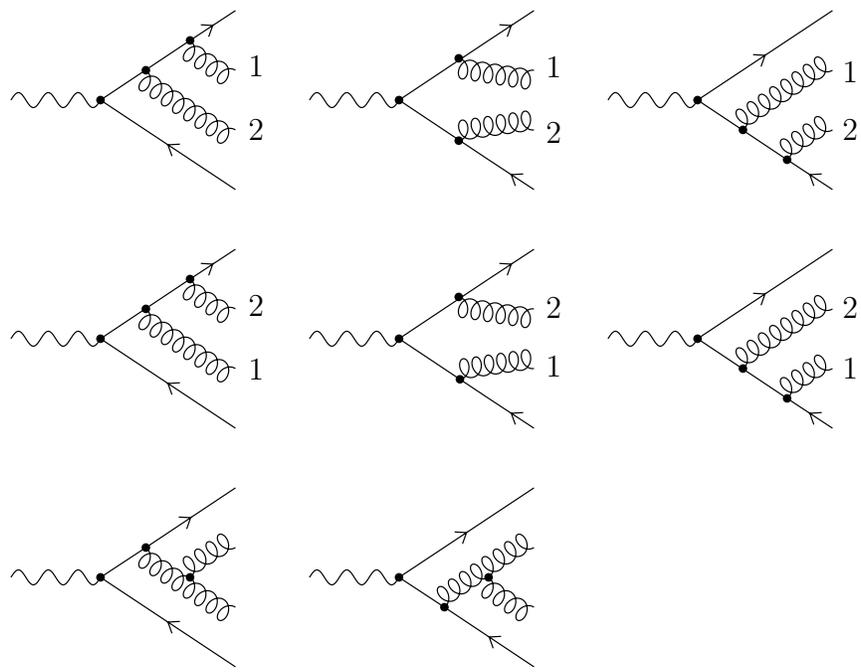
\begin{figure}
  \begin{center}
    \noindent
    \setlength{\unitlength}{25pt}
\renewcommand{\Init}{0.450 0.450 Scale Init}
\begin{picture}(13.500,10.800)
\put(0,0){\VectorBoson{1.000}{20.000}{4.000}{20.000}{}}
\put(0,0){\Fermion{8.520}{17.000}{4.000}{20.000}{+}}
\put(0,0){\Fermion{7.000}{22.000}{8.520}{23.000}{+}}
\put(0,0){\Fermion{4.000}{20.000}{7.000}{22.000}{}}
\put(0,0){\Gluon{7.000}{22.000}{8.520}{21.000}{}}
\put(0,0){\Gluon{5.520}{21.000}{8.520}{19.000}{}}
\put(0,0){\Bullet{4.000}{20.000}}
\put(0,0){\Bullet{5.520}{21.000}}
\put(0,0){\Bullet{7.000}{22.000}}
\put(0,0){\VectorBoson{1.000}{12.000}{4.000}{12.000}{}}
\put(0,0){\Fermion{4.000}{12.000}{7.000}{14.000}{}}
\put(0,0){\Fermion{8.520}{9.000}{4.000}{12.000}{+}}
\put(0,0){\Fermion{7.000}{14.000}{8.520}{15.000}{+}}
\put(0,0){\Gluon{7.000}{14.000}{8.520}{13.000}{}}
\put(0,0){\Gluon{5.520}{13.000}{8.520}{11.000}{}}
\put(0,0){\Bullet{4.000}{12.000}}
\put(0,0){\Bullet{5.520}{13.000}}
\put(0,0){\Bullet{7.000}{14.000}}
\put(0,0){\VectorBoson{11.000}{20.000}{14.000}{20.000}{}}
\put(0,0){\Fermion{14.000}{20.000}{17.000}{22.000}{}}
\put(0,0){\Fermion{17.000}{18.000}{14.000}{20.000}{}}
\put(0,0){\Fermion{17.000}{22.000}{18.520}{23.000}{+}}
\put(0,0){\Fermion{18.520}{17.000}{17.000}{18.000}{+}}
\put(0,0){\Gluon{16.000}{21.400}{18.520}{21.000}{}}
\put(0,0){\Gluon{18.520}{19.000}{16.000}{18.640}{}}
\put(0,0){\Bullet{16.000}{21.400}}
\put(0,0){\Bullet{14.000}{20.000}}
\put(0,0){\Bullet{16.000}{18.640}}
\put(0,0){\VectorBoson{11.000}{12.000}{14.000}{12.000}{}}
\put(0,0){\Fermion{17.000}{14.000}{18.520}{15.000}{+}}
\put(0,0){\Fermion{18.520}{9.000}{17.000}{10.000}{+}}
\put(0,0){\Fermion{14.000}{12.000}{17.000}{14.000}{}}
\put(0,0){\Fermion{17.000}{10.000}{14.000}{12.000}{}}
\put(0,0){\Gluon{16.000}{13.400}{18.520}{13.000}{}}
\put(0,0){\Gluon{18.520}{11.000}{16.040}{10.640}{}}
\put(0,0){\Bullet{14.000}{12.000}}
\put(0,0){\Bullet{16.000}{13.400}}
\put(0,0){\Bullet{16.040}{10.640}}
\put(0,0){\VectorBoson{21.000}{20.000}{24.000}{20.000}{}}
\put(0,0){\VectorBoson{21.000}{12.000}{24.000}{12.000}{}}
\put(0,0){\Fermion{24.000}{12.000}{28.520}{15.000}{+}}
\put(0,0){\Fermion{24.000}{20.000}{28.520}{23.000}{+}}
\put(0,0){\Fermion{28.520}{17.000}{27.000}{18.000}{+}}
\put(0,0){\Fermion{28.520}{9.000}{27.000}{10.000}{+}}
\put(0,0){\Fermion{27.000}{10.000}{24.000}{12.000}{}}
\put(0,0){\Fermion{27.000}{18.000}{24.000}{20.000}{}}
\put(0,0){\Gluon{28.520}{19.000}{27.000}{18.000}{}}
\put(0,0){\Gluon{28.520}{11.000}{27.000}{10.000}{}}
\put(0,0){\Gluon{28.520}{21.000}{25.520}{19.000}{}}
\put(0,0){\Gluon{28.520}{13.000}{25.520}{11.000}{}}
\put(0,0){\Bullet{24.000}{20.000}}
\put(0,0){\Bullet{25.520}{19.000}}
\put(0,0){\Bullet{27.000}{18.000}}
\put(0,0){\Bullet{24.000}{12.000}}
\put(0,0){\Bullet{25.520}{11.000}}
\put(0,0){\Bullet{27.000}{10.000}}
\put(0,0){\Fermion{14.000}{4.000}{18.520}{7.000}{+}}
\put(0,0){\Fermion{8.520}{1.000}{4.000}{4.000}{+}}
\put(0,0){\VectorBoson{1.000}{4.000}{4.000}{4.000}{}}
\put(0,0){\VectorBoson{11.000}{4.000}{14.000}{4.000}{}}
\put(0,0){\Gluon{17.000}{4.000}{18.520}{3.000}{}}
\put(0,0){\Gluon{7.000}{4.000}{8.520}{3.000}{}}
\put(0,0){\Gluon{5.520}{5.000}{7.000}{4.000}{}}
\put(0,0){\Gluon{15.520}{3.000}{17.000}{4.000}{+}}
\put(0,0){\Gluon{17.000}{4.000}{18.520}{5.000}{+}}
\put(0,0){\Gluon{7.000}{4.000}{8.520}{5.000}{+}}
\put(0,0){\Fermion{4.000}{4.000}{5.520}{5.000}{}}
\put(0,0){\Fermion{14.000}{4.000}{15.520}{3.000}{}}
\put(0,0){\Fermion{18.520}{1.000}{15.520}{3.000}{+}}
\put(0,0){\Fermion{5.520}{5.000}{8.520}{7.000}{+}}
\put(0,0){\Bullet{4.000}{4.000}}
\put(0,0){\Bullet{5.520}{5.000}}
\put(0,0){\Bullet{7.000}{4.000}}
\put(0,0){\Bullet{14.000}{4.000}}
\put(0,0){\Bullet{15.520}{3.000}}
\put(0,0){\Bullet{17.000}{4.000}}
\put(4.050,9.396){1}
\put(4.050,8.424){2}
\put(4.050,5.760){2}
\put(4.050,4.806){1}
\put(8.550,9.378){1}
\put(8.550,8.388){2}
\put(8.550,5.742){2}
\put(8.550,4.842){1}
\put(13.032,9.306){1}
\put(13.032,8.424){2}
\put(13.032,5.742){2}
\put(13.032,4.824){1}
\end{picture}
    \caption{\label{fig:f7}}
  \end{center}
\end{figure}
%
%
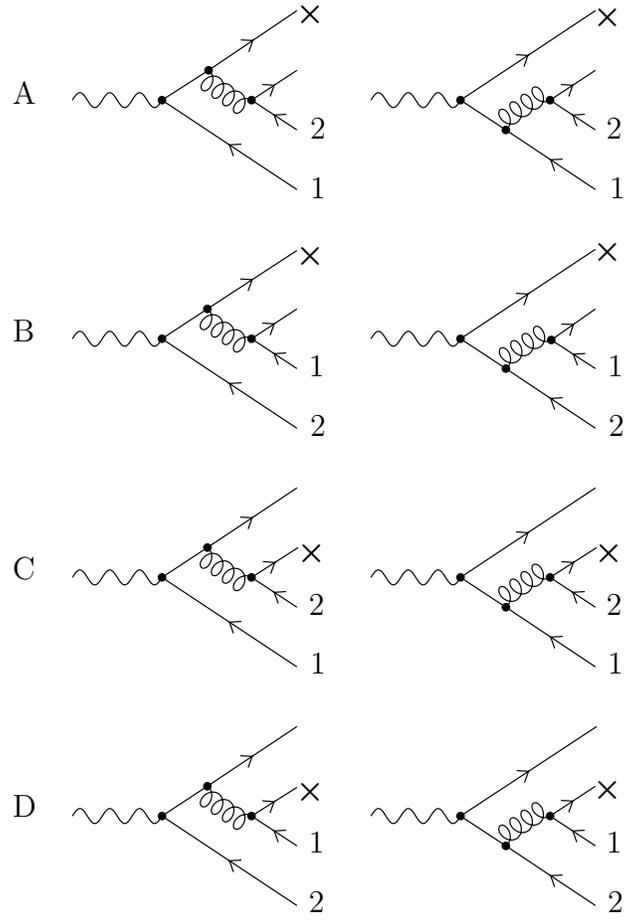
\begin{figure}
  \begin{center}
    \noindent
    \setlength{\unitlength}{25pt}
\renewcommand{\Init}{0.450 0.450 Scale Init}
\begin{picture}(9.900,14.400)
\put(0.450,12.600){A}
\put(0,0){\VectorBoson{3.000}{28.000}{6.000}{28.000}{}}
\put(0,0){\Fermion{10.520}{25.000}{6.000}{28.000}{+}}
\put(0,0){\Fermion{10.520}{27.000}{9.000}{28.000}{+}}
\put(0,0){\Fermion{9.000}{28.000}{10.520}{29.000}{+}}
\put(0,0){\Fermion{6.000}{28.000}{7.560}{29.000}{}}
\put(0,0){\Fermion{7.560}{29.000}{10.520}{31.000}{+}}
\put(0,0){\Gluon{7.560}{29.000}{9.000}{28.000}{}}
\put(0,0){\Bullet{6.000}{28.000}}
\put(0,0){\Bullet{7.560}{29.000}}
\put(0,0){\Bullet{9.000}{28.000}}
\put(0,0){\VectorBoson{13.000}{28.000}{16.000}{28.000}{}}
\put(0,0){\Fermion{16.000}{28.000}{20.520}{31.000}{+}}
\put(0,0){\Fermion{20.520}{25.000}{17.520}{27.000}{+}}
\put(0,0){\Fermion{17.520}{27.000}{16.000}{28.000}{}}
\put(0,0){\Fermion{20.520}{27.000}{19.000}{28.000}{+}}
\put(0,0){\Fermion{19.000}{28.000}{20.520}{29.000}{+}}
\put(0,0){\Gluon{19.000}{28.000}{17.520}{27.000}{}}
\put(0,0){\Bullet{16.000}{28.000}}
\put(0,0){\Bullet{17.520}{27.000}}
\put(0,0){\Bullet{19.000}{28.000}}
\put(4.950,11.160){1}
\put(4.950,12.060){2}
\put(9.486,11.160){1}
\put(9.432,12.060){2}
\put(0,0){\Fermion{16.000}{20.000}{20.520}{23.000}{+}}
\put(0,0){\VectorBoson{13.000}{20.000}{16.000}{20.000}{}}
\put(0,0){\VectorBoson{3.000}{20.000}{6.000}{20.000}{}}
\put(0.450,9.000){B}
\put(0,0){\Fermion{10.520}{17.000}{6.000}{20.000}{+}}
\put(0,0){\Fermion{7.520}{21.000}{10.520}{22.960}{+}}
\put(0,0){\Fermion{9.000}{20.000}{10.520}{21.000}{+}}
\put(0,0){\Fermion{10.520}{19.000}{9.000}{20.000}{+}}
\put(0,0){\Fermion{20.520}{17.000}{17.520}{19.000}{+}}
\put(0,0){\Fermion{19.040}{19.960}{20.520}{21.000}{+}}
\put(0,0){\Fermion{20.520}{19.000}{19.040}{19.960}{+}}
\put(0,0){\Fermion{17.520}{19.000}{16.000}{20.000}{}}
\put(0,0){\Fermion{6.000}{20.000}{7.520}{21.000}{}}
\put(0,0){\Gluon{7.520}{21.000}{9.000}{20.000}{}}
\put(0,0){\Gluon{19.040}{19.960}{17.520}{19.000}{}}
\put(0,0){\Bullet{6.000}{20.000}}
\put(0,0){\Bullet{7.520}{21.000}}
\put(0,0){\Bullet{9.000}{20.000}}
\put(0,0){\Bullet{16.000}{20.000}}
\put(0,0){\Bullet{17.520}{19.000}}
\put(0,0){\Bullet{19.040}{19.960}}
\put(4.950,7.560){2}
\put(4.932,8.460){1}
\put(9.468,7.560){2}
\put(9.450,8.460){1}
\put(0,0){\Fermion{16.000}{12.000}{20.520}{15.000}{+}}
\put(0,0){\Fermion{19.000}{12.000}{20.560}{13.000}{+}}
\put(0,0){\Fermion{20.520}{11.000}{19.000}{12.000}{+}}
\put(0,0){\Fermion{20.520}{9.000}{17.520}{11.000}{+}}
\put(0,0){\Fermion{10.520}{9.000}{6.000}{12.000}{+}}
\put(0,0){\Fermion{7.520}{13.000}{10.520}{15.000}{+}}
\put(0,0){\Fermion{9.000}{12.000}{10.560}{13.000}{+}}
\put(0,0){\Fermion{10.520}{11.000}{9.000}{12.000}{+}}
\put(0,0){\Fermion{6.000}{12.000}{7.520}{13.000}{}}
\put(0,0){\Fermion{16.000}{12.000}{17.520}{11.000}{}}
\put(0,0){\VectorBoson{3.000}{12.000}{6.000}{12.000}{}}
\put(0,0){\VectorBoson{13.000}{12.000}{16.000}{12.000}{}}
\put(0,0){\Gluon{7.520}{13.000}{9.000}{12.000}{}}
\put(0,0){\Gluon{19.000}{12.000}{17.520}{11.000}{}}
\put(0,0){\Bullet{6.000}{12.000}}
\put(0,0){\Bullet{7.520}{13.000}}
\put(0,0){\Bullet{9.000}{12.000}}
\put(0,0){\Bullet{16.000}{12.000}}
\put(0,0){\Bullet{17.520}{11.000}}
\put(0,0){\Bullet{19.000}{12.000}}
\put(0.450,5.400){C}
\put(4.950,3.960){1}
\put(4.932,4.860){2}
\put(9.450,3.960){1}
\put(9.432,4.860){2}
\put(0,0){\Fermion{16.000}{4.000}{20.560}{7.000}{+}}
\put(0,0){\Fermion{20.520}{1.000}{17.520}{3.000}{+}}
\put(0,0){\Fermion{19.000}{4.000}{20.520}{5.000}{+}}
\put(0,0){\Fermion{20.520}{3.000}{19.000}{4.000}{+}}
\put(0,0){\Fermion{10.520}{1.000}{6.000}{4.000}{+}}
\put(0,0){\Fermion{7.520}{5.000}{10.520}{7.000}{+}}
\put(0,0){\Fermion{9.000}{4.000}{10.520}{4.960}{+}}
\put(0,0){\Fermion{10.520}{3.000}{9.000}{4.000}{+}}
\put(0,0){\Fermion{6.000}{4.000}{7.520}{5.000}{}}
\put(0,0){\Fermion{16.000}{4.000}{17.520}{3.000}{}}
\put(0,0){\VectorBoson{13.000}{4.000}{16.000}{4.000}{}}
\put(0,0){\VectorBoson{3.000}{4.000}{6.000}{4.000}{}}
\put(0,0){\Gluon{7.520}{5.000}{9.000}{4.000}{}}
\put(0,0){\Gluon{19.000}{4.000}{17.520}{3.000}{}}
\put(0,0){\Bullet{16.000}{4.000}}
\put(0,0){\Bullet{17.520}{3.000}}
\put(0,0){\Bullet{19.000}{4.000}}
\put(0,0){\Bullet{6.000}{4.000}}
\put(0,0){\Bullet{7.520}{5.000}}
\put(0,0){\Bullet{9.000}{4.000}}
\put(0.450,1.800){D}
\put(4.932,0.360){2}
\put(4.932,1.260){1}
\put(9.450,0.360){2}
\put(9.432,1.260){1}
\put(4.734,13.824){\boldmath$\times$}
\put(4.734,10.170){\boldmath$\times$}
\put(9.234,13.788){\boldmath$\times$}
\put(9.234,10.224){\boldmath$\times$}
\put(4.752,5.670){\boldmath$\times$}
\put(4.734,2.052){\boldmath$\times$}
\put(9.234,2.088){\boldmath$\times$}
\put(9.252,5.670){\boldmath$\times$}
\end{picture}
    \caption{\label{fig:f8}}
  \end{center}
\end{figure}
%
%
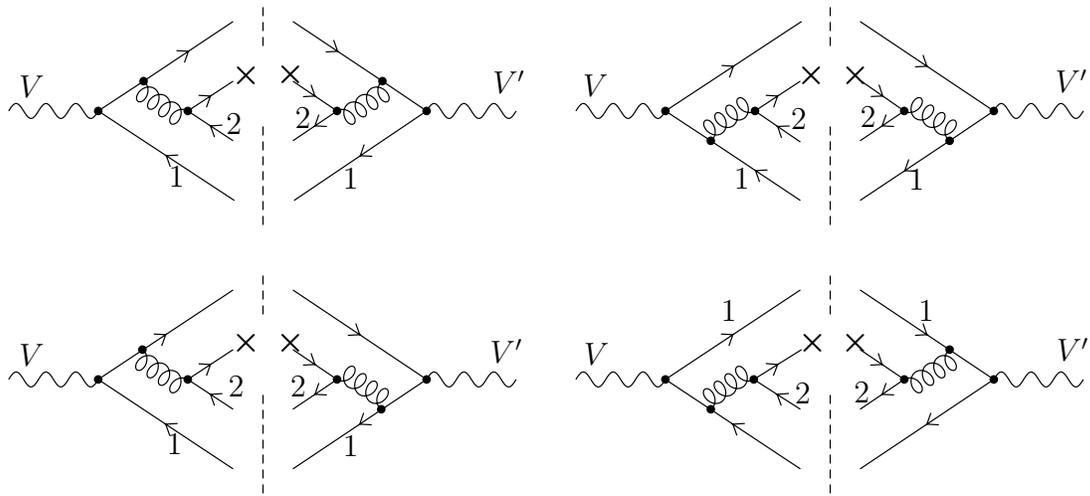
\begin{figure}
  \begin{center}
    \noindent
    \setlength{\unitlength}{25pt}
\renewcommand{\Init}{0.450 0.450 Scale Init}
\begin{picture}(17.100,8.550)
\put(0,0){\VectorBoson{1.000}{14.000}{4.000}{14.000}{}}
\put(0,0){\Fermion{4.000}{14.000}{5.520}{15.000}{}}
\put(0,0){\Fermion{8.560}{11.000}{4.000}{14.000}{+}}
\put(0,0){\Fermion{15.000}{14.000}{10.560}{11.000}{+}}
\put(0,0){\VectorBoson{15.000}{14.000}{18.000}{14.000}{}}
\put(0,0){\Fermion{15.000}{14.000}{13.520}{15.000}{}}
\put(0,0){\Bullet{4.000}{14.000}}
\put(0,0){\Bullet{15.000}{14.000}}
\put(2.880,5.184){1}
\put(5.508,5.148){1}
\put(0,0){\VectorBoson{20.000}{14.000}{23.000}{14.000}{}}
\put(0,0){\VectorBoson{34.000}{14.000}{37.000}{14.000}{}}
\put(0,0){\Fermion{23.000}{14.000}{27.560}{17.000}{+}}
\put(0,0){\Fermion{24.520}{13.000}{23.000}{14.000}{}}
\put(0,0){\Fermion{29.520}{17.000}{34.000}{14.000}{+}}
\put(0,0){\Fermion{34.000}{14.000}{32.520}{13.000}{}}
\put(0,0){\Bullet{34.000}{14.000}}
\put(0,0){\Bullet{23.000}{14.000}}
\put(0,0){\Fermion{4.000}{5.000}{8.520}{8.000}{+}}
\put(0,0){\Fermion{8.520}{2.000}{4.000}{5.000}{+}}
\put(0,0){\VectorBoson{1.000}{5.000}{4.000}{5.000}{}}
\put(0,0){\Gluon{5.480}{6.000}{7.000}{5.000}{}}
\put(0,0){\Fermion{7.000}{5.000}{8.520}{6.000}{+}}
\put(0,0){\Fermion{8.520}{4.000}{7.000}{5.000}{+}}
\put(3.834,2.700){\boldmath$\times$}
\put(0,0){\ScaGho{9.520}{8.480}{9.520}{7.000}{}}
\put(0,0){\ScaGho{9.520}{4.480}{9.520}{1.000}{}}
\put(0,0){\VectorBoson{15.000}{5.000}{18.000}{5.000}{}}
\put(0,0){\Fermion{10.520}{8.000}{15.000}{5.000}{+}}
\put(0,0){\Fermion{15.000}{5.000}{10.520}{2.000}{+}}
\put(0,0){\Gluon{13.480}{4.000}{12.000}{5.000}{}}
\put(0,0){\Fermion{12.000}{5.000}{10.520}{4.000}{+}}
\put(0,0){\Fermion{10.520}{6.000}{12.000}{5.000}{+}}
\put(4.500,2.700){\boldmath$\times$}
\put(0,0){\Bullet{4.000}{5.000}}
\put(0,0){\Bullet{5.480}{6.000}}
\put(0,0){\Bullet{7.000}{5.000}}
\put(0,0){\Bullet{12.000}{5.000}}
\put(0,0){\Bullet{13.480}{4.000}}
\put(0,0){\Bullet{15.000}{5.000}}
\put(2.844,1.134){1}
\put(5.508,1.098){1}
\put(3.780,1.998){2}
\put(4.716,1.998){2}
\put(0,0){\VectorBoson{20.000}{5.000}{23.000}{5.000}{}}
\put(0,0){\VectorBoson{34.000}{5.000}{37.000}{5.000}{}}
\put(0,0){\Fermion{23.000}{5.000}{27.520}{8.000}{+}}
\put(0,0){\Fermion{29.520}{8.000}{34.000}{5.000}{+}}
\put(0,0){\Fermion{34.000}{5.000}{29.520}{2.000}{+}}
\put(0,0){\Fermion{27.520}{2.000}{23.000}{5.000}{+}}
\put(0,0){\Gluon{25.960}{5.000}{24.520}{4.000}{}}
\put(0,0){\Gluon{31.000}{5.000}{32.520}{6.000}{}}
\put(0,0){\Fermion{25.960}{5.000}{27.560}{6.000}{+}}
\put(0,0){\Fermion{27.520}{4.000}{25.960}{5.000}{+}}
\put(0,0){\Fermion{29.520}{6.000}{31.000}{5.000}{+}}
\put(0,0){\Fermion{31.000}{5.000}{29.520}{4.000}{+}}
\put(0,0){\ScaGho{28.520}{4.480}{28.520}{1.000}{}}
\put(0,0){\ScaGho{28.520}{8.480}{28.520}{7.000}{}}
\put(0,0){\Bullet{23.000}{5.000}}
\put(0,0){\Bullet{24.520}{4.000}}
\put(0,0){\Bullet{25.960}{5.000}}
\put(0,0){\Bullet{31.000}{5.000}}
\put(0,0){\Bullet{32.520}{6.000}}
\put(0,0){\Bullet{34.000}{5.000}}
\put(12.402,2.700){\boldmath$\times$}
\put(13.050,2.700){\boldmath$\times$}
\put(11.232,3.150){1}
\put(14.220,3.150){1}
\put(12.348,1.890){2}
\put(13.248,1.908){2}
\put(0,0){\Fermion{5.520}{15.000}{8.520}{17.000}{+}}
\put(0,0){\Fermion{10.520}{17.000}{13.520}{15.000}{+}}
\put(0,0){\Fermion{7.000}{14.000}{8.520}{15.000}{+}}
\put(0,0){\Fermion{8.520}{13.000}{7.000}{14.000}{+}}
\put(0,0){\Fermion{10.520}{15.000}{12.000}{14.000}{+}}
\put(0,0){\Fermion{12.000}{14.000}{10.520}{13.000}{+}}
\put(0,0){\ScaGho{9.520}{17.480}{9.520}{16.000}{}}
\put(0,0){\ScaGho{9.520}{13.480}{9.520}{10.000}{}}
\put(3.834,6.750){\boldmath$\times$}
\put(4.500,6.768){\boldmath$\times$}
\put(3.744,5.976){2}
\put(4.788,6.066){2}
\put(0,0){\Fermion{27.520}{11.000}{24.520}{13.000}{+}}
\put(0,0){\Fermion{32.520}{13.000}{29.520}{11.000}{+}}
\put(0,0){\Fermion{26.000}{14.000}{27.520}{15.000}{+}}
\put(0,0){\Fermion{27.520}{12.960}{26.000}{14.000}{+}}
\put(0,0){\Gluon{5.520}{15.000}{7.000}{14.000}{}}
\put(0,0){\Gluon{12.000}{14.000}{13.520}{15.000}{}}
\put(0,0){\Bullet{5.520}{15.000}}
\put(0,0){\Bullet{7.000}{14.000}}
\put(0,0){\Bullet{12.000}{14.000}}
\put(0,0){\Bullet{13.520}{15.000}}
\put(0,0){\Fermion{29.560}{15.000}{31.000}{14.000}{+}}
\put(0,0){\Fermion{31.000}{14.000}{29.520}{13.000}{+}}
\put(0,0){\ScaGho{28.520}{17.480}{28.520}{16.000}{}}
\put(0,0){\ScaGho{28.520}{13.480}{28.520}{10.000}{}}
\put(0,0){\Gluon{26.000}{14.000}{24.520}{13.000}{}}
\put(0,0){\Gluon{32.520}{13.000}{31.000}{14.000}{}}
\put(0,0){\Bullet{24.520}{13.000}}
\put(0,0){\Bullet{26.000}{14.000}}
\put(0,0){\Bullet{31.000}{14.000}}
\put(0,0){\Bullet{32.520}{13.000}}
\put(12.384,6.750){\boldmath$\times$}
\put(13.050,6.750){\boldmath$\times$}
\put(12.294,6.048){2}
\put(13.338,6.084){2}
\put(11.430,5.148){1}
\put(14.076,5.130){1}
\put(0.612,6.534){$V$}
\put(0.612,2.466){$V$}
\put(7.776,6.588){$V'$}
\put(7.740,2.520){$V'$}
\put(9.144,6.552){$V$}
\put(9.162,2.484){$V$}
\put(16.290,6.606){$V'$}
\put(16.308,2.538){$V'$}
\end{picture}
    \caption{\label{fig:f9}}
  \end{center}
\end{figure}
%
%
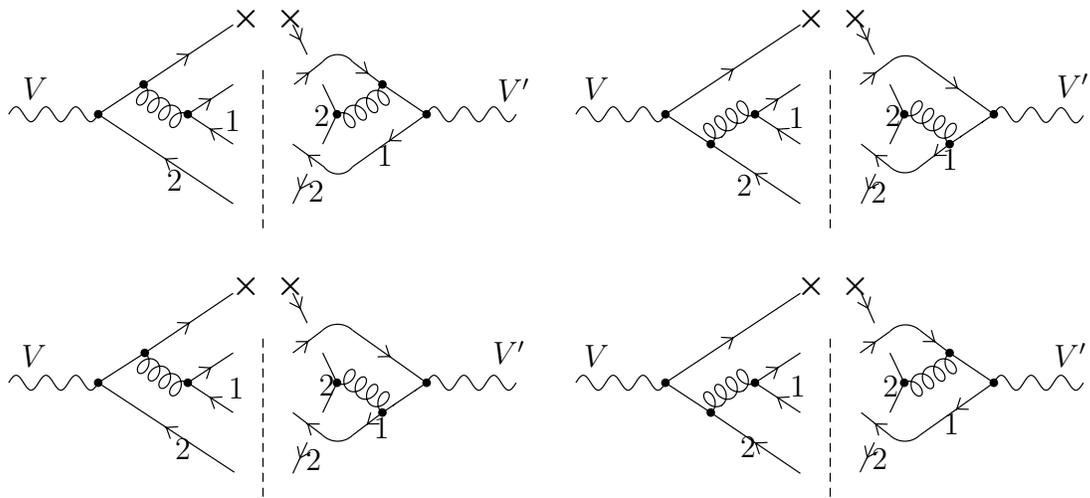
\begin{figure}
  \begin{center}
    \noindent
    \setlength{\unitlength}{25pt}
\renewcommand{\Init}{0.450 0.450 Scale Init}
\begin{picture}(17.100,9.000)
\put(0,0){\VectorBoson{1.000}{15.000}{4.000}{15.000}{}}
\put(0,0){\VectorBoson{15.000}{15.000}{18.000}{15.000}{}}
\put(0,0){\Fermion{8.520}{12.000}{4.000}{15.000}{+}}
\put(0,0){\Fermion{5.520}{16.000}{8.520}{18.000}{+}}
\put(0,0){\Fermion{7.000}{15.000}{8.520}{16.000}{+}}
\put(0,0){\Fermion{8.520}{14.000}{7.000}{15.000}{+}}
\put(0,0){\Fermion{12.520}{16.760}{13.520}{16.000}{+}}
\put(0,0){\Fermion{15.000}{15.000}{12.520}{13.240}{+}}
\put(0,0){\Fermion{11.560}{13.240}{10.520}{14.000}{+}}
\put(0,0){\Fermion{10.560}{16.000}{11.520}{16.760}{+}}
\put(0,0){\Fermion{11.000}{13.000}{10.520}{12.000}{+}}
\put(0,0){\Fermion{10.520}{18.000}{11.000}{17.000}{+}}
\put(0,0){\Fermion{4.000}{15.000}{5.520}{16.000}{}}
\put(0,0){\Fermion{11.520}{16.000}{12.000}{15.000}{}}
\put(0,0){\Fermion{12.000}{15.000}{11.520}{14.000}{}}
\put(0,0){\Fermion{13.520}{16.000}{15.000}{15.000}{}}
\put(0,0){\Gluon{5.520}{16.000}{7.000}{15.000}{}}
\put(0,0){\Gluon{12.000}{15.000}{13.520}{16.000}{}}
\put(0,0){\FermionArc{11.560}{13.240}{12.520}{13.240}{-106.260}{}}
\put(0,0){\FermionArc{11.520}{16.760}{12.520}{16.760}{102.564}{}}
\put(0,0){\ScaGho{9.520}{16.480}{9.520}{11.000}{}}
\put(0,0){\VectorBoson{20.000}{15.000}{23.000}{15.000}{}}
\put(0,0){\VectorBoson{34.000}{15.000}{37.000}{14.960}{}}
\put(0,0){\Fermion{23.000}{15.000}{27.520}{18.000}{+}}
\put(0,0){\Fermion{29.520}{18.000}{30.000}{17.000}{+}}
\put(0,0){\Fermion{31.560}{16.760}{34.000}{15.000}{+}}
\put(0,0){\Fermion{32.520}{14.000}{31.520}{13.240}{+}}
\put(0,0){\Fermion{30.520}{13.240}{29.560}{14.000}{+}}
\put(0,0){\Fermion{29.520}{16.000}{30.520}{16.760}{+}}
\put(0,0){\Fermion{26.000}{15.000}{27.520}{16.000}{+}}
\put(0,0){\Fermion{27.520}{14.000}{26.000}{15.000}{+}}
\put(0,0){\Fermion{27.520}{12.000}{24.520}{14.000}{+}}
\put(0,0){\Fermion{30.040}{13.000}{29.560}{12.000}{+}}
\put(0,0){\Fermion{23.000}{15.000}{24.520}{14.000}{}}
\put(0,0){\Fermion{34.000}{15.000}{32.520}{14.000}{}}
\put(0,0){\Fermion{30.520}{16.000}{31.000}{15.000}{}}
\put(0,0){\Fermion{31.000}{15.000}{30.520}{14.000}{}}
\put(0,0){\FermionArc{30.520}{16.760}{31.560}{16.760}{84.150}{}}
\put(0,0){\FermionArc{31.520}{13.240}{30.520}{13.240}{87.206}{}}
\put(0,0){\ScaGho{28.520}{16.480}{28.520}{11.000}{}}
\put(0,0){\Gluon{26.000}{15.000}{24.520}{14.000}{}}
\put(0,0){\Gluon{32.520}{14.000}{31.000}{15.000}{}}
\put(0,0){\VectorBoson{1.000}{6.000}{4.000}{6.000}{}}
\put(0,0){\VectorBoson{15.000}{6.000}{18.000}{6.000}{}}
\put(0,0){\VectorBoson{20.000}{6.000}{23.000}{6.000}{}}
\put(0,0){\VectorBoson{34.000}{6.000}{37.000}{6.000}{}}
\put(0,0){\Fermion{5.560}{7.000}{8.520}{9.000}{+}}
\put(0,0){\Fermion{8.560}{3.000}{4.000}{6.000}{+}}
\put(0,0){\Fermion{7.000}{6.000}{8.520}{7.000}{+}}
\put(0,0){\Fermion{8.520}{5.000}{7.000}{6.000}{+}}
\put(0,0){\Fermion{12.520}{7.760}{15.000}{6.000}{+}}
\put(0,0){\Fermion{13.520}{5.000}{12.520}{4.240}{+}}
\put(0,0){\Fermion{11.520}{4.240}{10.520}{5.000}{+}}
\put(0,0){\Fermion{11.000}{4.000}{10.520}{2.960}{+}}
\put(0,0){\Fermion{10.520}{9.000}{11.000}{8.000}{+}}
\put(0,0){\Fermion{10.520}{7.000}{11.520}{7.760}{+}}
\put(0,0){\Fermion{4.000}{6.000}{5.560}{7.000}{}}
\put(0,0){\Fermion{15.000}{6.000}{13.520}{5.000}{}}
\put(0,0){\Fermion{12.000}{6.000}{11.520}{5.000}{}}
\put(0,0){\Fermion{11.520}{7.000}{12.000}{6.000}{}}
\put(0,0){\FermionArc{11.520}{7.760}{12.520}{7.760}{87.206}{}}
\put(0,0){\FermionArc{12.520}{4.240}{11.520}{4.240}{87.206}{}}
\put(0,0){\Gluon{5.560}{7.000}{7.000}{6.000}{}}
\put(0,0){\Gluon{13.520}{5.000}{12.000}{6.000}{}}
\put(0,0){\ScaGho{9.520}{7.480}{9.520}{2.000}{}}
\put(0,0){\Fermion{23.000}{6.000}{27.520}{9.000}{+}}
\put(0,0){\Fermion{27.520}{2.960}{24.520}{5.000}{+}}
\put(0,0){\Fermion{26.000}{6.000}{27.520}{7.000}{+}}
\put(0,0){\Fermion{27.520}{5.000}{26.000}{6.000}{+}}
\put(0,0){\Fermion{29.560}{9.000}{30.000}{8.000}{+}}
\put(0,0){\Fermion{30.000}{4.000}{29.520}{3.000}{+}}
\put(0,0){\Fermion{34.000}{6.000}{31.520}{4.240}{+}}
\put(0,0){\Fermion{30.520}{4.240}{29.560}{5.000}{+}}
\put(0,0){\Fermion{29.520}{7.000}{30.520}{7.720}{+}}
\put(0,0){\Fermion{23.000}{6.000}{24.520}{5.000}{}}
\put(0,0){\Fermion{30.520}{7.000}{31.000}{6.000}{}}
\put(0,0){\Fermion{31.000}{6.000}{30.520}{5.000}{}}
\put(0,0){\Fermion{31.520}{7.760}{32.520}{7.000}{+}}
\put(0,0){\Fermion{32.520}{7.000}{34.000}{6.000}{}}
\put(0,0){\Gluon{26.000}{6.000}{24.520}{5.000}{}}
\put(0,0){\Gluon{31.000}{6.000}{32.520}{7.000}{}}
\put(0,0){\FermionArc{30.520}{7.720}{31.520}{7.760}{87.079}{}}
\put(0,0){\FermionArc{31.520}{4.240}{30.520}{4.240}{87.206}{}}
\put(0,0){\ScaGho{28.520}{7.480}{28.520}{2.000}{}}
\put(0,0){\Bullet{4.000}{15.000}}
\put(0,0){\Bullet{5.520}{16.000}}
\put(0,0){\Bullet{7.000}{15.000}}
\put(0,0){\Bullet{12.000}{15.000}}
\put(0,0){\Bullet{15.000}{15.000}}
\put(0,0){\Bullet{13.520}{16.000}}
\put(0,0){\Bullet{15.000}{6.000}}
\put(0,0){\Bullet{13.520}{5.000}}
\put(0,0){\Bullet{12.000}{6.000}}
\put(0,0){\Bullet{7.000}{6.000}}
\put(0,0){\Bullet{5.560}{7.000}}
\put(0,0){\Bullet{4.000}{6.000}}
\put(0,0){\Bullet{26.000}{15.000}}
\put(0,0){\Bullet{24.520}{14.000}}
\put(0,0){\Bullet{23.000}{15.000}}
\put(0,0){\Bullet{31.000}{15.000}}
\put(0,0){\Bullet{32.520}{14.000}}
\put(0,0){\Bullet{34.000}{15.000}}
\put(0,0){\Bullet{34.000}{6.000}}
\put(0,0){\Bullet{32.520}{7.000}}
\put(0,0){\Bullet{31.000}{6.000}}
\put(0,0){\Bullet{26.000}{6.000}}
\put(0,0){\Bullet{24.520}{5.000}}
\put(0,0){\Bullet{23.000}{6.000}}
\put(3.834,8.100){\boldmath$\times$}
\put(4.500,8.100){\boldmath$\times$}
\put(3.834,4.050){\boldmath$\times$}
\put(4.500,4.050){\boldmath$\times$}
\put(12.384,8.100){\boldmath$\times$}
\put(13.050,8.100){\boldmath$\times$}
\put(12.384,4.050){\boldmath$\times$}
\put(13.050,4.050){\boldmath$\times$}
\put(3.726,6.516){1}
\put(6.030,5.994){1}
\put(12.258,6.570){1}
\put(14.580,5.940){1}
\put(3.780,2.466){1}
\put(5.976,1.872){1}
\put(12.276,2.502){1}
\put(14.598,1.926){1}
\put(2.844,5.616){2}
\put(4.986,5.454){2}
\put(5.076,6.588){2}
\put(11.466,5.544){2}
\put(13.500,5.436){2}
\put(13.716,6.588){2}
\put(2.970,1.548){2}
\put(4.950,1.386){2}
\put(5.148,2.520){2}
\put(11.520,1.584){2}
\put(13.518,1.404){2}
\put(13.680,2.502){2}
\put(0.666,7.002){$V$}
\put(0.648,2.952){$V$}
\put(7.848,6.966){$V'$}
\put(7.776,2.988){$V'$}
\put(9.162,7.002){$V$}
\put(9.162,2.952){$V$}
\put(16.308,7.038){$V'$}
\put(16.272,2.952){$V'$}
\end{picture}
    \caption{\label{fig:f10}}
  \end{center}
\end{figure}
%
%
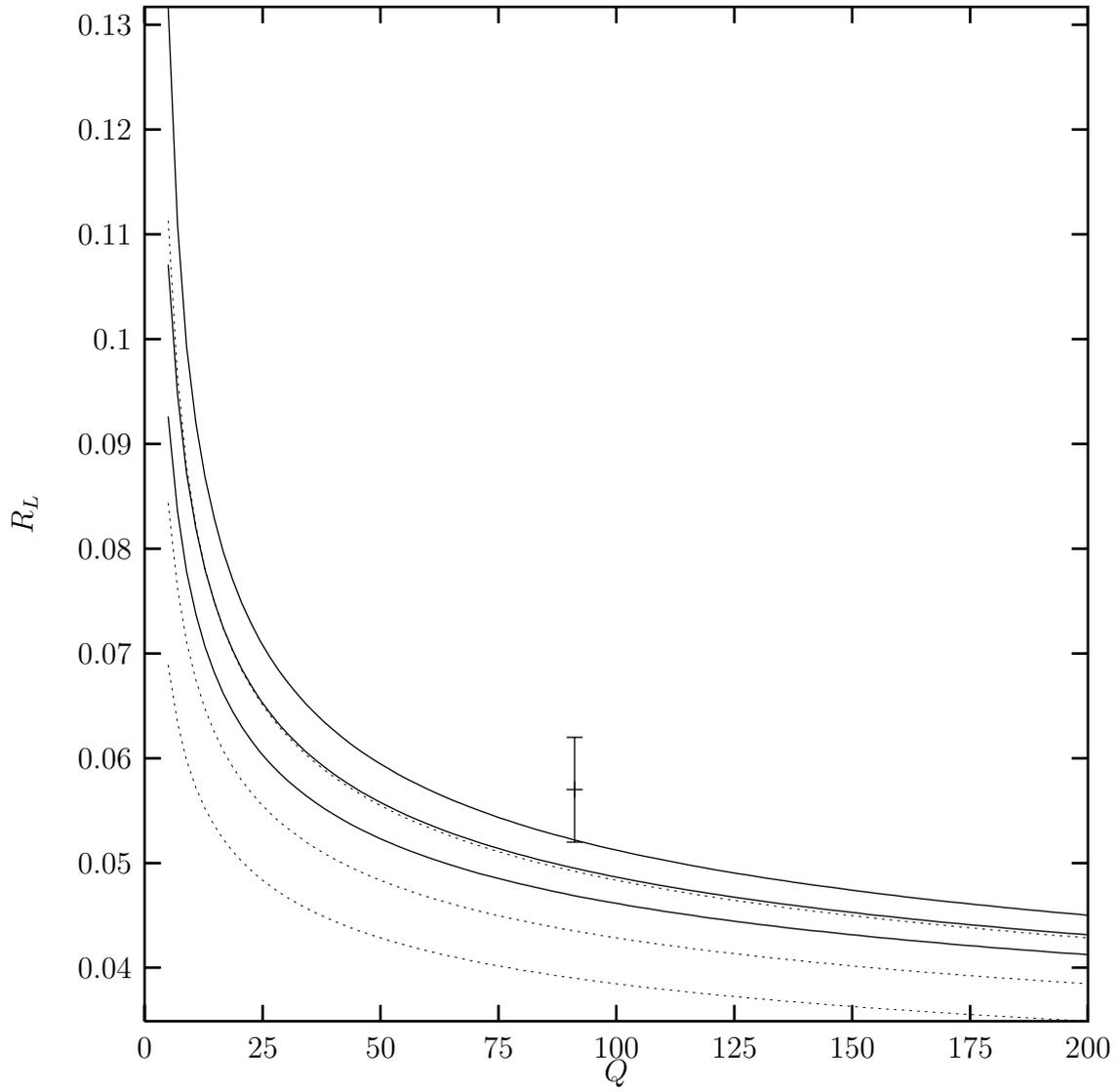
\begin{figure}
  \begin{center}
    \input figure1
  \end{center}
  \caption{\label{fig:f11}}
\end{figure}
%
%
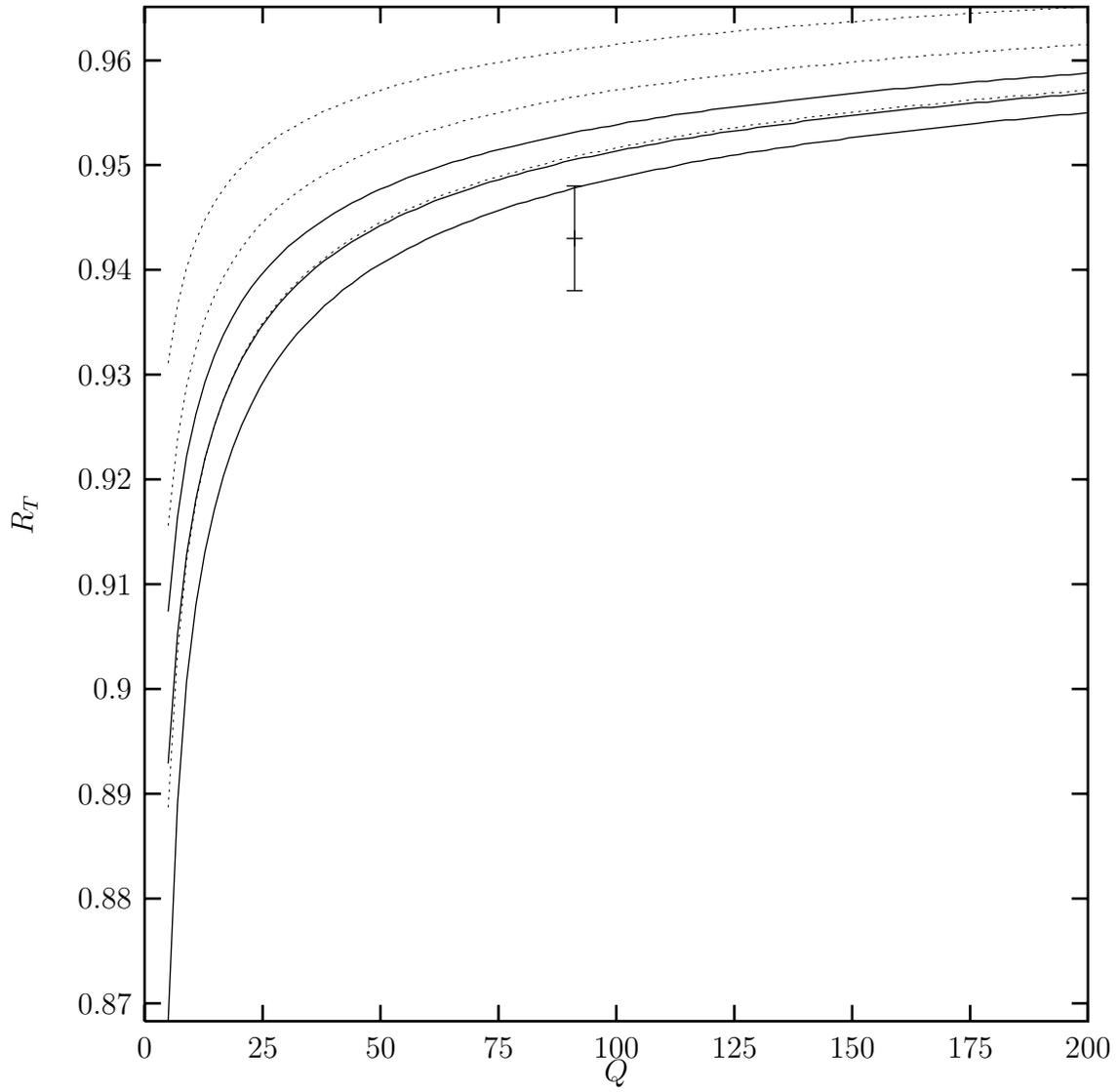
\begin{figure}
  \begin{center}
    \input figure2
  \end{center}
  \caption{\label{fig:f12}}
\end{figure}
%
%
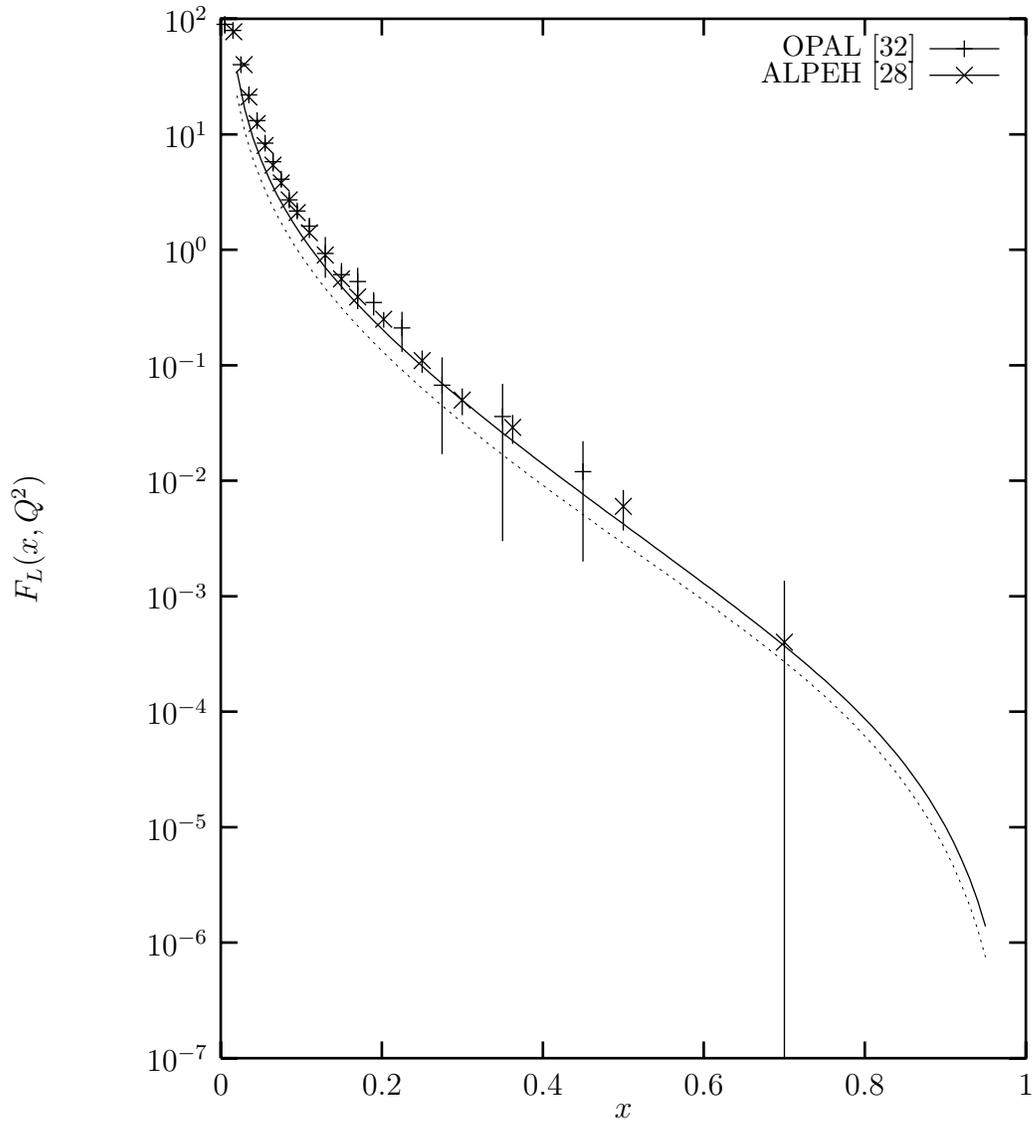
\begin{figure}
  \begin{center}
    \input figure3
  \end{center}
  \caption{\label{fig:f13}}
\end{figure}
%
%
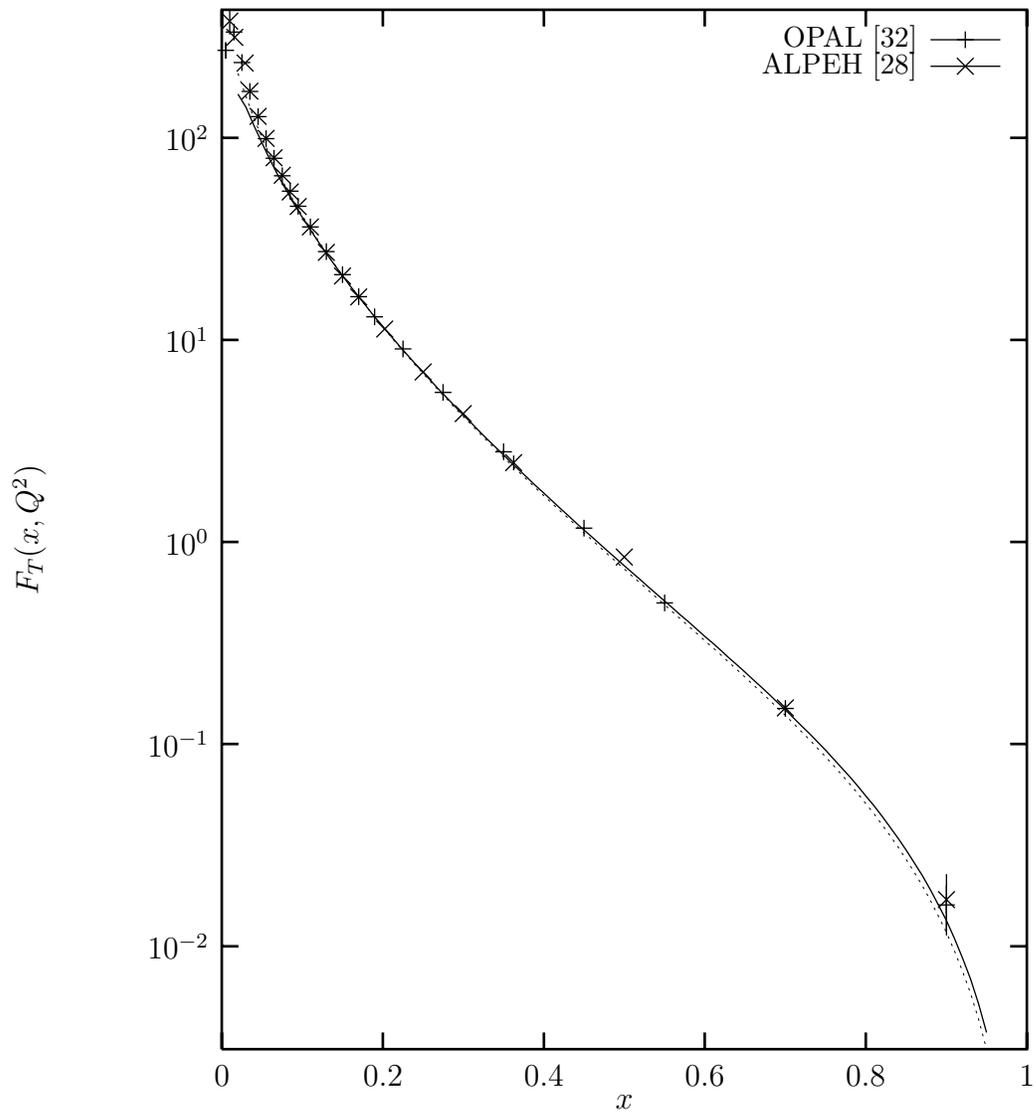
\begin{figure}
  \begin{center}
    \input figure4
  \end{center}
  \caption{\label{fig:f14}}
\end{figure}
%
%
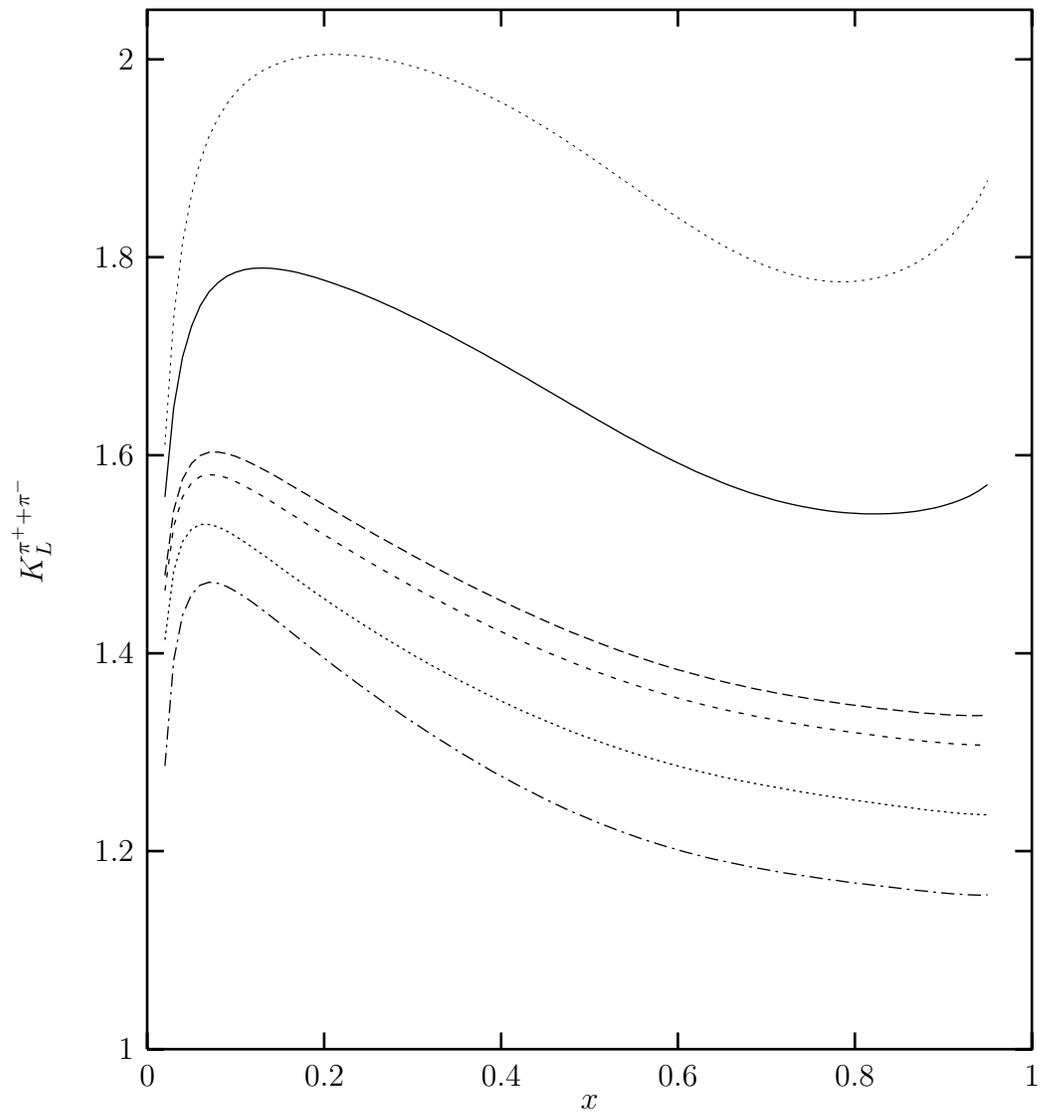
\begin{figure}
  \begin{center}
    \input figure5
  \end{center}
  \caption{\label{fig:f15}}
\end{figure}
%
%
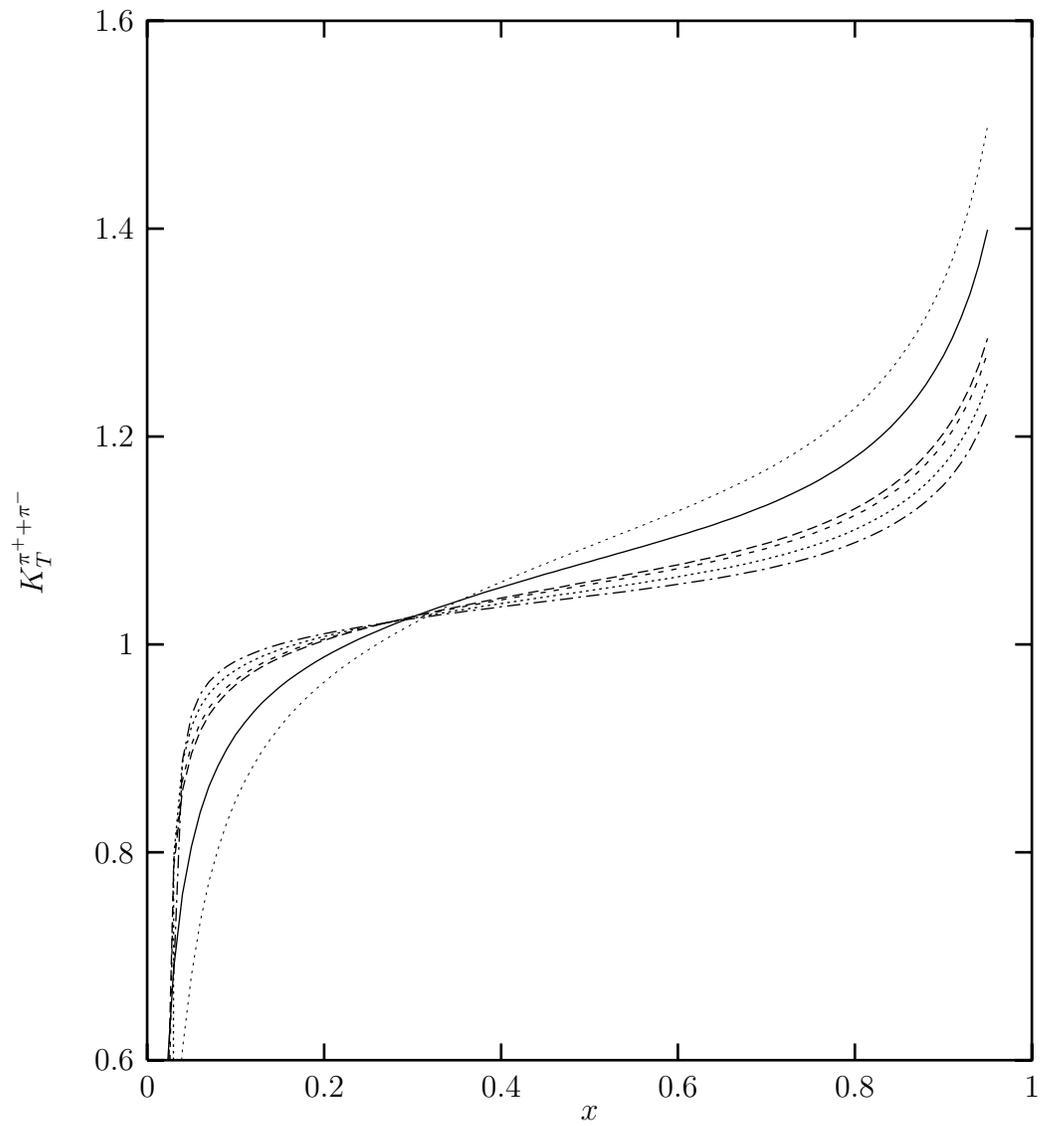
\begin{figure}
  \begin{center}
    \input figure6
  \end{center}
  \caption{\label{fig:f16}}
\end{figure}
%
%
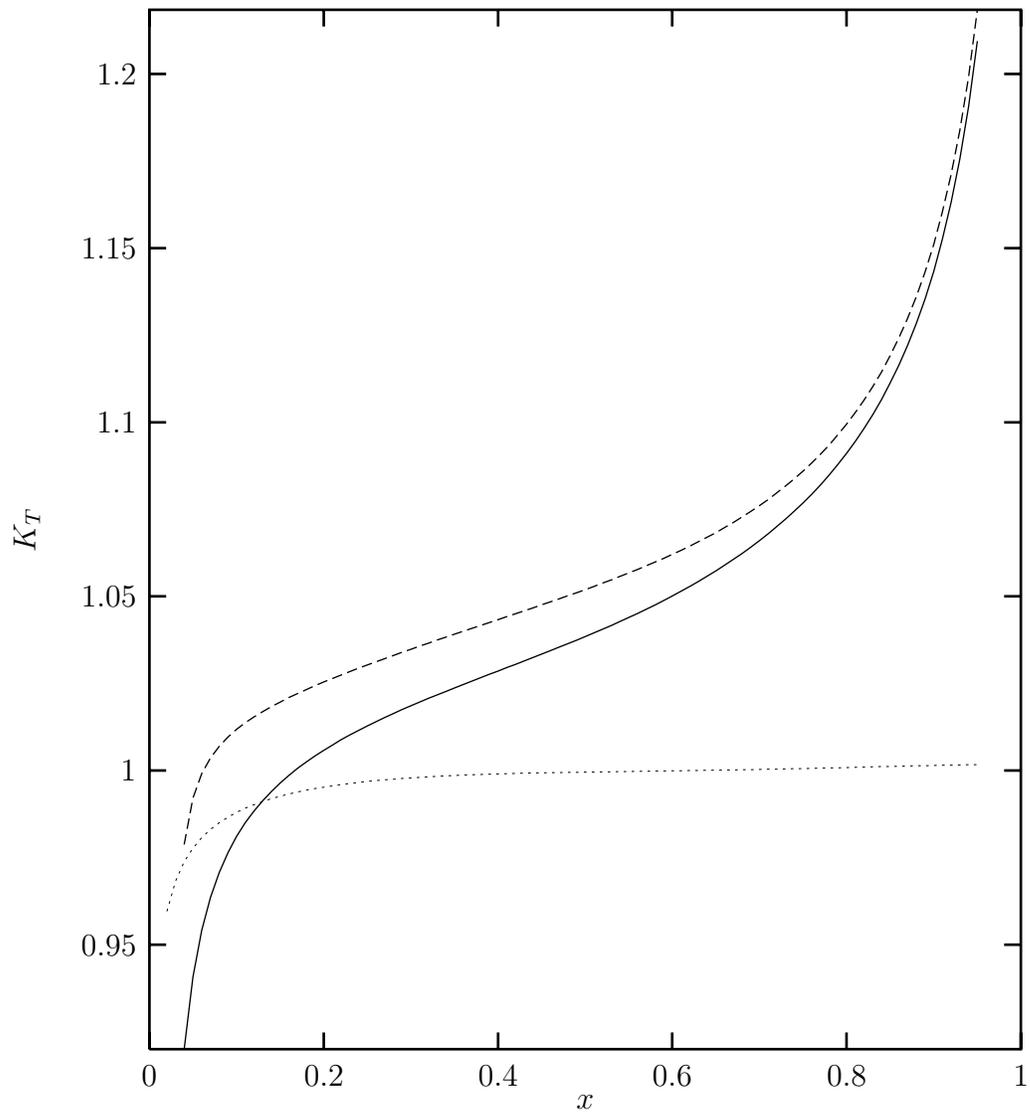
\begin{figure}
  \begin{center}
    \input figure13
  \end{center}
  \caption{\label{fig:f17}}
\end{figure}
%
%
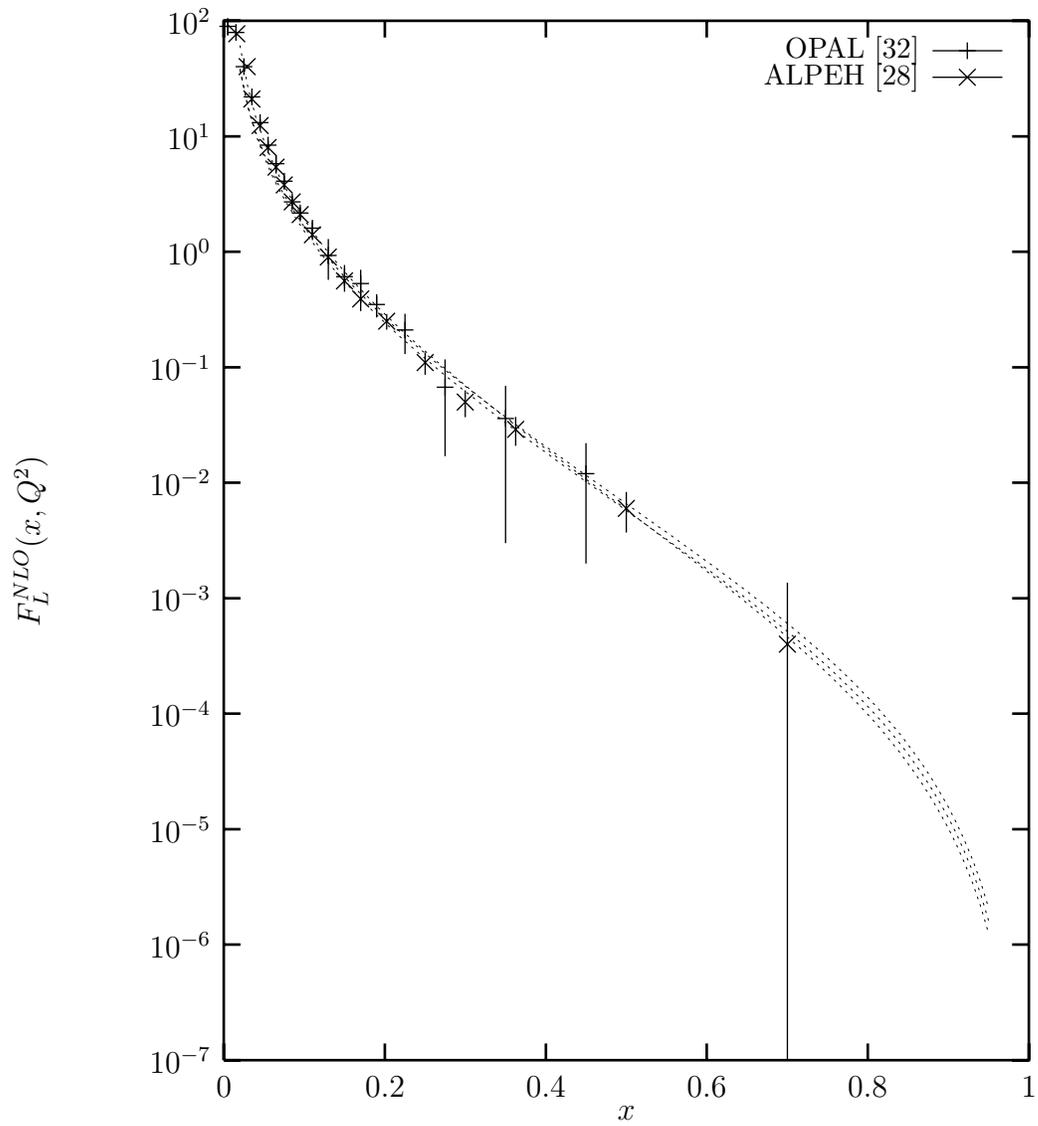
\begin{figure}
  \begin{center}
    \input figure7
  \end{center}
  \caption{\label{fig:f18}}
\end{figure}
\clearpage
%
%
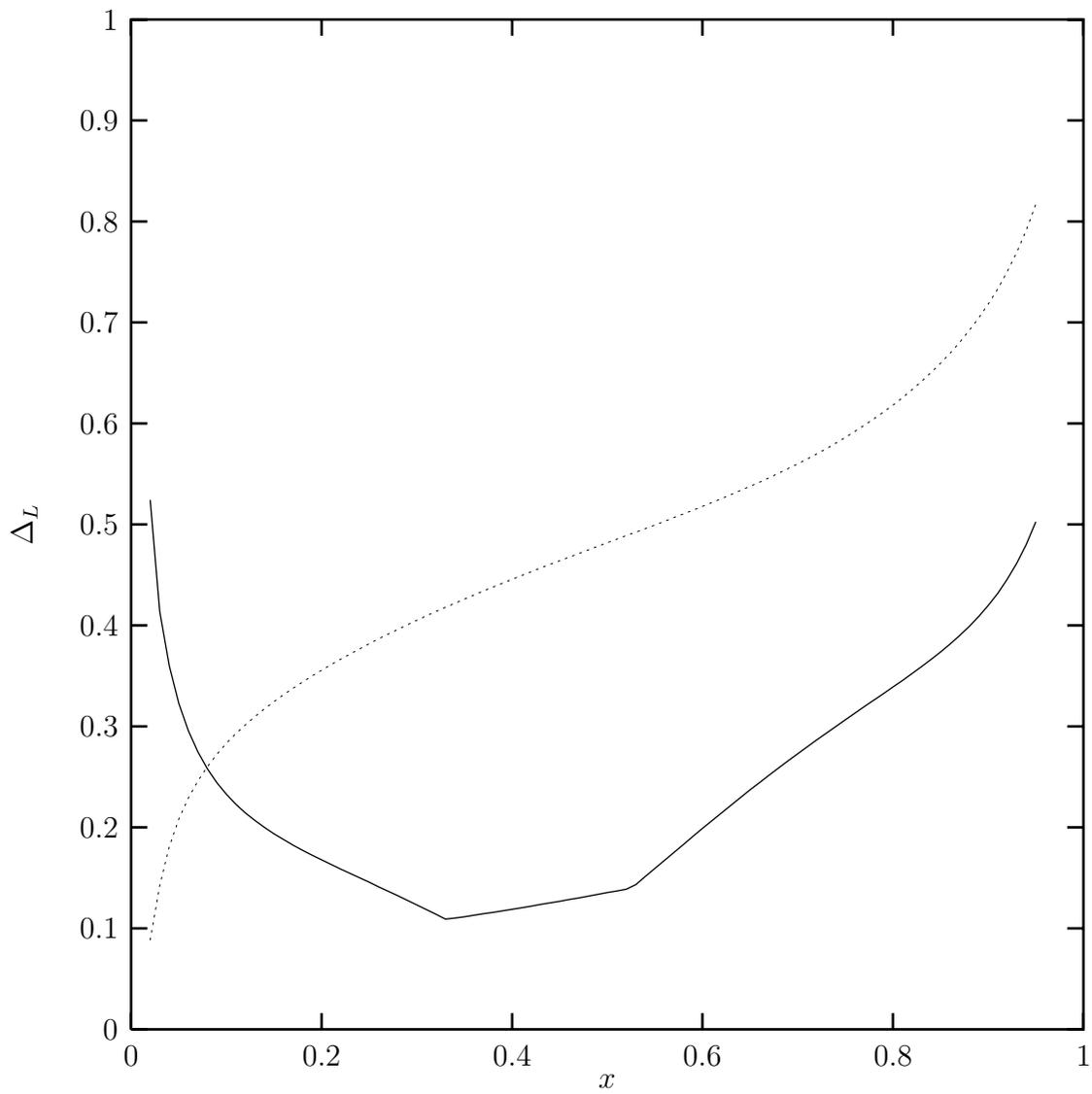
\begin{figure}
  \begin{center}
    \input figure8
  \end{center}
  \caption{\label{fig:f19}}
\end{figure}
%
%
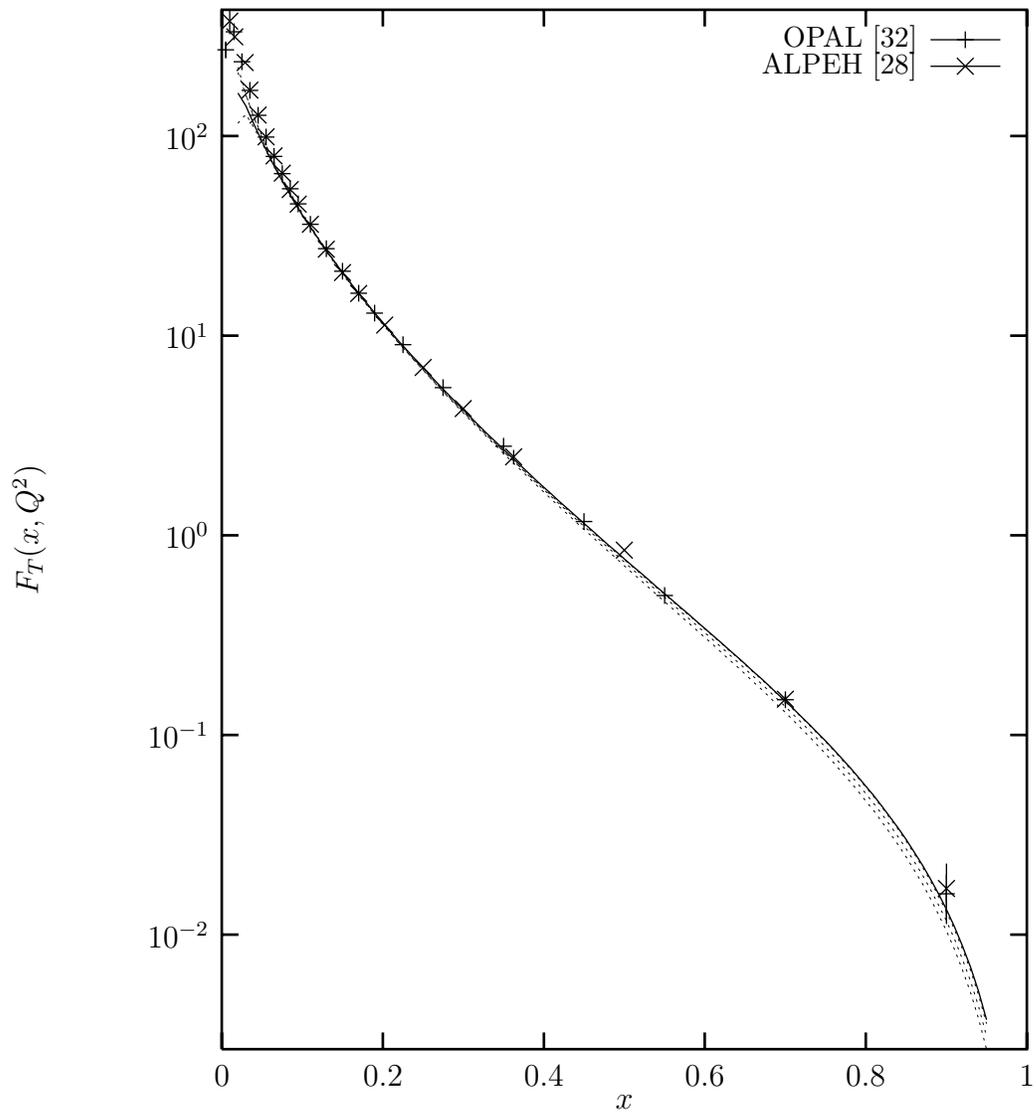
\begin{figure}
  \begin{center}
    \input figure9
  \end{center}
  \caption{\label{fig:f20}}
\end{figure}
%
%
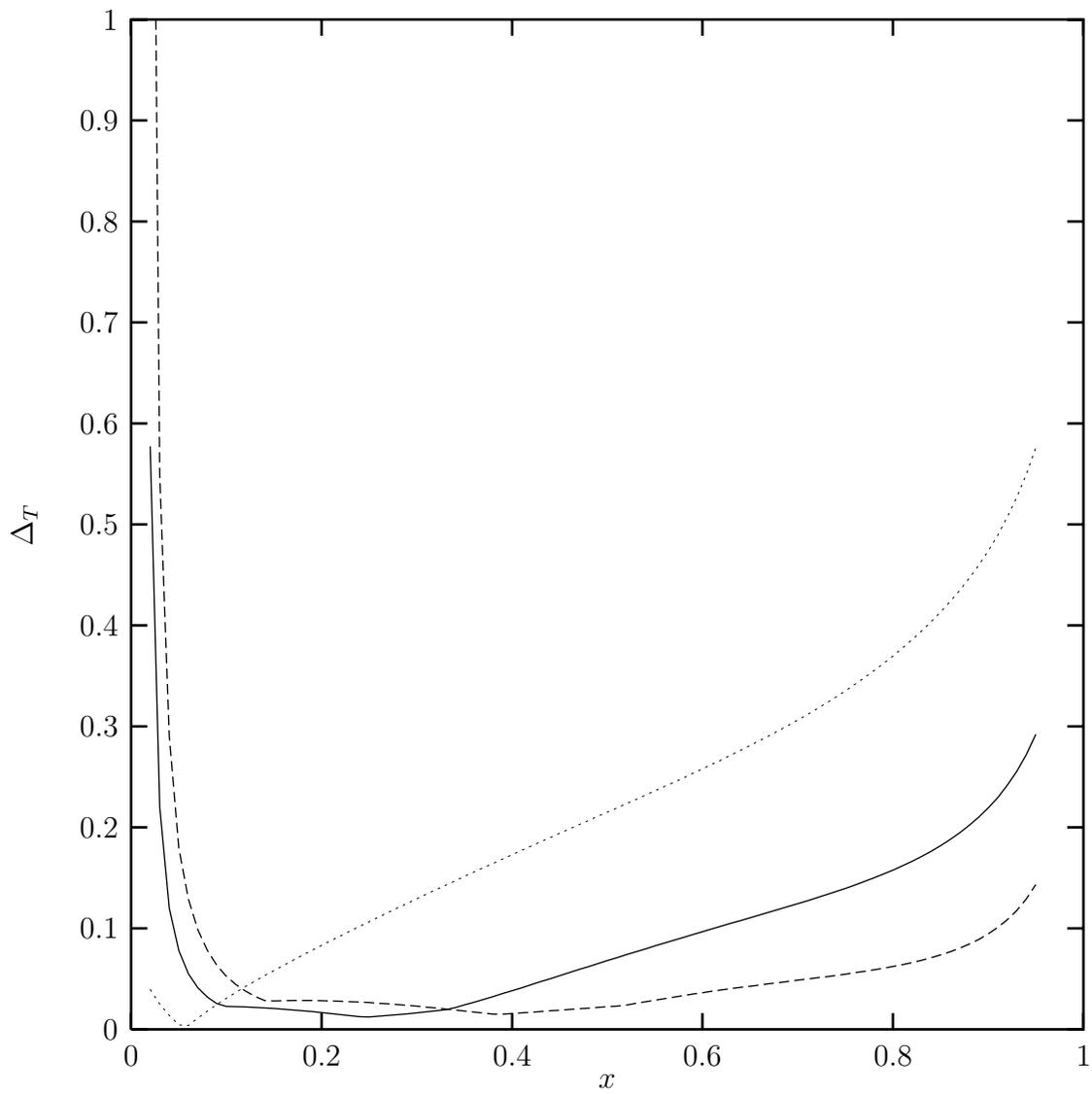
\begin{figure}
  \begin{center}
    \input figure10
  \end{center}
  \caption{\label{fig:f21}}
\end{figure}
%
%
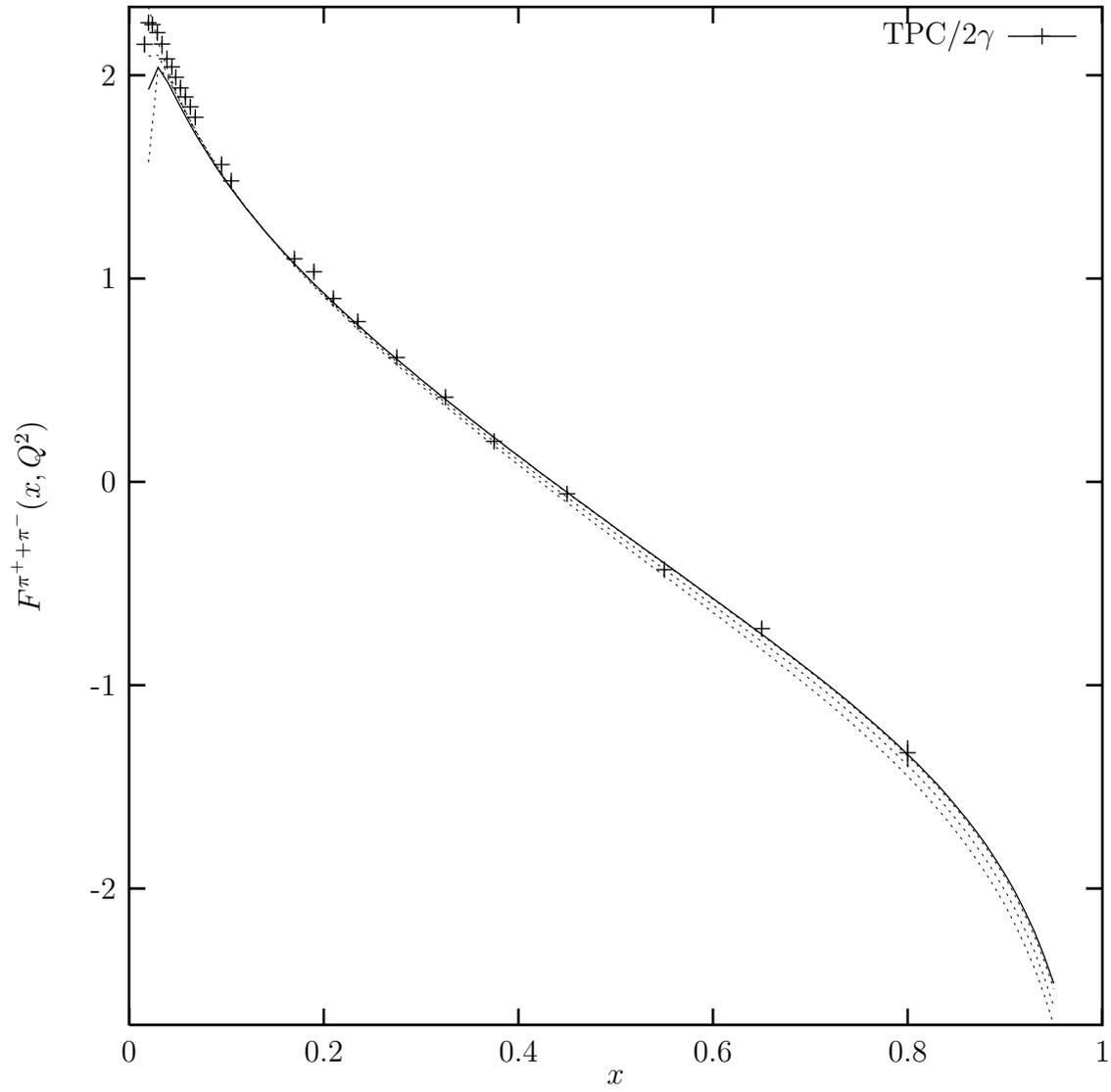
\begin{figure}
  \begin{center}
    \input figure11
  \end{center}
  \caption{\label{fig:f22}}
\end{figure}
%
%
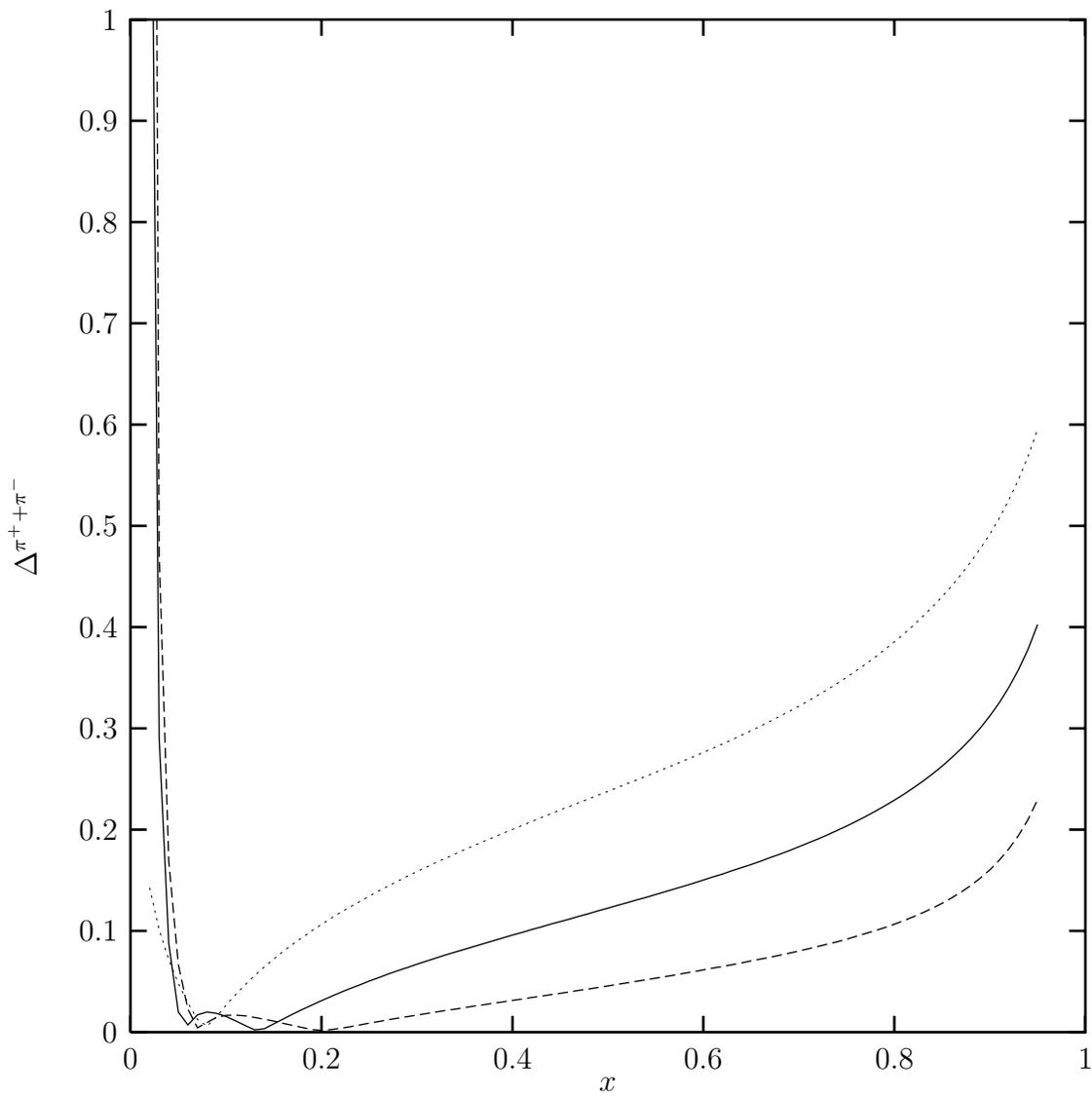
\begin{figure}
  \begin{center}
    \input figure12
  \end{center}
  \caption{\label{fig:f23}}
\end{figure}
\clearpage
\setdefaultlengths

%
\end{document}